\documentclass[useAMS,usenatbib]{mn2e}
\usepackage{times, graphicx, setspace, latexsym, amssymb, amsmath, fancyhdr, caption}

\newcommand{\degree}{\ensuremath{^\circ}}
\title[Polarimetry of HTRU long-period pulsars]{The High Time Resolution Universe Survey - IX: Polarimetry of long-period pulsars}
\author
[C. Tiburzi et~al.]{C. Tiburzi$^{1,2}$\thanks{E-mail:ctiburzi@oa-cagliari.inaf.it}, S. Johnston$^{3}$, M. Bailes$^{4,5}$, S. D. Bates$^{7}$, N. D. R. Bhat$^{4,5}$, M. Burgay$^{1}$,\newauthor S. Burke-Spolaor$^{8}$, D. Champion$^{9}$, P. Coster$^{4,3}$, N. D'Amico$^{1,2}$,  M. J. Keith$^{3}$,  M. Kramer$^{9,6}$,\newauthor L. Levin$^{7}$, S. Milia$^{1}$, C. Ng$^{9}$, A. Possenti$^{1}$, B. W. Stappers$^{6}$, D. Thornton$^{6,3}$, W. van Straten$^{4,5}$\\
\\
$^{1}$INAF - Osservatorio Astronomico di Cagliari, Via della Scienza, 09047 Selargius (CA), Italy\\ 
$^{2}$Dipartimento di Fisica, Universit\`a di Cagliari, Cittadella Universitaria 09042 Monserrato (CA), Italy \\
$^{3}$CSIRO Astronomy \& Space Science, Australia Telescope National Facility, PO Box 76, Epping, NSW 1710, Australia\\
$^{4}$Centre for Astrophysics and Supercomputing, Swinburne University of Technology, Mail H39, PO Box 218, VIC 3122, Australia\\
$^{5}$ARC Centre of Excellence for All-sky Astrophysics\\
$^{6}$Jodrell Bank Centre for Astrophysics, University of Manchester, Alan Turing Building, Oxford Road, Manchester M13 9PL, United Kingdom\\
$^{7}$Department of Physics, West Virginia University, Morgantown, WV 26506, USA\\
$^{8}$Jet Propulsion Laboratory, California Institute of Technology, 4800 Oak
Grove Drive, Pasadena CA 91104 USA\\
$^{9}$Max Planck Institut f\"{u}r Radioastronomie, Auf dem H\"{u}gel 69, 53121 Bonn, Germany\\
}

\begin{document}

\date{}
\pagerange{\pageref{firstpage}--\pageref{lastpage}} \pubyear{}

\maketitle
\label{firstpage}

\begin{abstract}
We present a polarimetric analysis of 49 long-period pulsars discovered as part of the High Time Resolution Universe (HTRU) southern survey. The sources exhibit the typical characteristics of ``old'' pulsars, with low fractional linear and circular polarisation and narrow, multicomponent profiles. Although the position angle swings are generally complex, for two of the analysed pulsars (J1622$-$3751 and J1710$-$2616) we obtained an indication of the geometry via the rotating vector model. We were able to determine a value of the rotation measure (${\rm RM}$) for 34 of the sources which, when combined with their dispersion measures (${\rm DM}$), yields an integrated magnetic field strength along the line of sight. With the data presented here, the total number of values of ${\rm RM}$ associated to pulsars discovered during the HTRU southern survey sums to 51. The ${\rm RMs}$ are not consistent with the hypothesis of a counter-clockwise direction of the Galactic magnetic field within an annulus included between 4 and 6 kpc from the Galactic centre. A partial agreement with a counter-clockwise sense of the Galactic magnetic field within the spiral arms is however found in the area of the Carina-Sagittarius arm.
\end{abstract}


\section{Introduction}
Pulsars are rapidly rotating neutron stars, characterised by an extremely intense magnetic field, whose first order term is a dipole. They are in general broadband emitters powered by their spin-down energy. In particular, at radio wavelengths they are observed to be faint pulsating sources: two emission beams are generated from the magnetic poles due to streams of charged particles that are accelerated along the magnetic field lines.
Assuming a misalignment between the rotational and the magnetic axes, the two radiation beams sweep across the sky. If the observer's line of sight crosses one or both of them, pulsations can be observed.
Although this model is well supported by observations, many details are still not completely understood, e.g. the exact structure of the beams, their time evolution and how the emitting regions are distributed within the beams.

One method used for determining the geometry of the star and the beam structure is polarisation analysis. The radio emission from pulsars is among the most polarised we receive from celestial objects, showing an average degree of linear polarisation $L$ up to 20\% in the sources with a spin-down luminosity ($\dot{E}$) less than $5\times10^{33}$~erg~s$^{-1}$ and exceeding 50\% in those where $\dot{E}>2\times10^{35}$~erg~s$^{-1}$ (see \citealt{gl98,hlk98,wj08}).
Unlike the degree of circular polarisation (smaller but still significant, around 10\%), $L$ is typically anti-correlated with observing frequency (e.g. \citealt{jk08}).

In the ``Rotating Vector Model'' (RVM, \citealp{rc69a}), the linear polarisation orientation (observed as the position angle, ${\rm PA}$) is determined by the orientation of the magnetic field lines. While observing a pulse, the line of sight crosses the magnetic field lines with a continuously changing orientation. Hence, an S-shaped swing in the ${\rm PA}$ value is expected. 
The RVM provides the formalism, whereby the swing shape is determined by the pulsar's geometric parameters (the impact parameter between the line of sight and the magnetic axis and the misalignment between the spin and the magnetic axes, the angle $\alpha$). In turn then, knowledge of the ${\rm PA}$ swing can, in principle, allow a
determination of the geometry (see e.g. \citealt{ew01}).
Moreover, from the delay in longitude between the ${\rm PA}$ swing and the total power peak due to retardation and aberration effects \citep{bcw91,ha01,gg03}, it is possible to infer the emission altitude \citep{hx97a,jw06}.
However, it is rare that the RVM can be applied: often the narrow longitude pulse window prevents a unique determination of the geometry.
Beside this, many ``forbidden'' ${\rm PA}$ profiles are observed, e.g. characterised by flat trends or complex variations in the ${\rm PA}$ profile as abrupt jumps of about $\sim90\degree$. This has been explained (\citealt{brc76}, \citealt{crb78}) with the presence of two orthogonal polarisation modes (OPM, \citealt{ms00}) in pulsar emissions. The changes in the relative amplitude between the two modes induce sudden variations in the ${\rm PA}$ profile. Several cases of non-orthogonal ${\rm PA}$ jumps have also been reported \citep{kjm05}. The occurrence of OPM (and also non orthogonal) ${\rm PA}$ jumps induces a linear depolarisation, due to the incoherent mix of the modes. In particular, old pulsars and millisecond pulsars often display complicated ${\rm PA}$ profiles that are inconsistent with the RVM predictions, see e.g. \citet{jw06,jk08,xkj+98,stc99,ymv11}

Circular polarisation is usually brighter in the centre (or core) of a pulse profile (\citealt{ran93}, \citealt{gl98}). It often shows handedness variability as a function of the pulse longitude \citep{rr90}, and in many cases hand reversal also occurs near the profile centre. This gives a further hint about the location of the magnetic axis, and a possible relationship between the ${\rm PA}$ swing sense and the handedness of the circular polarisation is debated \citep{hmxq98}.
A more reliable connection appears to be between the circularly polarised emission and the OPM jumps. In fact \citet{jk04} showed that a strong correlation occurs between the presence of ${\rm PA}$ jumps and the variation in the handedness of the circular polarisation.
In spite of the complex nature of pulsar polarisation, polarimetry of
pulsars gives a unique insight into the three dimensional structure
of the beam above the polar caps \citep{ran83,lm88,hm01,kj07,bp12}.

Another use of polarisation analysis is in probing the magnetic field structure of the medium crossed by the radiation. A polarised signal that propagates through an ionised and magnetised medium (see \S3) undergoes differential propagation velocity between its (right- and left-handed) components. This effect, known as Faraday rotation, that is a birefringence phenomenon, induces a rotation in the ${\rm PA}$. This is quantified through the rotation measure (${\rm RM}$) parameter, which depends on the ionised medium density and the magnetic field component along the line of sight. For pulsars, the polarised signal passes across three different kinds of ionised and magnetised medium: the pulsar magnetosphere, the Milky Way interstellar medium and Earth ionosphere. In pulsars, we can also quantify the average density of the ionised medium along the line of sight via the dispersion measure (${\rm DM}$) parameter and a combination of the ${\rm RM}$ and ${\rm DM}$ allows a direct measurement of the magnetic field along the line of sight.

Several attempts have been made to apply pulsar polarisation analysis to probe the Galactic magnetic field structure \citep{man72,mt77,tn80,ls89,wck+04}. In particular, the results obtained by \citet{hq94,hmq99,hmlq02,hml+06} and \citet{nj08} suggest that the large scale in the magnetic field structure of the Milky Way disk is compatible with a bi-symmetric spiral, where the magnetic field in the spiral arms is mainly counter-clockwise if seen from the Galactic north, and the field in between the arms is chiefly clockwise \citep{sf83}. On the other hand, the work of \citet{val05} supports a general clockwise orientation of the large scale Galactic magnetic field, with the presence of a counter-clockwise annulus included between 4 and 6 kpc from the Galactic Centre. It is clearly necessary to increase the ${\rm RM}$ sample in order to discriminate among the various hypotheses, and to guard against interstellar medium fluctuations and local turbulence in the magnetic field that could bias the ${\rm RM}$ estimation. An additional complication in this framework is ${\rm RM}$ fluctuation as a function of the pulse longitude. In particular, three sources of additional ${\rm PA}$ rotation beyond the large scale Galactic magnetic field have been identified \citep{lh03,rbr+04,k09,nk09}: the incoherent superposition of quasi-orthogonal polarisation modes, the pulsar magnetosphere and scattering in the interstellar medium. In particular the latter is indicated as the most probable reason for the detected fluctuations.

The southern component of the High Time Resolution Universe survey for pulsars and fast transients (HTRU, \citealt{htruI}) is being carried out at the 64-metre Parkes radio telescope. It is divided into three parts with different integration times depending on the Galactic latitude: low, medium and high. Since its beginning, it has led to the discovery of more than one hundred pulsars. Among them there is a remarkable sample of millisecond pulsars (\citealt{htruII}, \citealt{htruIV}, \citealt{htruVII}, Thornton et al. in preparation, Ng et al. in preparation). However, the majority of them are normal pulsars (\citealt{htruVI}, Ng et al. in preparation). Following the presentation of the millisecond pulsar polarimetry (\citealt{htruIV}, \citealt{htruVII}), in this work we present a systematic polarisation analysis of 48 long-period pulsars discovered in the medium latitude part of the survey, and one discovered in the high latitude part.

\section{Observations and Analysis}
We present the polarisation analysis of a sample of 49 long-period pulsars, whose spin periods range from a few hundred milliseconds to about two and a half seconds. They were all discovered during the mid-latitude part of the HTRU survey \citep{htruI,htruVI} apart from PSR~J1846--4249 (that has been discovered in the high latitude survey and it will be presented in one of the next papers of the HTRU series). PSR~J1237--6725 and PSR~J1539--4835 were originally thought to be new discoveries of the mid-latitude part of the HTRU survey but were first published by \citet{kbm+03} and Eatough et al. before 2010.

After discovery and confirmation, the pulsars were followed-up with with the third Parkes Digital Filterbank, observing them for at least one year to allow the determination of a complete timing solution. The typical length of the timing observations ranges from $\sim100$ to $\sim600$ seconds. The data were acquired over a 256 MHz bandwidth centred at 1369 MHz, split into 1024 frequency channels, each 0.25 MHz wide. The collected samples were folded on-line forming pulse profiles with 1024 bins for all four Stokes parameters in each frequency channel. To calibrate the target pointings for the differential gain and phase between the linear feeds, we made observations of noise diode coupled to the receptors in the feeds.

We reduced the data using the PSRCHIVE software package \citep{hvm04}.
For each individual observation, we first removed the radio frequency
interference from the data. The observations were polarisation-calibrated
using a square wave signal in order to produce true Stokes parameters, and flux-calibrated using an averaged observation of Hydra A. In addition, corrections were made to the polarisation impurity of the feed following the method in \citet{vs04}. Finally, the observations were aligned using the best-fit ephemeris and optimally summed weighting them according to the signal-to-noise ratio (${\rm SNR}$): 

\begin{equation}
\rm wt = \frac{SNR(\it{I})}{rms(\it{I})}
\end{equation}

where ${\rm wt}$ is the weight we applied and ${\rm SNR}(I)$ and ${\rm rms(\it{I})}$ are the signal to noise ratio and the off pulse root mean square of the total intensity profile, respectively.

This resulted in a higher ${\rm SNR}$ for the final profiles with respect to adding profiles based solely on the integration time. The resultant final profile still retained the full frequency information.

The next step was to compute the ${\rm RM}$ when possible.
We collapsed the frequency channels to four, in order to enhance the ${\rm SNR}$, and we computed an average ${\rm PA}$ across the bins of the pulse for each channel as in \citet{nj08}:
\begin{equation}
\rm{PA_{ave}}= \frac{1}{2} \arctan \left(\frac{\sum_{i=n_{\rm{start}}}^{n_{\rm{end}}} U_i}{\sum_{i=n_{\rm{start}}}^{n_{\rm{end}}} Q_i}\right)
\end{equation}
where $Q_i$ and $U_i$ are the Stokes parameters $Q$ and $U$ for the $i_{th}$ bin, and $n_{start}$ and $n_{end}$ are the bins of the pulse edges. In order to calculate the ${\rm PA}$ error bars we first measured the linear polarisation as:
\begin{equation}
L_{\mathrm {meas}} = \sqrt{\left(\sum_{i=n_{\rm{start}}}^{n_{\rm{end}}} U_{i}\right)^2 + \left(\sum_{i=n_{\rm{start}}}^{n_{\rm{end}}} Q_{i}\right)^2 }.
\label{eq:Lmeas}
\end{equation}
Since it is a positive definite quantity, the average value ${\it L}_{\rm meas}$ is biased. We followed the method of \citet{wk74} in order to obtain a better determination of the value of the linear polarisation ${\it L}_{\rm{true}}$:
\begin{equation}
L_{\mathrm {true}}=\begin{cases} 0.0 & \mbox{if }p_0 < 2.0 \\ \sqrt{L_{\mathrm {meas}}^2 - (\mathrm{rms}(I)\sqrt{n})^2} & \mbox{else}
\end{cases}
\end{equation}
where $\it{p_0} = \it{L_{\rm{meas}}/\rm{rms}(\it{I})\sqrt{n_{\rm pulse}}}$, where $\it{n}_{\rm pulse}$ is the on pulse number of bins\footnote{For a handful of pulsars of our sample, we accepted a lower threshold for $p_0$ either in agreement with \citet{ew01} or by visually inspecting that the PAs in the frequency channels where $p_0$ resulted less than 2 follow the trend predicted by Eq. \ref{eq:pafit}.}.
\citet{ss85} showed that this is the best method to be applied whenever $p_0$ is greater than $0.7$ (see also \citealt{nj08}), as it happens in all the pulsars of our sample.
As for the estimates of the uncertainties on ${\rm PA_{ave}}$, $\sigma_{\rm{PA}_{ave}}$, for high values ($P_0 > 10$) of $P_0 = L_{\rm true}/{\rm rms}(I)\sqrt{n_{\rm pulse}}$, we used the formula from \citet{ew01}: 
\begin{equation}
\sigma_{\rm{PA_{ave}}}=\frac{1}{2 P_0}
\end{equation}
For lower values of $P_0$, we numerically computed the error integrating the ${\rm PA}$ probability distribution between $\pm \sigma_{\rm{PA_{ave}}}$ in order to obtain 0.68, as in \citet{nc93} and \citet{ew01}:
\begin{eqnarray}
G(\rm{PA}-\rm{PA_{true}};{\it P_0})&& =  \frac{1}{\sqrt{\pi}}\left\{\frac{1}{\sqrt{\pi}} + \eta_0 e^{\eta_0^2}[1 + \rm{erf}(\eta_0)]\right\}\nonumber  \\
&& \times e^{-(P_0^2/2)}
\end{eqnarray}
where ${\rm PA_{true} = PA_{ave}}$ in our case, $\eta_0 = (P_0\sqrt{2})\cos 2(\rm{PA-PA_{\rm true}})$, ${\rm erf}$ is the Gaussian error function. We obtained the ${\rm RM}$ and its error implementing a least squares fit through the following equation:
\begin{equation}
{\rm PA}({\it f})={\rm PA_{ref} + RM}{\it c}^2\times \left(\frac{1}{{\it f}^2} - \frac{1}{{\it f}_{\rm ref}^2}\right)
\label{eq:pafit}
\end{equation}
where ${\rm PA}({\it f})$ is the ${\rm PA}$ at a certain frequency $f$, ${\rm PA_{\rm ref}}$ is the ${\rm PA}$ at a reference frequency ${\it f}_{\rm ref}$ and $c$ is the speed of light.
For pulsars with two recognisable components in the linear polarised profile, we fit for the ${\rm RM}$ separately for each component and we compared the obtained results \textit{a posteriori}.

Once the ${\rm RM}$ was obtained, we summed over the frequency channels to produce
the final integrated profile.
A combination of low signal-to-noise and/or low polarisation fraction meant
that we were unable to compute the ${\rm RM}$ for a number of pulsars in our sample.
In these cases, we simply set the ${\rm RM}$ to zero before summing over frequency.\\

A further useful quantity when considering pulsar polarisation is the
total amount of circular polarisation irrespective of the handedness. The measured quantity $|V|_{\rm meas}$ is biased because it is positive definite. We followed \citet{kj+06} to obtain an unbiased value via:
\begin{equation}
|V|_{\mathrm {true}}=\begin{cases} 0.0 & \mbox{if }|V|_{\mathrm {meas}}/b < 2.0 \\ \sqrt{|V|_{\mathrm {meas}}^2 - b^2} & \mbox{else}
\end{cases}
\end{equation}
where:
\begin{equation}
b=\sqrt{\frac{2}{\pi}} \times {\rm{rms({\it V})}}.
\end{equation}
Here, $\rm rms({\it V})$ is the root mean square of the off pulse $V$ profile.
\\[5pt]
In order to quantify the luminosity and the percentage of polarisation of the pulsars in our sample (showed in Table \ref{Tab:RMpar} and \ref{Tab:noRMpar}), we computed the quantities ${\it S_0}$, ${\it L}\%$, ${\it V}\%$ and $|{\it V}|\%$:
\begin{equation}
\begin{aligned}
& S_0 = \sum_{i=n_{\rm{start}}}^{n_{\rm{end}}} I_i \times \frac{1}{n_{bins}}\\
& L\% = \sum_{i=n_{\rm{start}}}^{n_{\rm{end}}} L_{true,i} \times \frac{S_0}{100}\\
& V\% = \sum_{i=n_{\rm{start}}}^{n_{\rm{end}}} V_i \times \frac{S_0}{100}\\
& |V|\% = \sum_{i=n_{\rm{start}}}^{n_{\rm{end}}} |V|_{true,i} \times \frac{S_0}{100}\\
\end{aligned}
\end{equation}
where $n_{bins}$ is the total number of phase bins that are present in the observations, ${\it L}_{{\it{i}}, {\rm{true}}}$, ${\it V_i}$ and $|{\it V}|_{{\it{i}},{\rm{true}}}$ are the (unbiased, in the cases of ${\it L}_{{\it{i}}, {\rm{true}}}$ and $|{\it V}|_{{\it{i}},{\rm{true}}}$) values of linear, circular and absolute circular polarisation in the i-th phase bin. These definitions are consistent with those adopted by \citet{gl98}.

\section{Polarimetric Results}
The main results are given in Tables \ref{Tab:RMpar} and \ref{Tab:noRMpar}
along with the full Stokes profiles of the pulsars in Figure~1.
Table \ref{Tab:RMpar} includes information for pulsars for which we are
able to determine the ${\rm RM}$ and Table \ref{Tab:noRMpar} contains the sample
for which ${\rm RM}$s were not constrained. We can notice from Table \ref{Tab:RMpar} and Table \ref{Tab:noRMpar} that only $\sim9\%$ of the pulsars with a computable ${\rm RM}$ shows a ${\rm DM}$ value higher than 200 pc~cm$^{-3}$. On the other hand, $\sim 33\%$ of the pulsars for which we have not been able to compute a ${\rm RM}$ exhibits ${\rm DM}>200$ pc~cm$^{-3}$. This is not totally unexpected: in fact, large values of ${\rm DM}$ can be associated with high values of ${\rm RM}$, provided a uniform field is  present along the line of sight. Collapsing the total bandwidth in 4 sub-bands, as we did to obtain the ${\rm RM}$, will depolarise the signal if the ${\rm RM}$ is large enough (note that the low ${\rm SNR}$ of the pulsars in our sample force us to not increase the number of sub-bands). For the high-${\rm DM}$ pulsars, we attempted a different method for computing ${\rm RM}$ which involved a search in ${\rm RM}$ space to maximise for linearly polarised flux. Unfortunately, the low ${\rm SNR}$ of the pulsars meant that we were unable to determine a reliable ${\rm RM}$ in any of the cases.

Below, we briefly give a qualitative description of the profiles in total power, linear and circular polarisation and the ${\rm PA}$ curves of the analysed pulsars, except for PSRs~\textbf{J0919--6040, J1054--5946, J1143--5536, J1539--4835, J1625--4913, J1634--5640, J1647--3607} and \textbf{J1700--4422}, for which we were not able to obtain a ${\rm RM}$ value, and that show very low linear and circular polarisation and no interesting features.\\
\\
\begin{figure*}
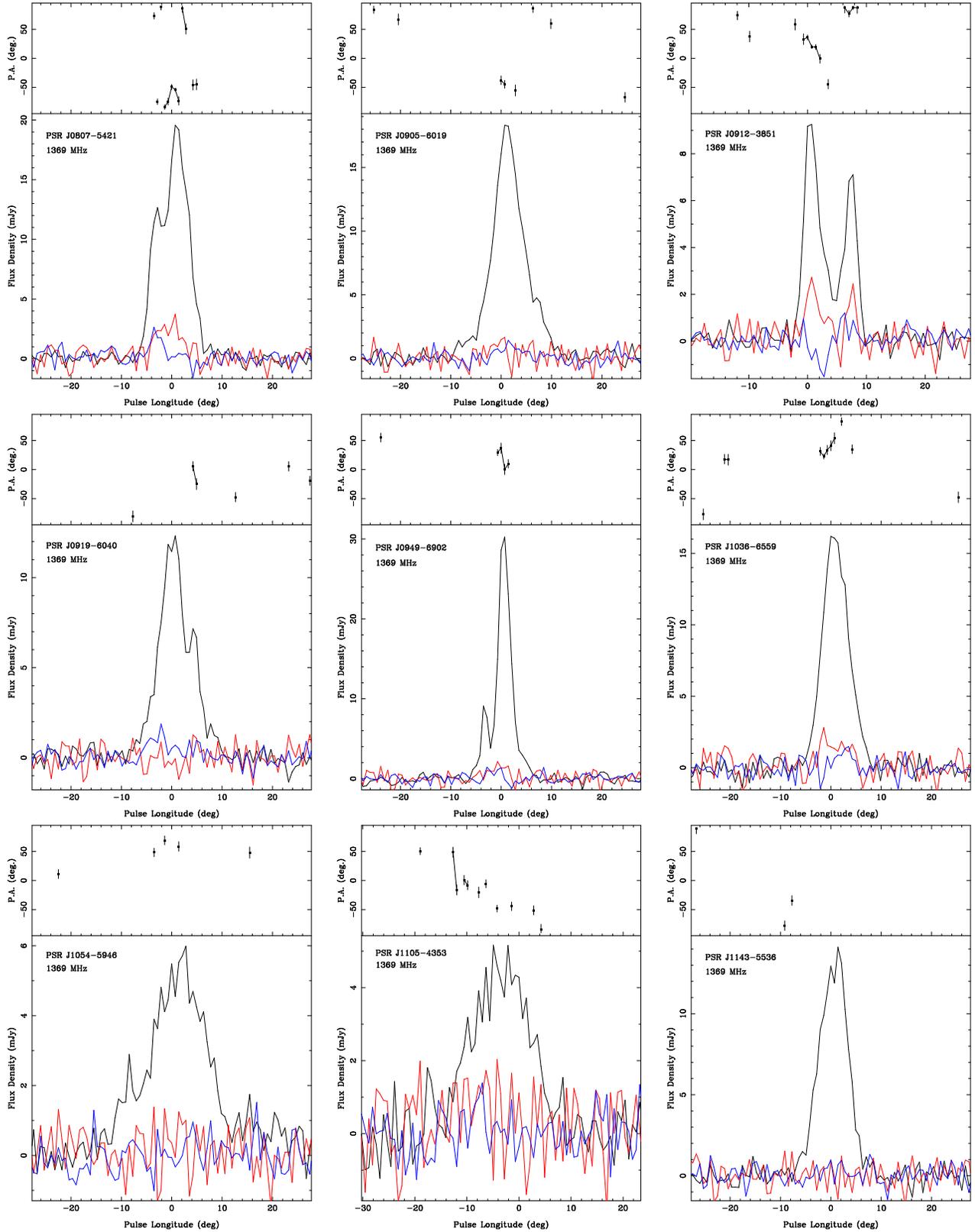

\begin{center}
\begin{tabular}{ccc}
\resizebox{0.3\hsize}{!}{\includegraphics{./J0807-5421.ps}}&
\resizebox{0.3\hsize}{!}{\includegraphics{./J0905-6019.ps}}&
\resizebox{0.3\hsize}{!}{\includegraphics{./J0912-3851.ps}}\\
\resizebox{0.3\hsize}{!}{\includegraphics{./J0919-6040.ps}}&
\resizebox{0.3\hsize}{!}{\includegraphics{./J0949-6902.ps}}&
\resizebox{0.3\hsize}{!}{\includegraphics{./J1036-6559.ps}}\\
\resizebox{0.3\hsize}{!}{\includegraphics{./J1054-5946.ps}}&
\resizebox{0.3\hsize}{!}{\includegraphics{./J1105-4353.ps}}&
\resizebox{0.3\hsize}{!}{\includegraphics{./J1143-5536.ps}}\\
\end{tabular}
\end{center}
\caption{Polarisation profiles at 1369 MHz for the pulsars in the sample.
The top panel of each plot shows the ${\rm PA}$ variation with respect to
celestial north as a function of longitude. The ${\rm PA}$s are corrected for ${\rm RM}$ and represent the (frequency independent) value at the pulsar, and are plotted if the linear polarisation is above 2 $\sigma$. The lower panel shows the integrated profile in total intensity (thick black line), linear polarisation (red line) and circular polarisation (blue line).}
\end{figure*}

\addtocounter{figure}{-1}
\begin{figure*}
\begin{center}
\begin{tabular}{ccc}
\resizebox{0.3\hsize}{!}{\includegraphics{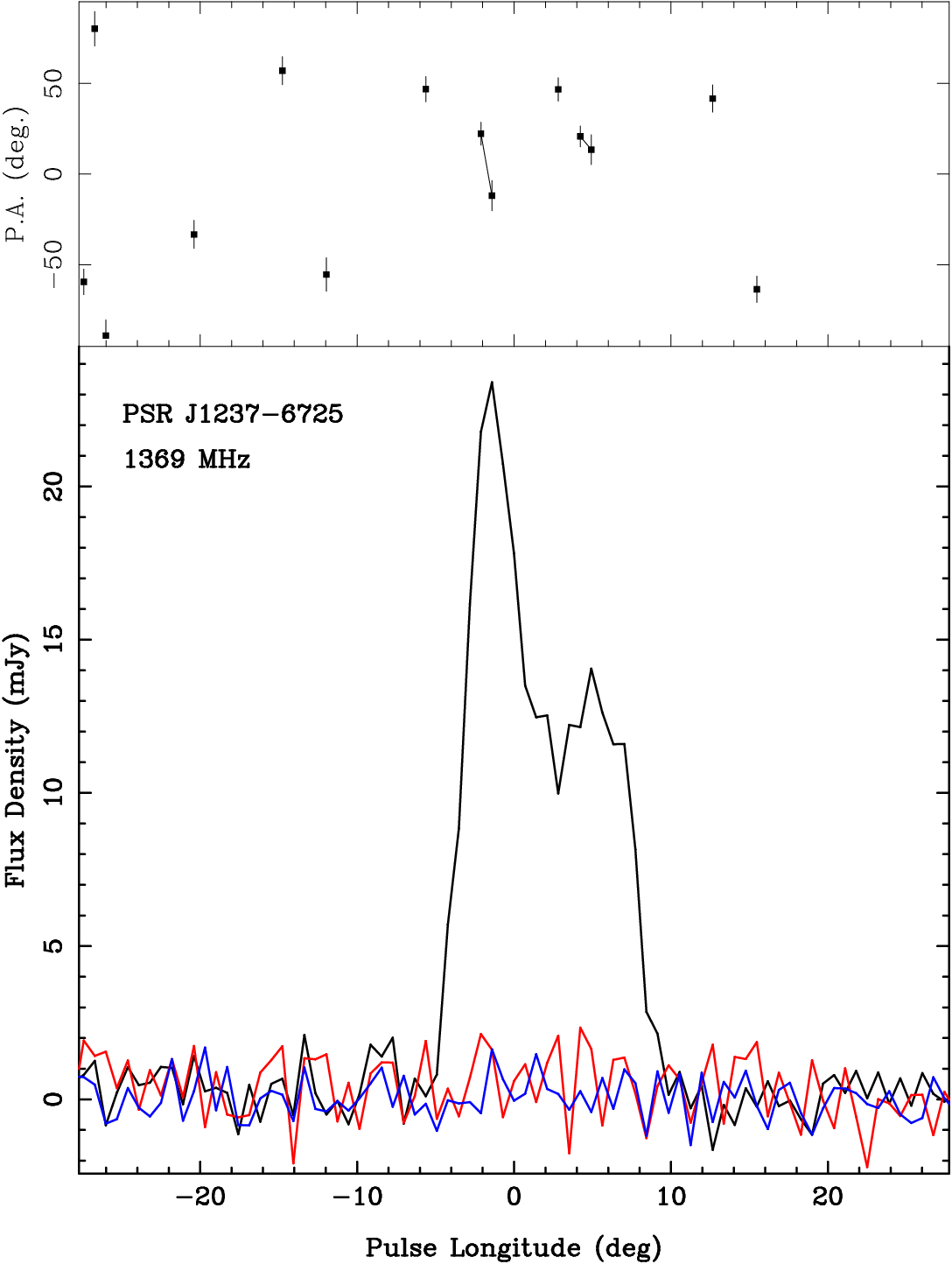}}&
\resizebox{0.3\hsize}{!}{\includegraphics{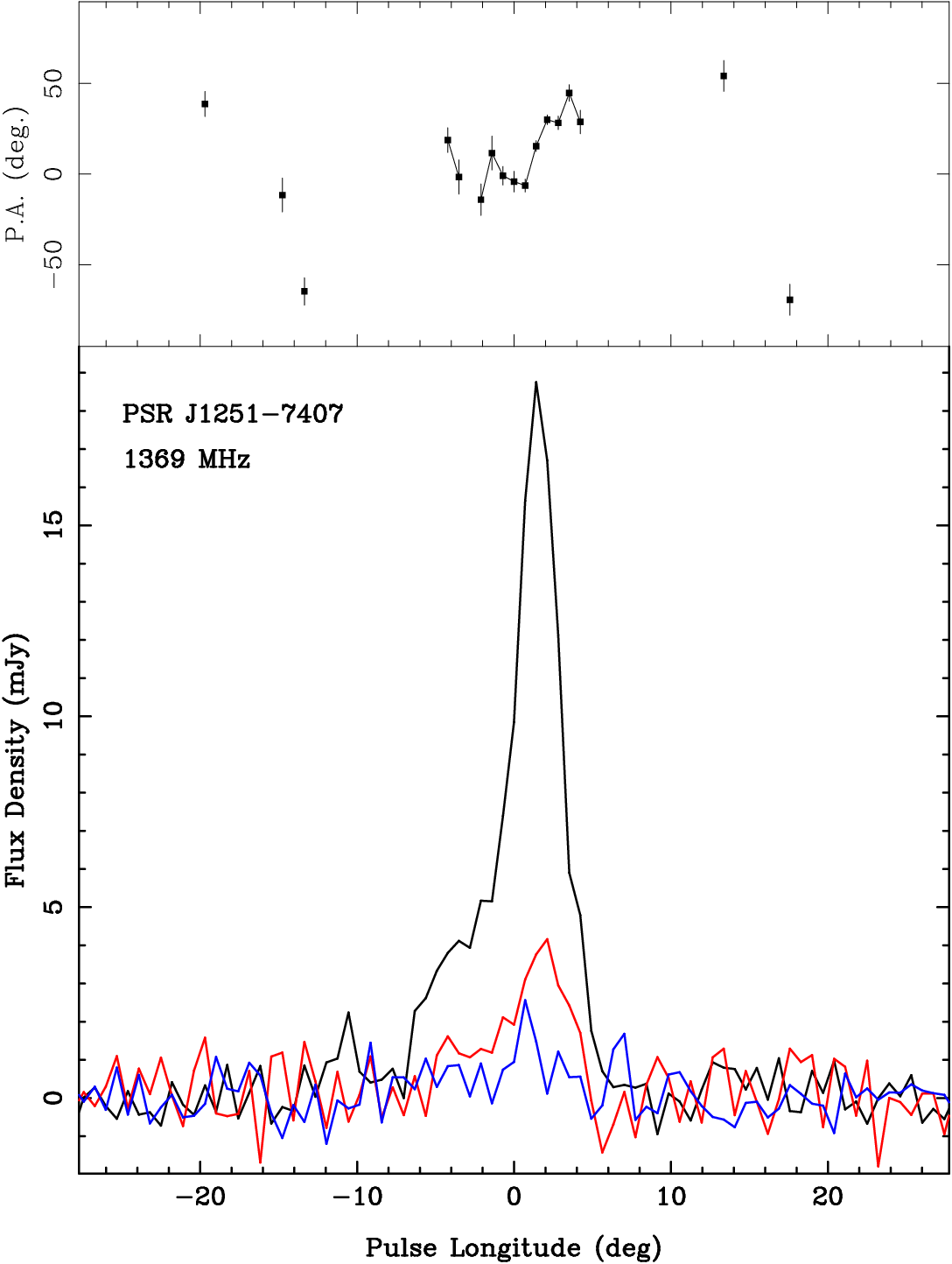}}&
\resizebox{0.3\hsize}{!}{\includegraphics{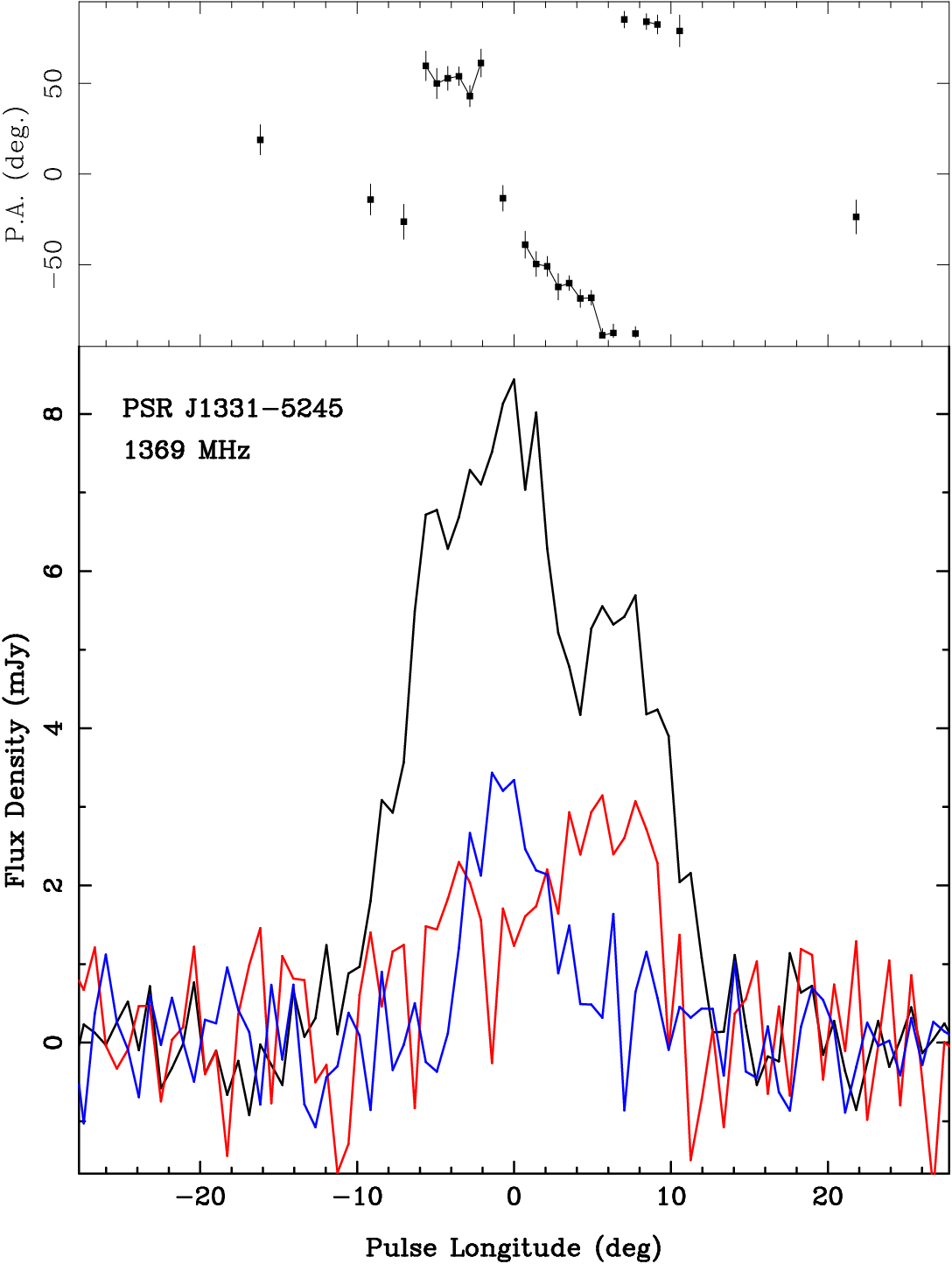}}\\
\resizebox{0.3\hsize}{!}{\includegraphics{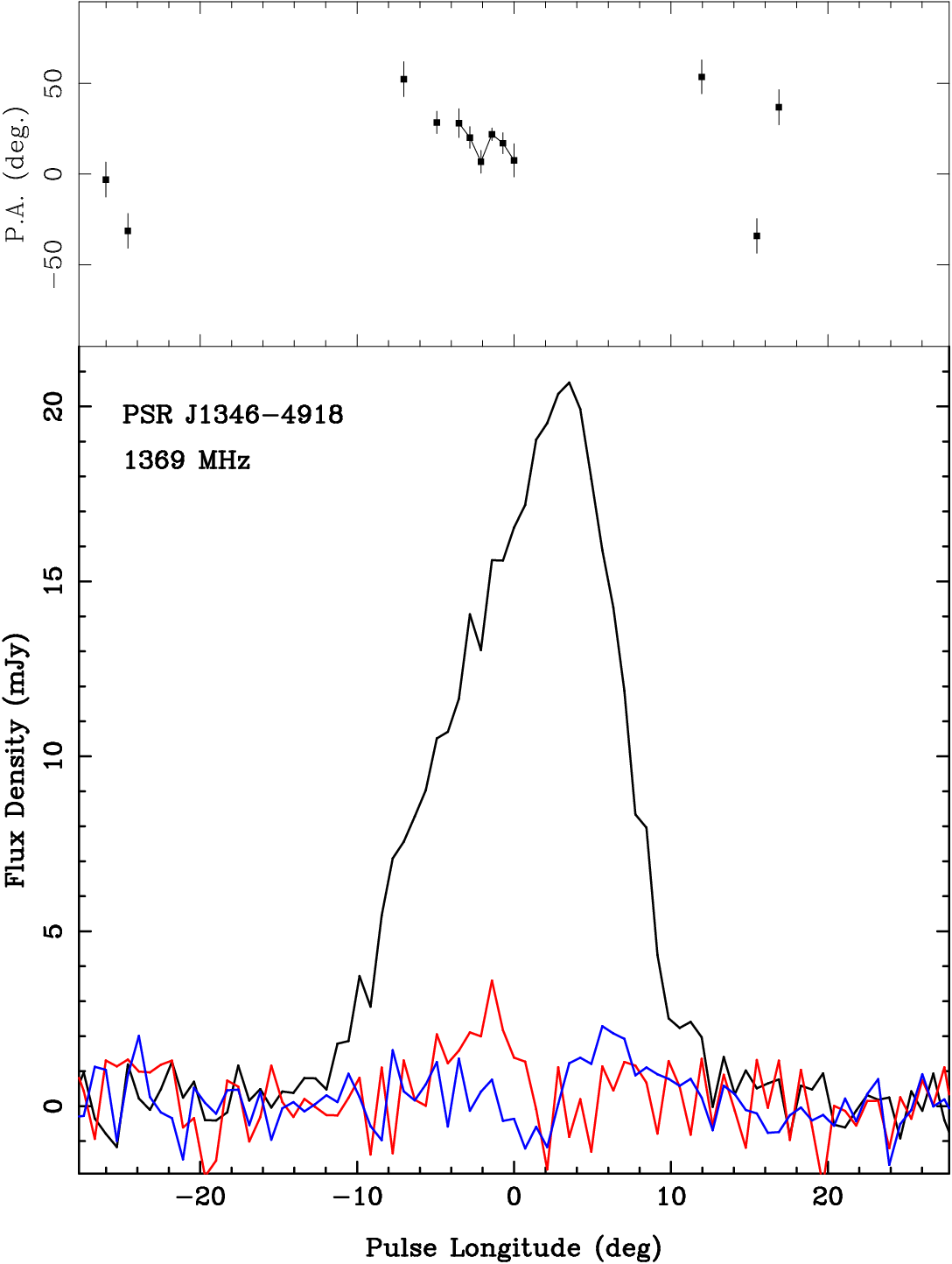}}&
\resizebox{0.3\hsize}{!}{\includegraphics{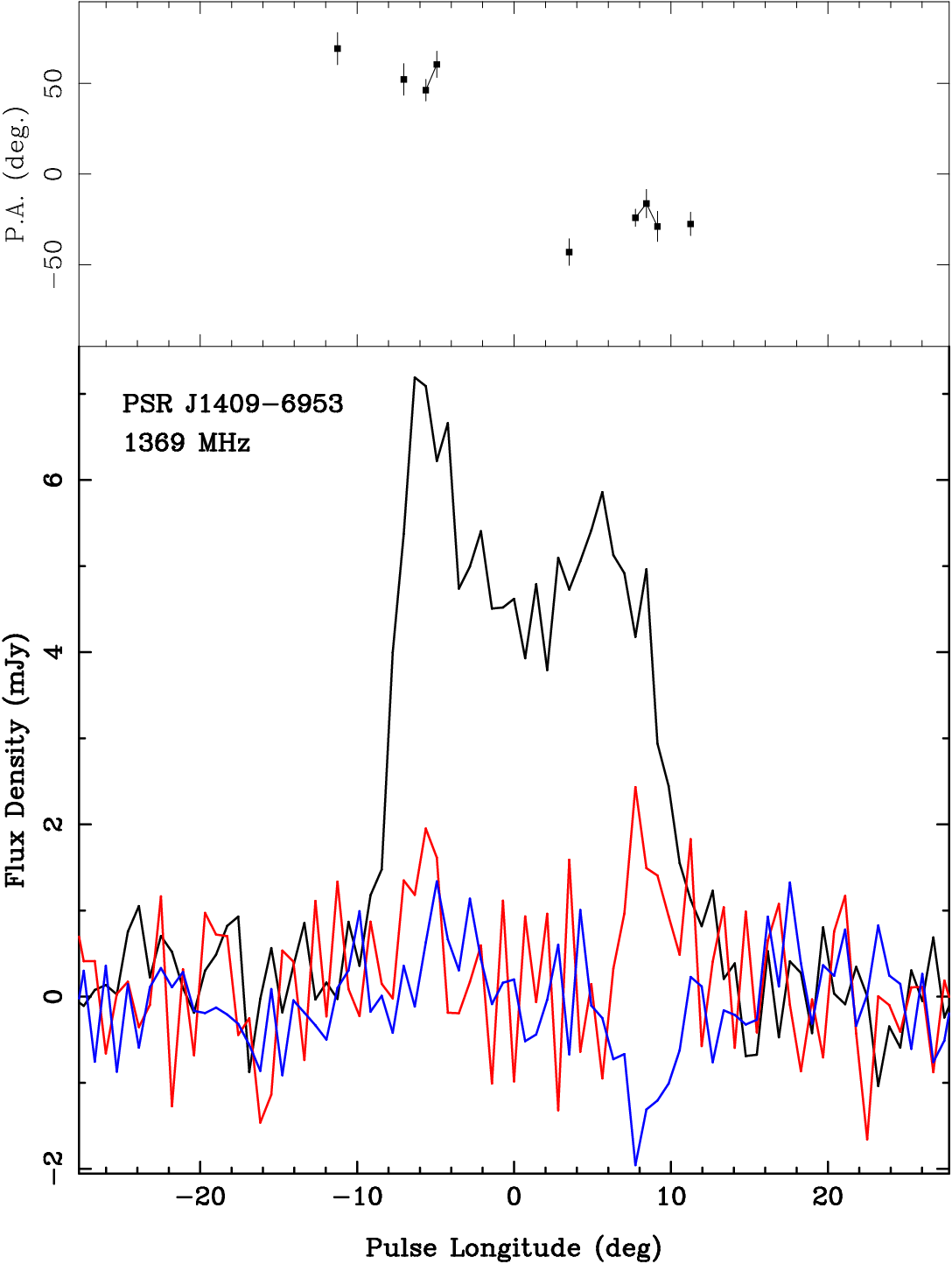}}&
\resizebox{0.3\hsize}{!}{\includegraphics{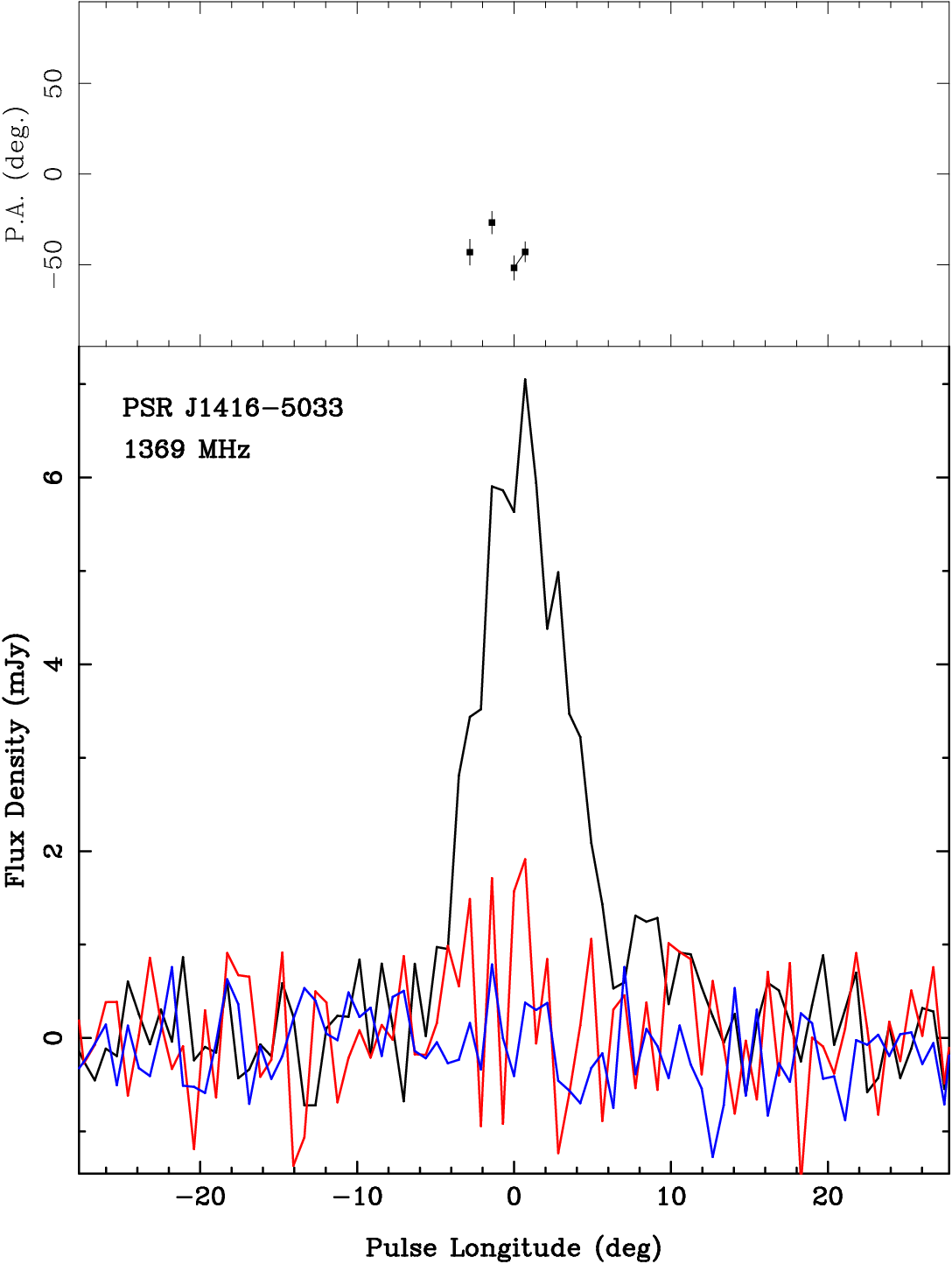}}\\
\resizebox{0.3\hsize}{!}{\includegraphics{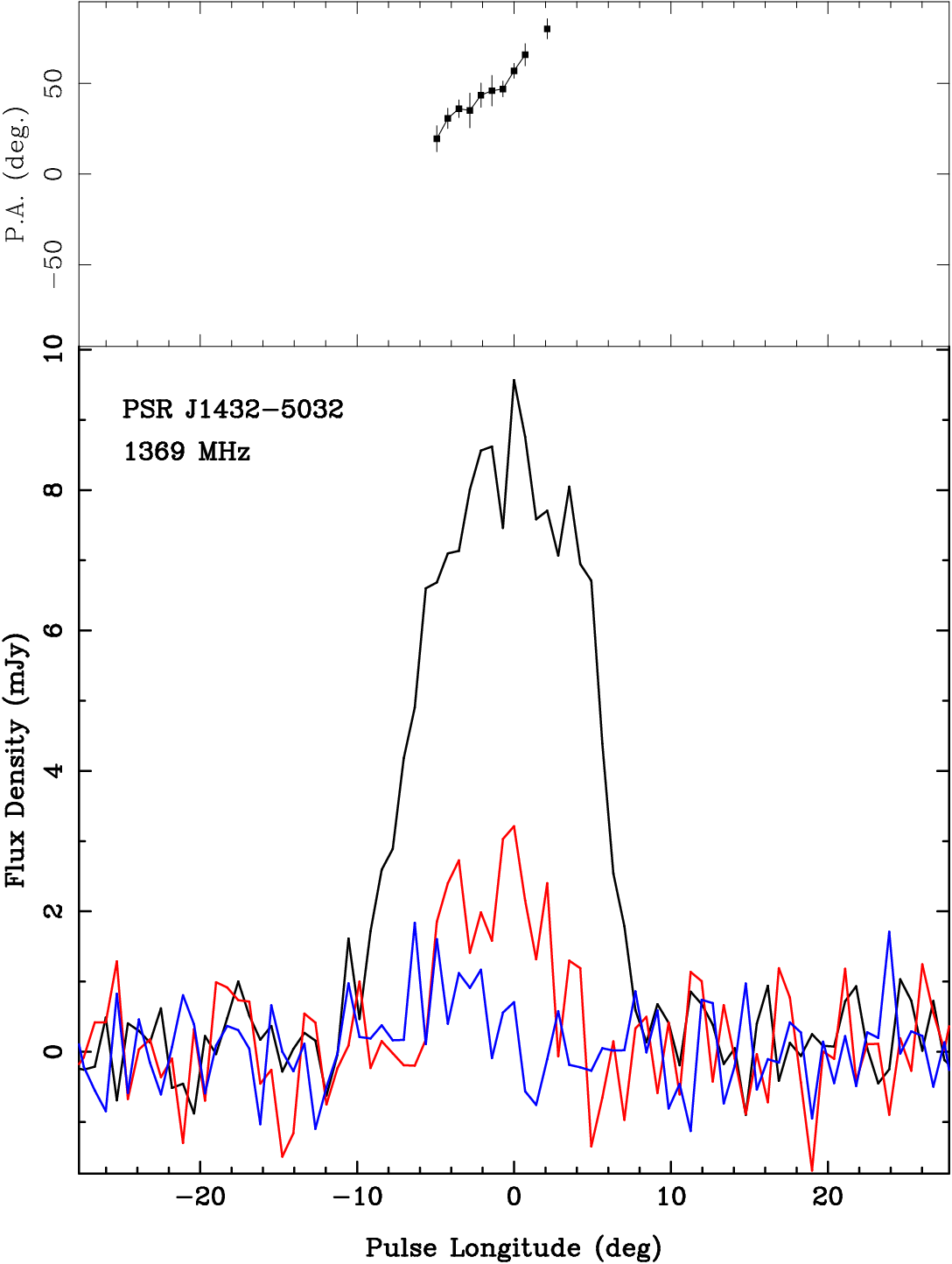}}&
\resizebox{0.3\hsize}{!}{\includegraphics{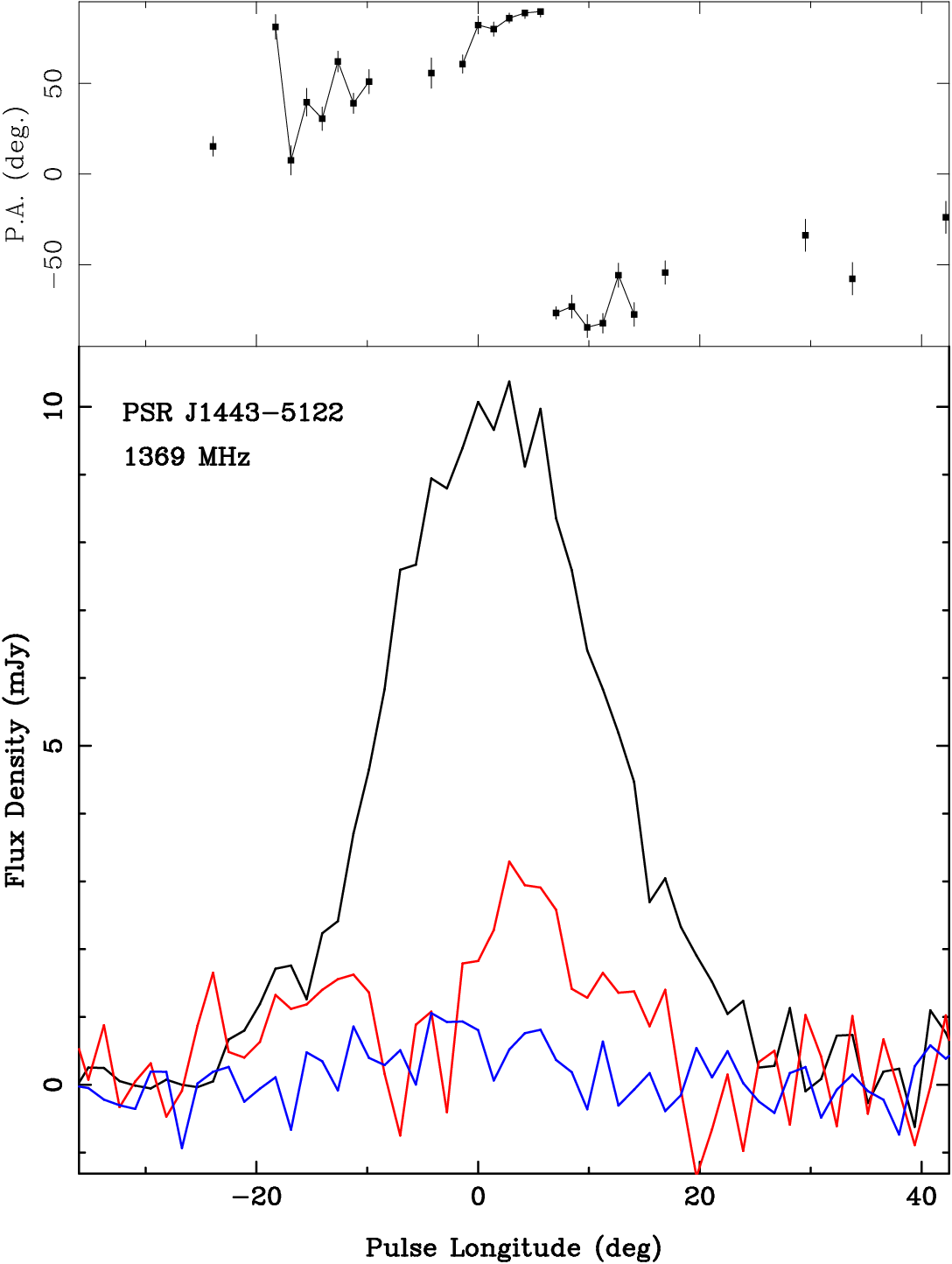}}&
\resizebox{0.3\hsize}{!}{\includegraphics{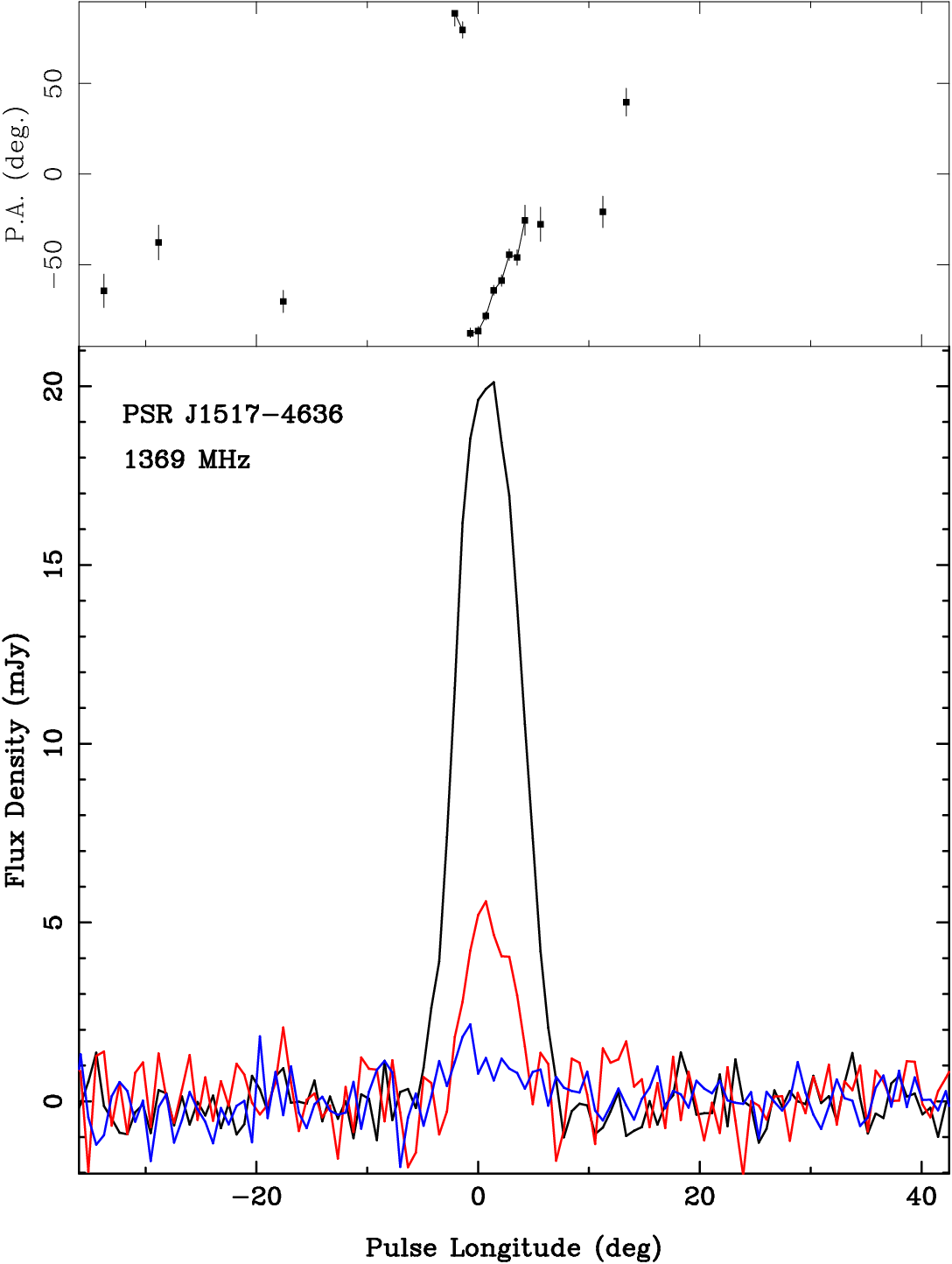}}\\
\end{tabular}
\end{center}
\caption[]{(continued)}
\end{figure*}

\addtocounter{figure}{-1}
\begin{figure*}
\begin{center}
\begin{tabular}{ccc}
\resizebox{0.3\hsize}{!}{\includegraphics{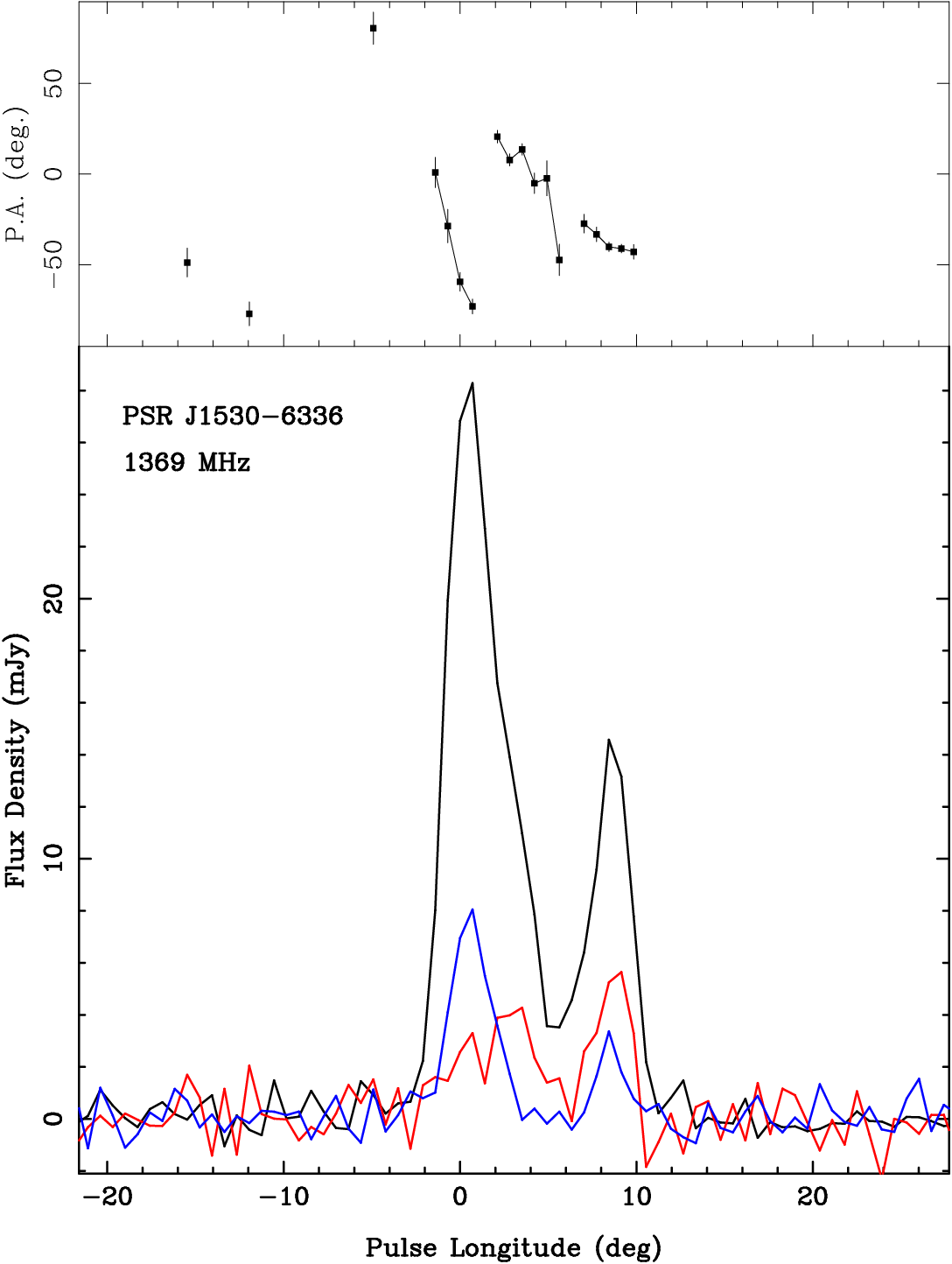}}&
\resizebox{0.3\hsize}{!}{\includegraphics{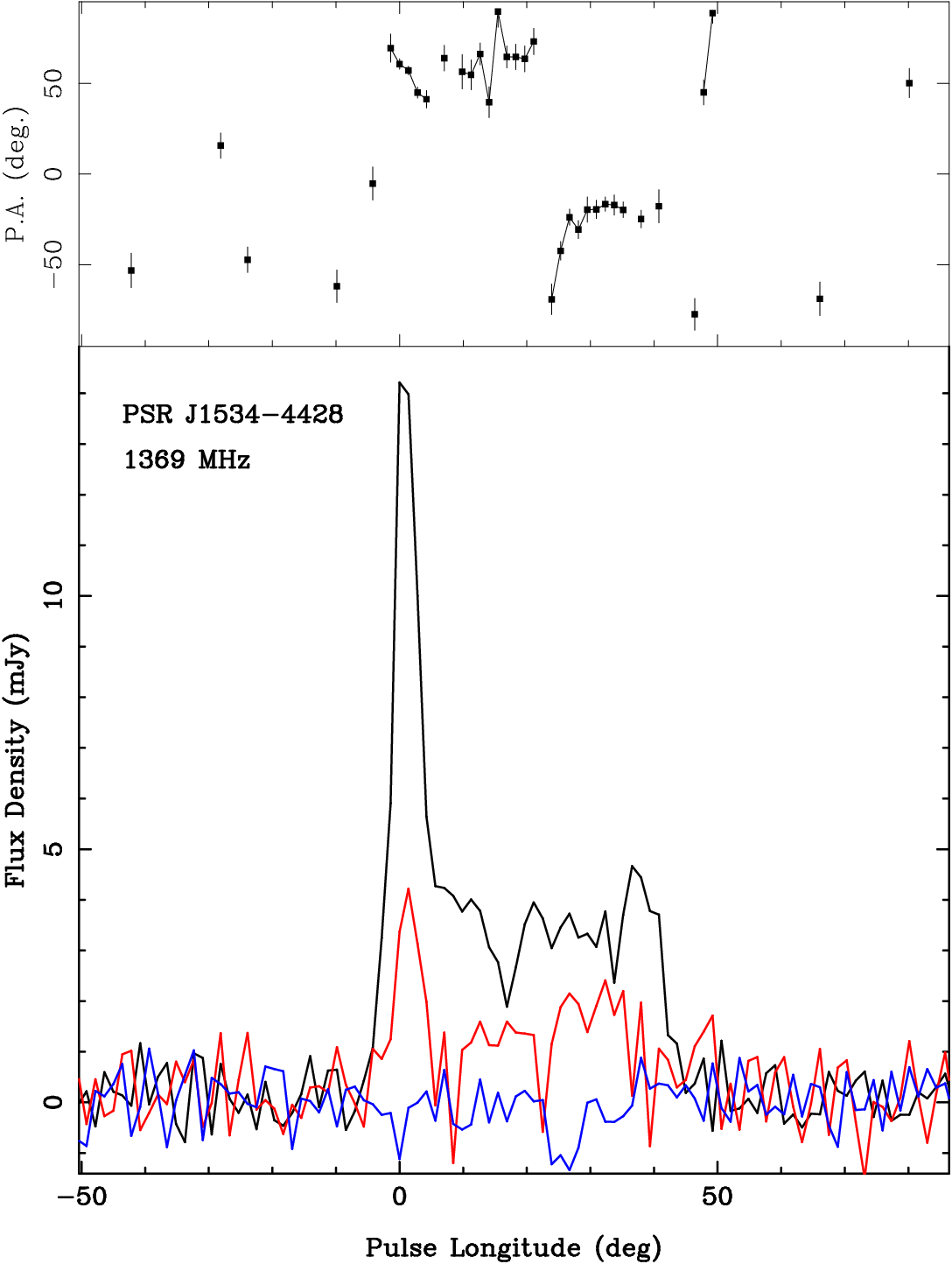}}&
\resizebox{0.3\hsize}{!}{\includegraphics{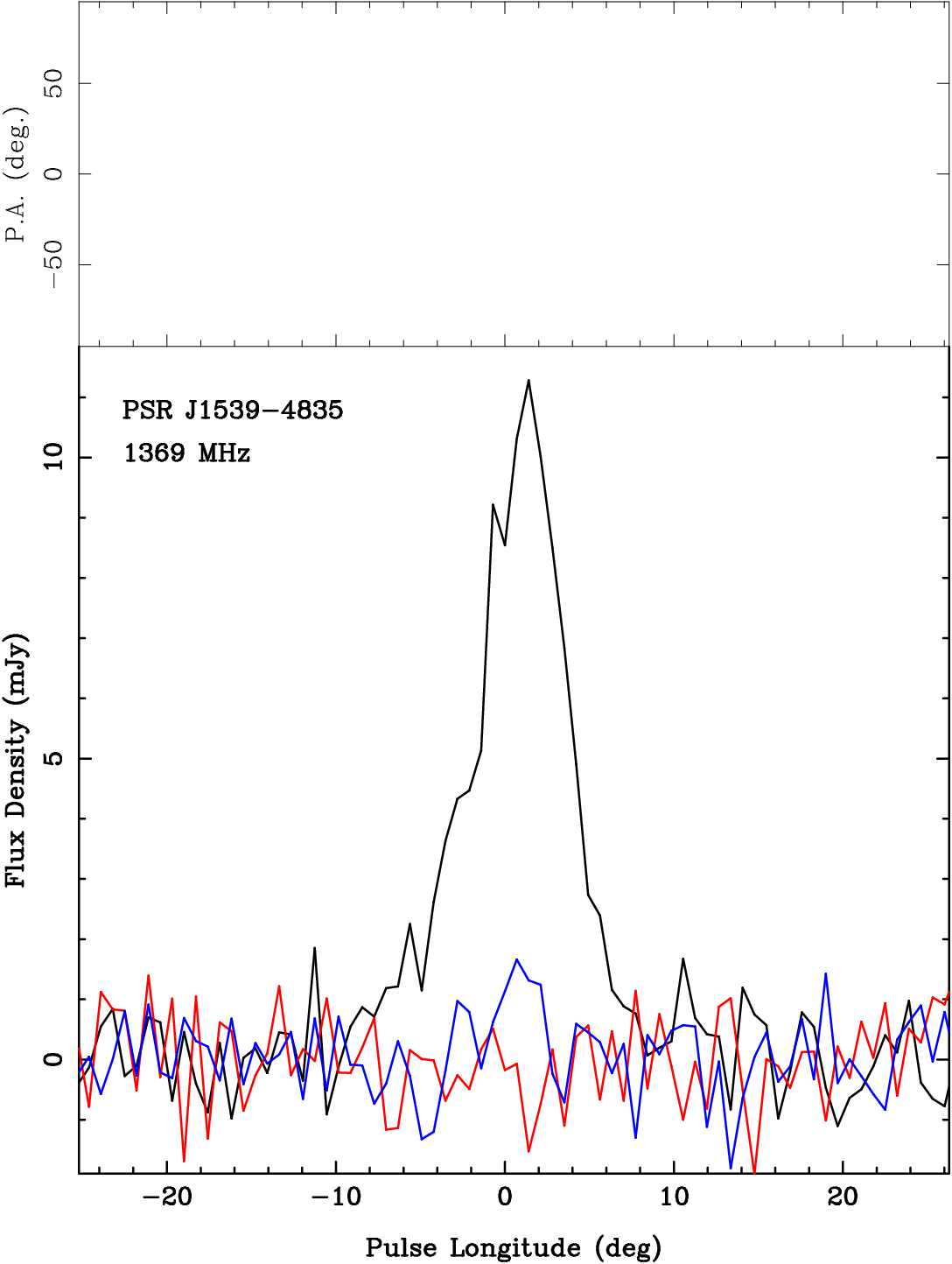}}\\
\resizebox{0.3\hsize}{!}{\includegraphics{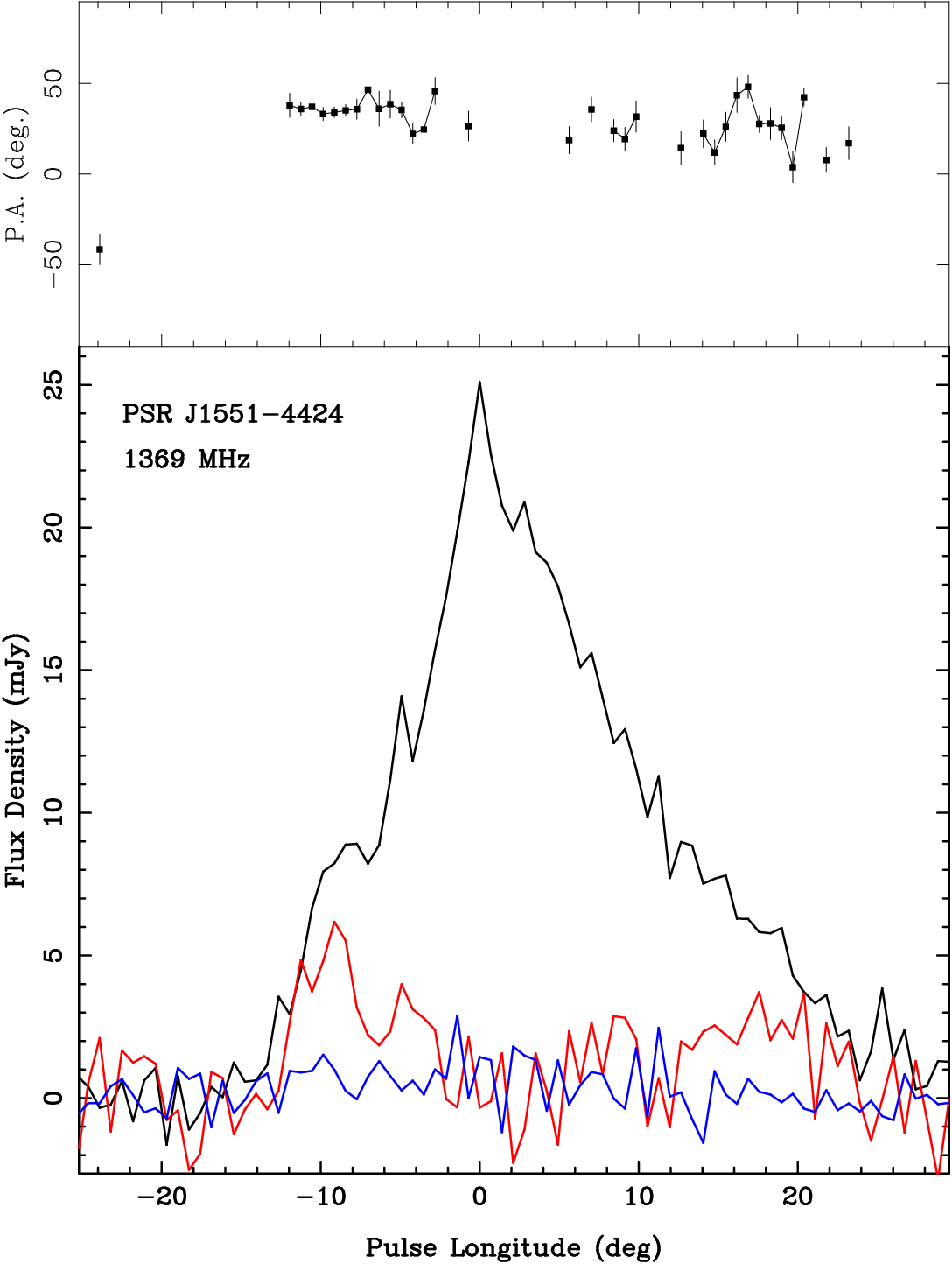}}&
\resizebox{0.3\hsize}{!}{\includegraphics{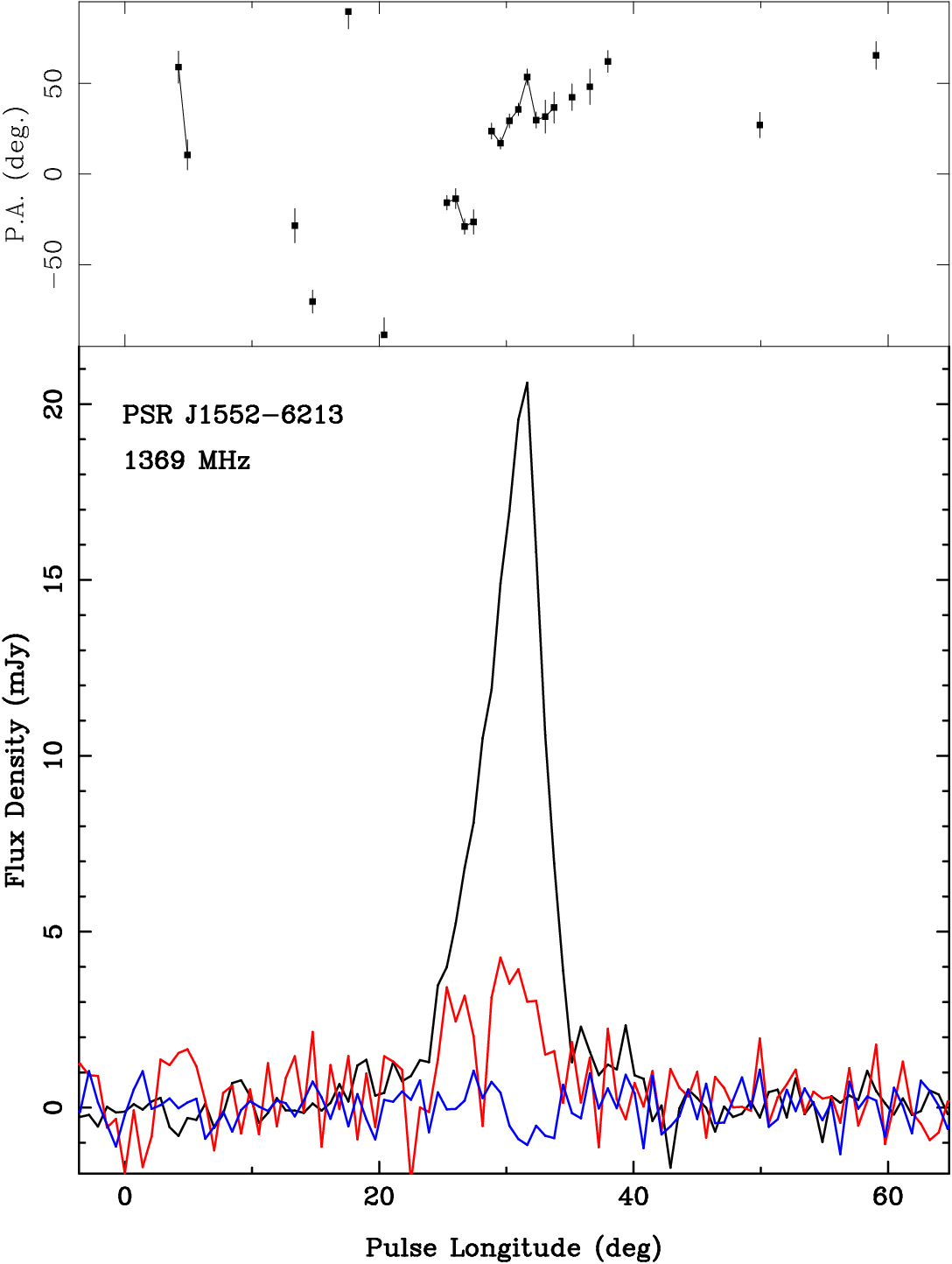}}&
\resizebox{0.3\hsize}{!}{\includegraphics{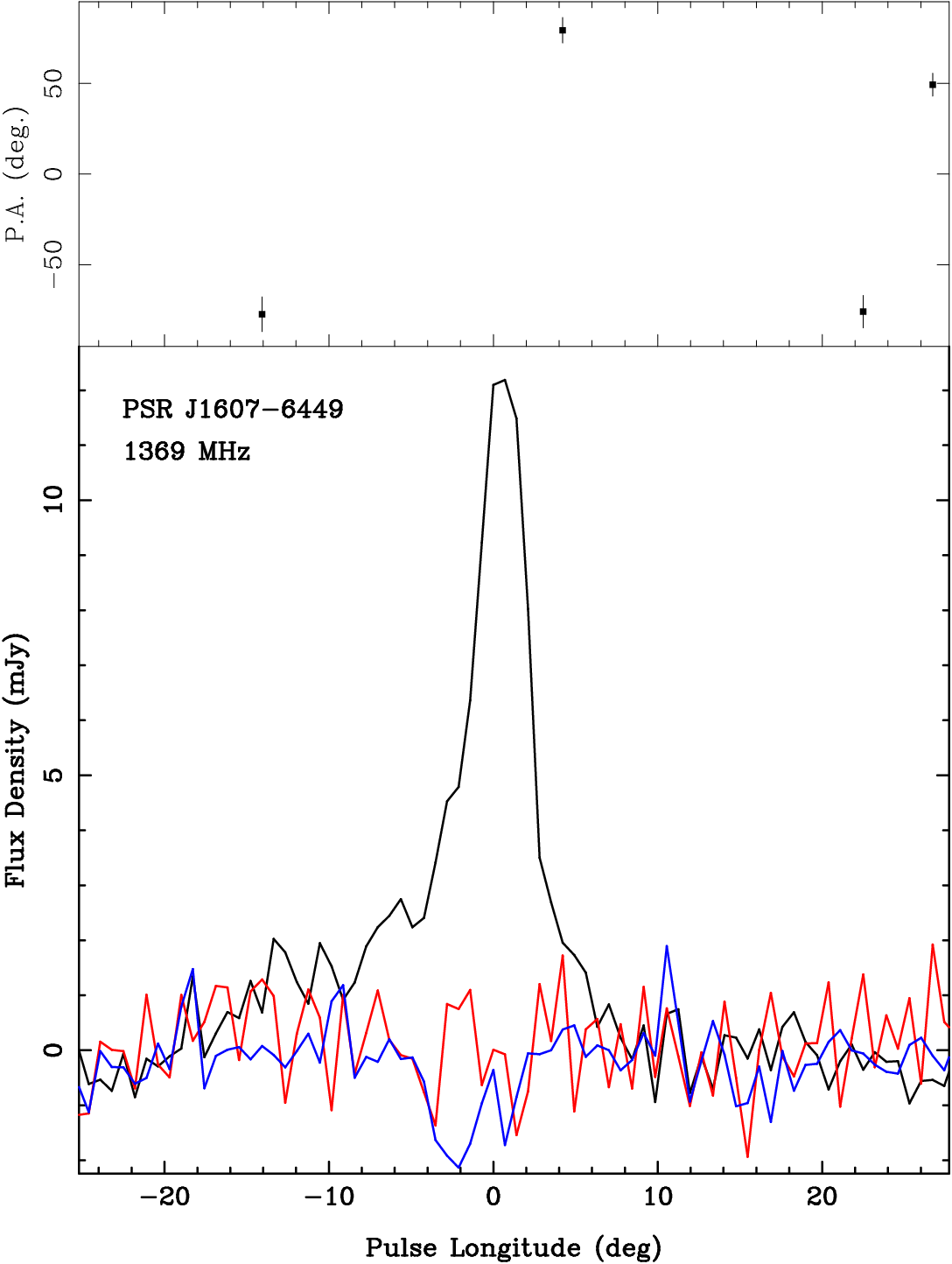}}\\
\resizebox{0.3\hsize}{!}{\includegraphics{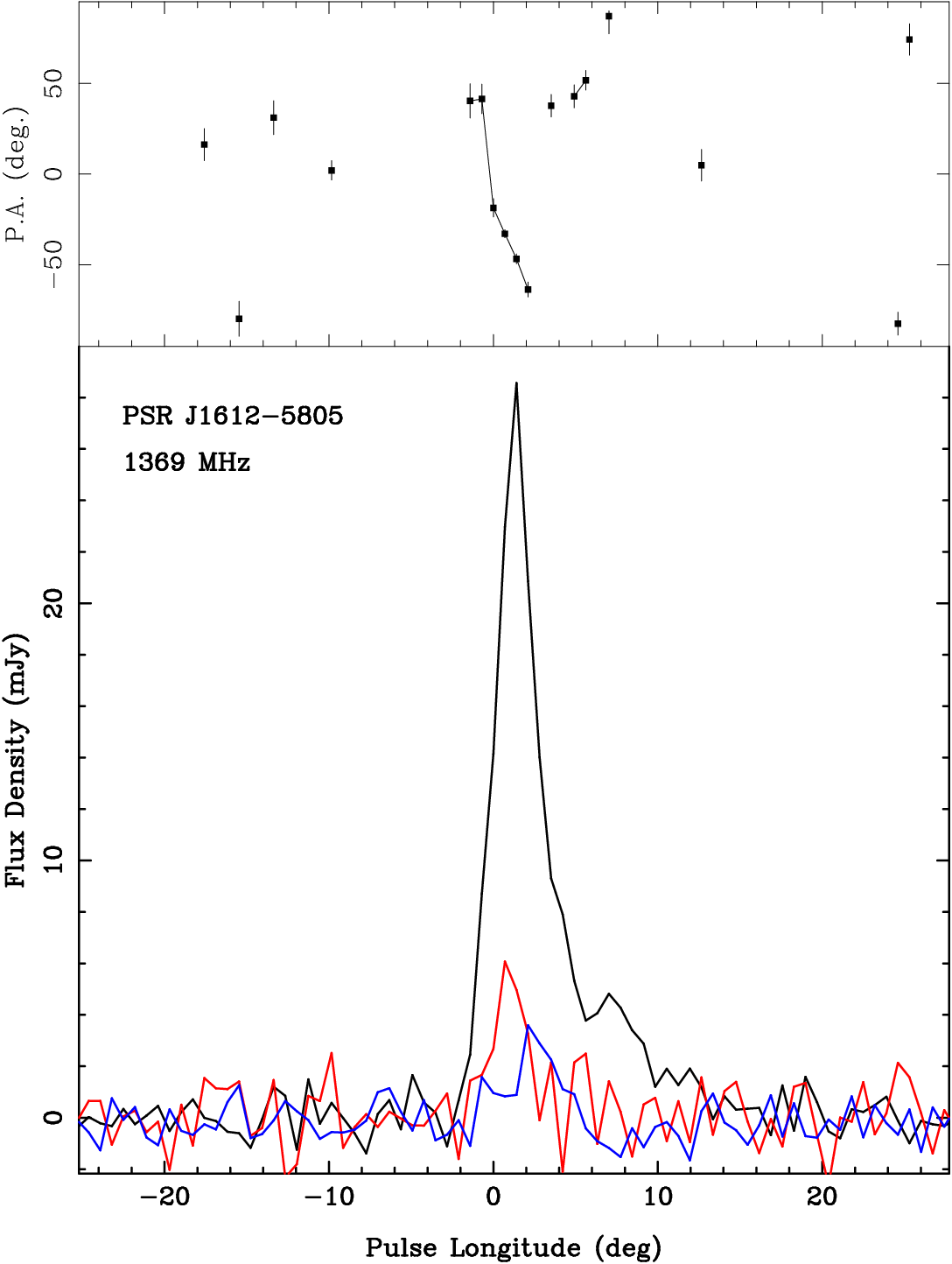}}&
\resizebox{0.3\hsize}{!}{\includegraphics{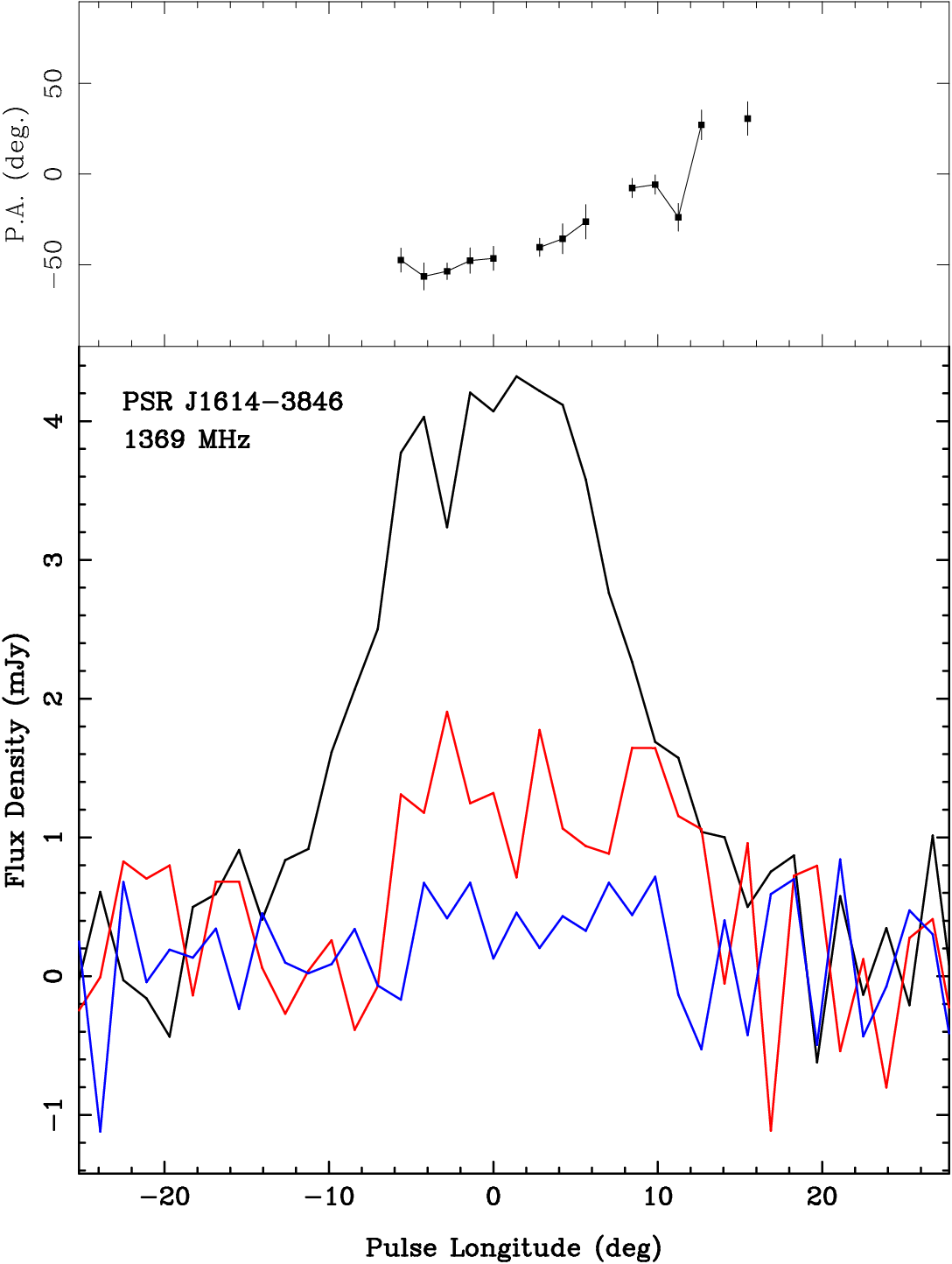}}&
\resizebox{0.3\hsize}{!}{\includegraphics{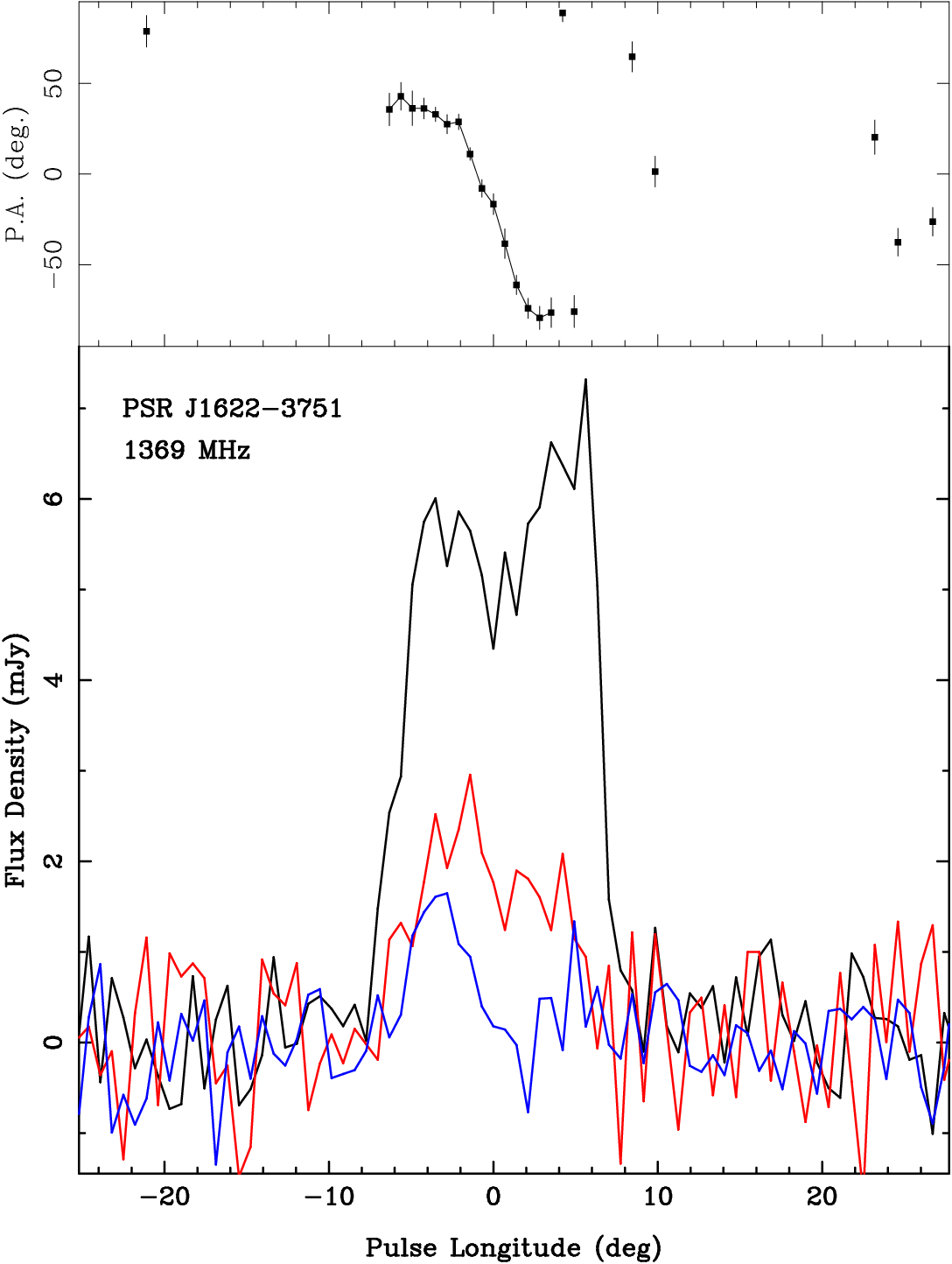}}\\
\end{tabular}
\end{center}
\caption[]{(continued)}
\end{figure*}

\addtocounter{figure}{-1}
\begin{figure*}
\begin{center}
\begin{tabular}{ccc}
\resizebox{0.3\hsize}{!}{\includegraphics{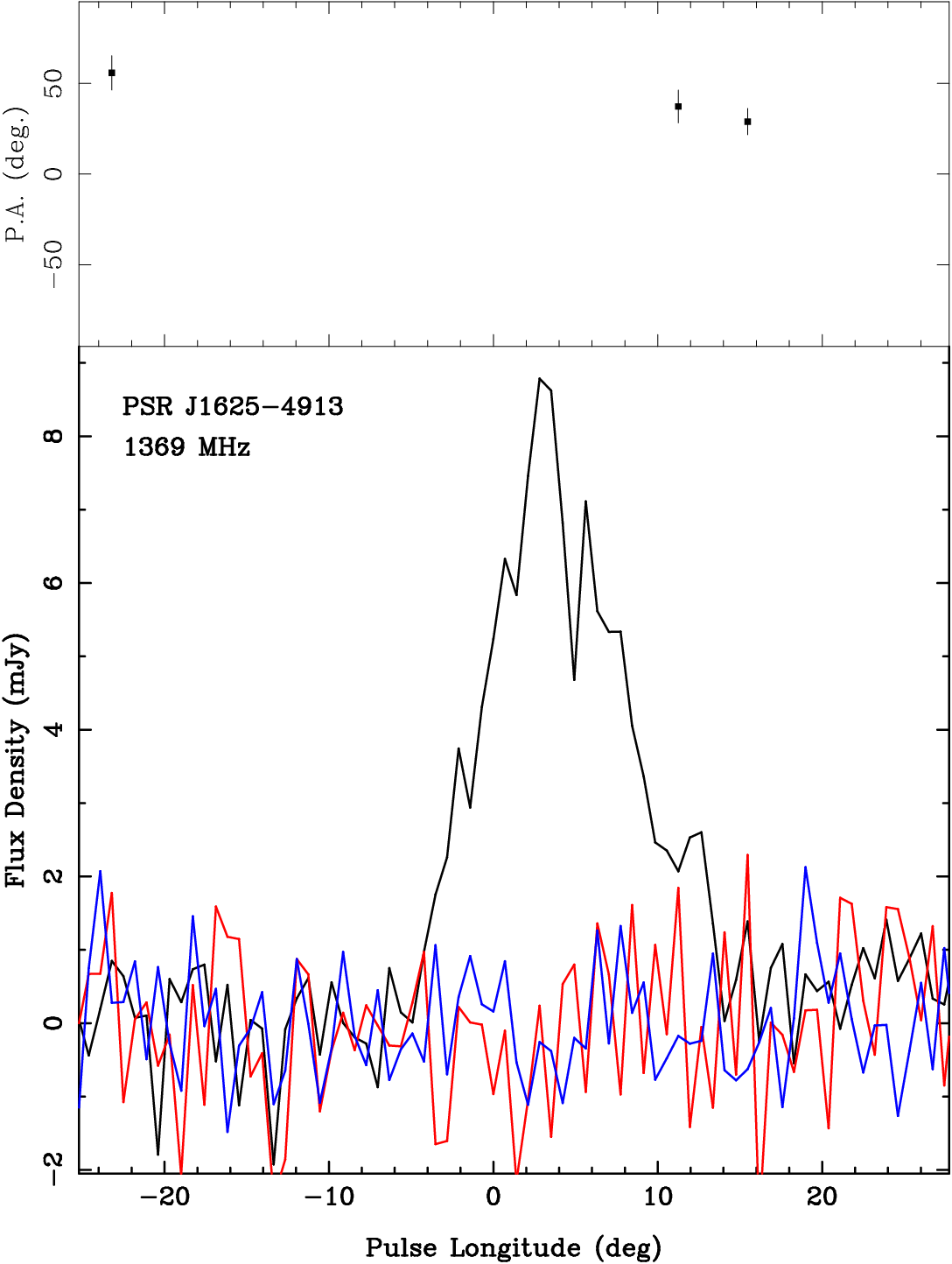}}&
\resizebox{0.3\hsize}{!}{\includegraphics{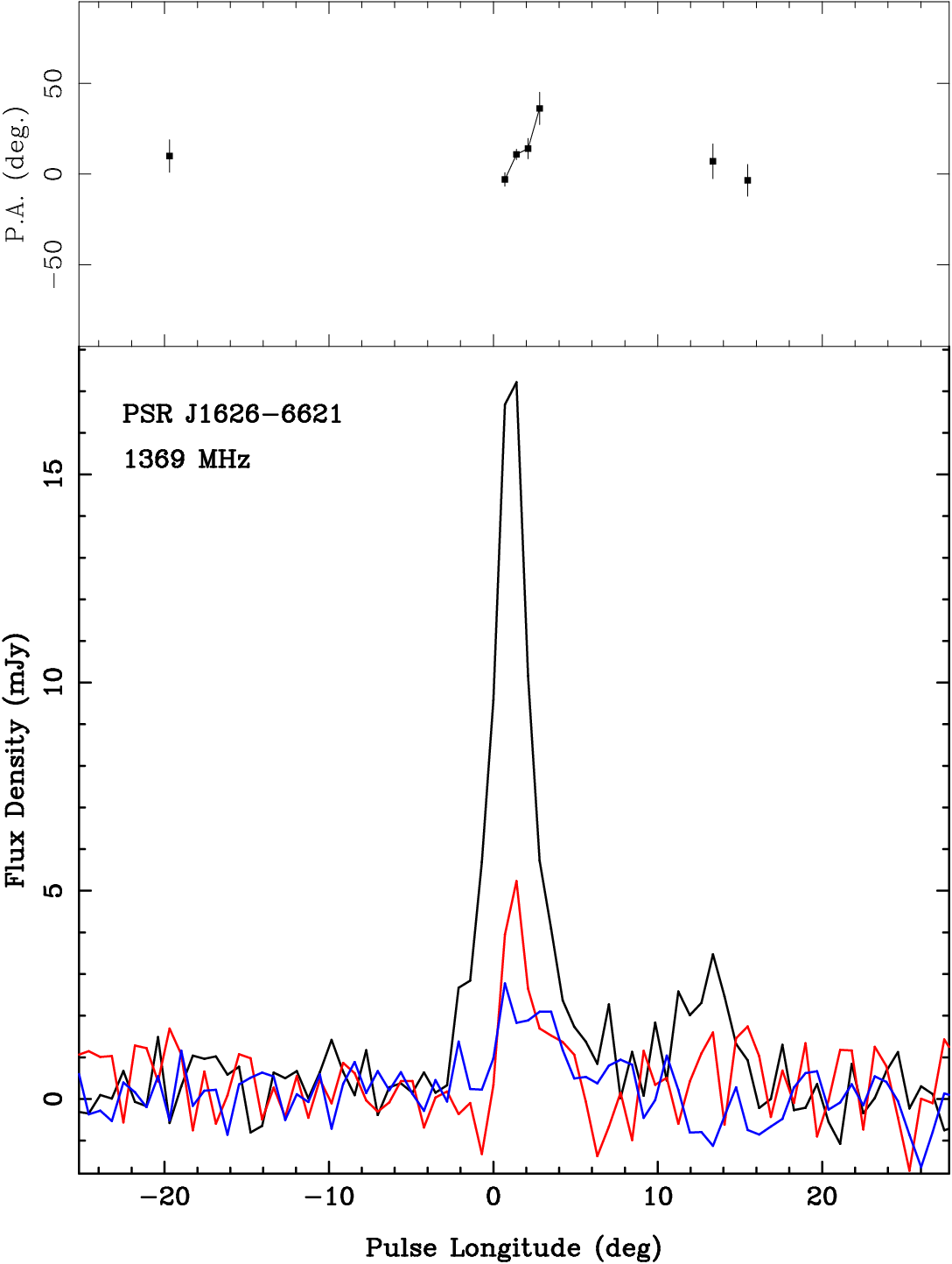}}&
\resizebox{0.3\hsize}{!}{\includegraphics{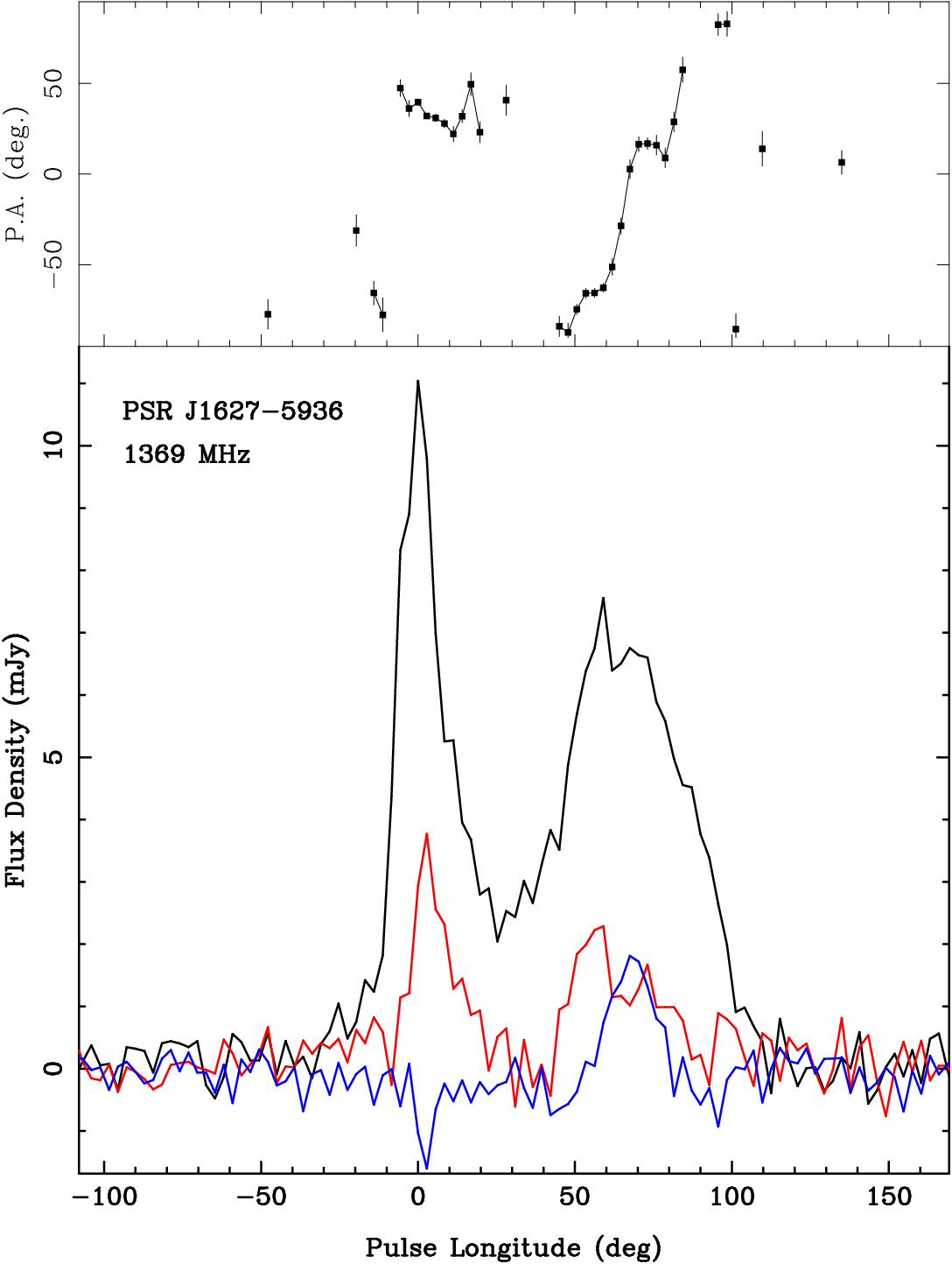}}\\
\resizebox{0.3\hsize}{!}{\includegraphics{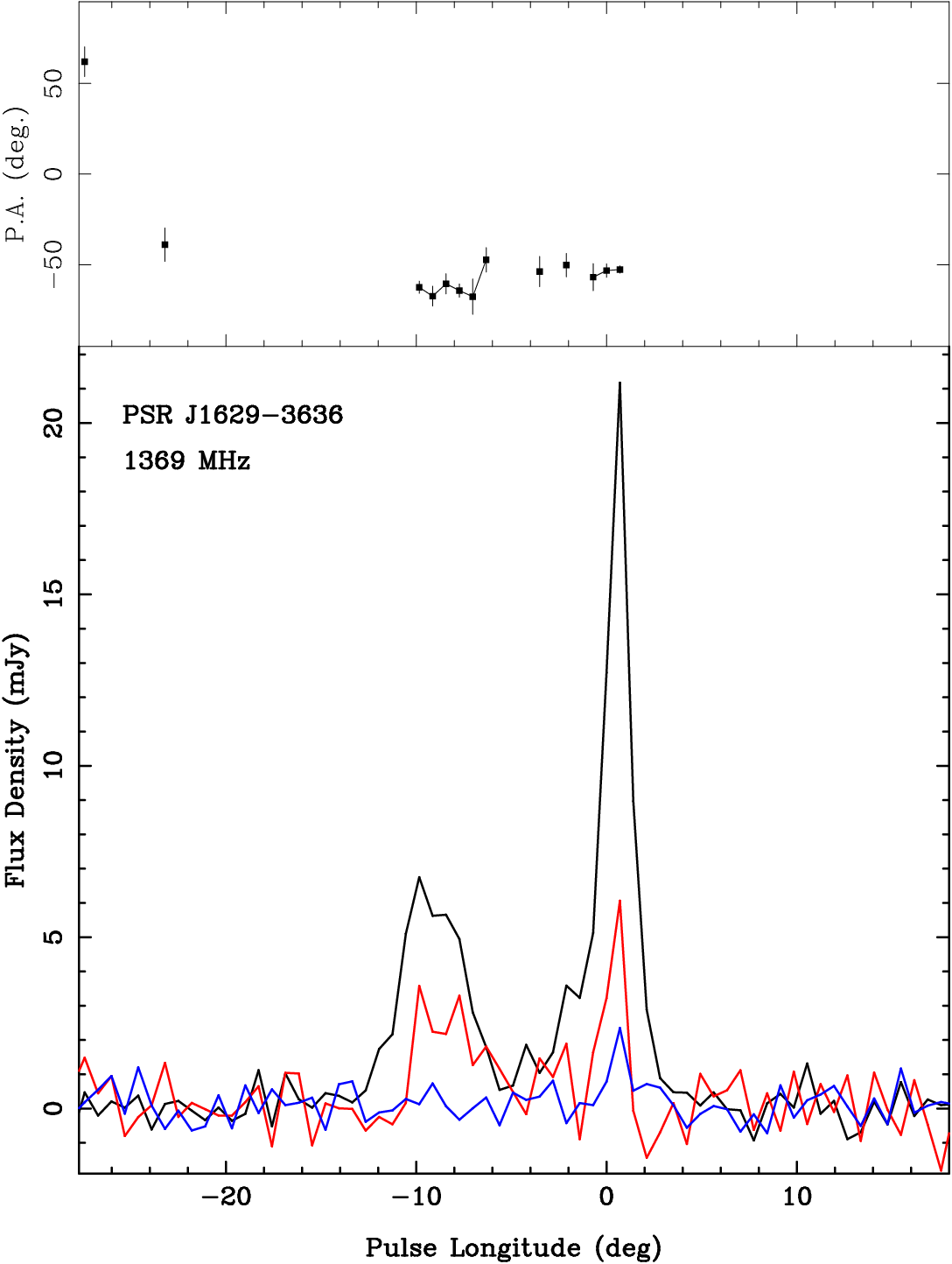}}&
\resizebox{0.3\hsize}{!}{\includegraphics{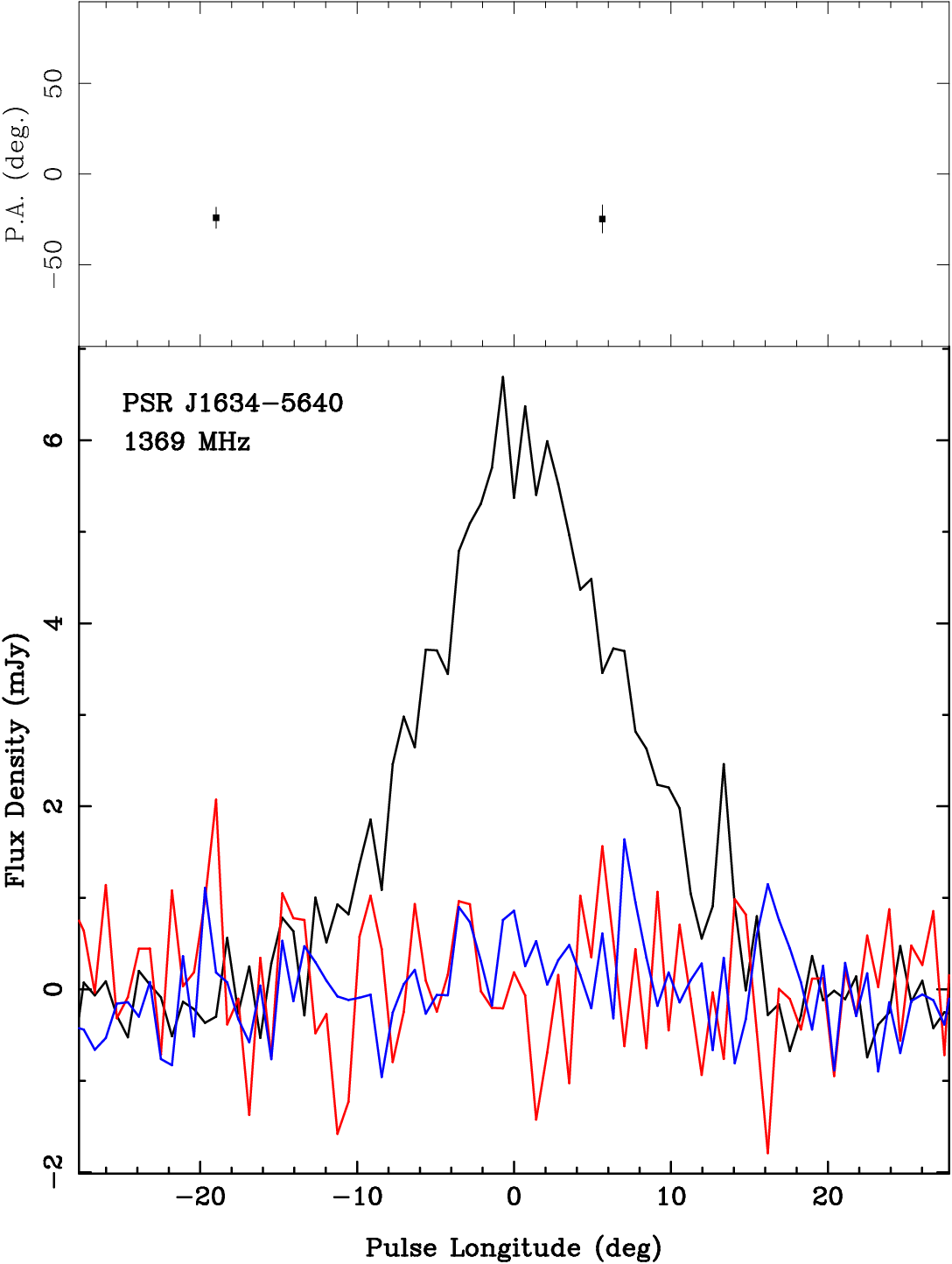}}&
\resizebox{0.3\hsize}{!}{\includegraphics{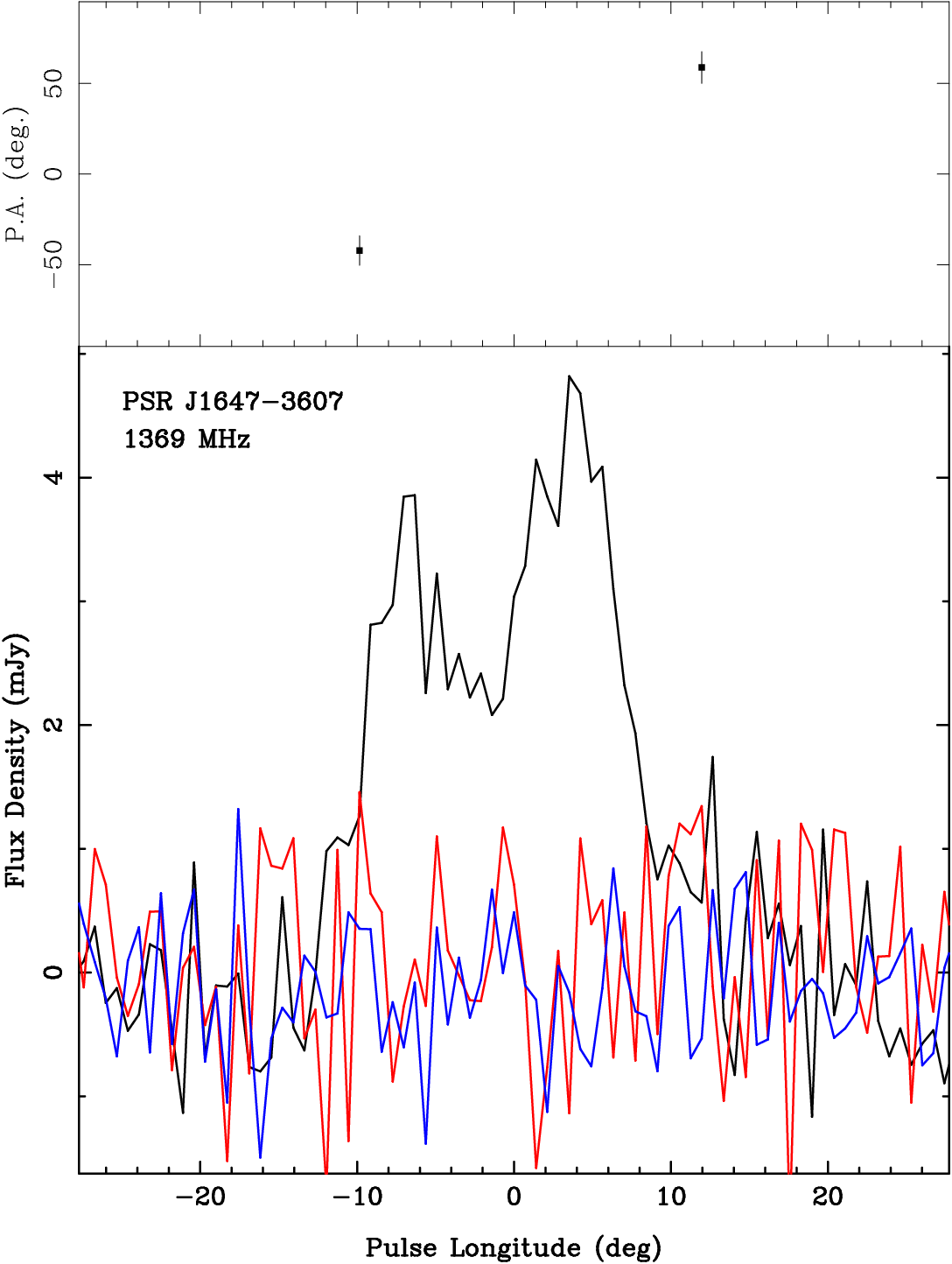}}\\
\resizebox{0.3\hsize}{!}{\includegraphics{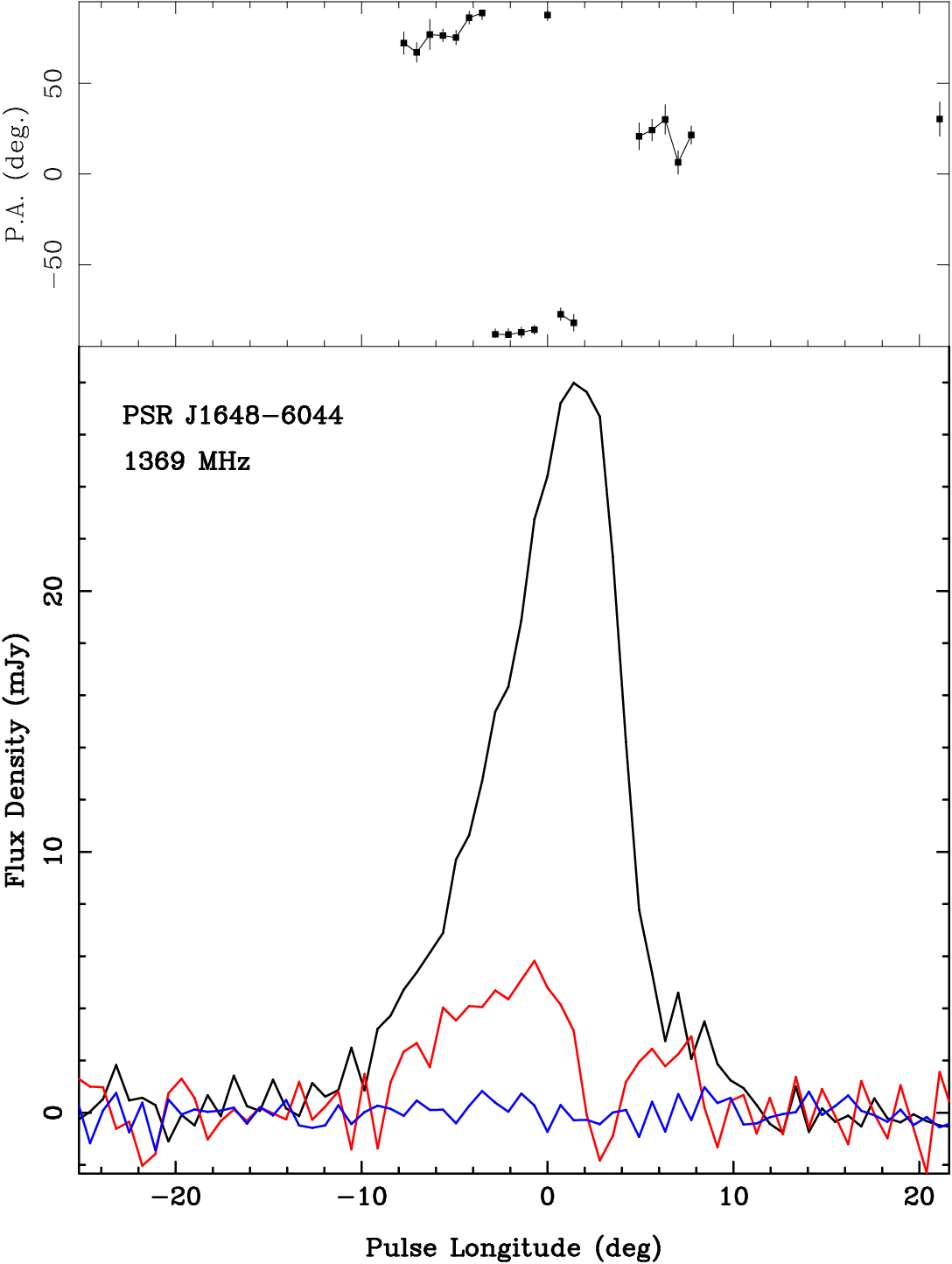}}&
\resizebox{0.3\hsize}{!}{\includegraphics{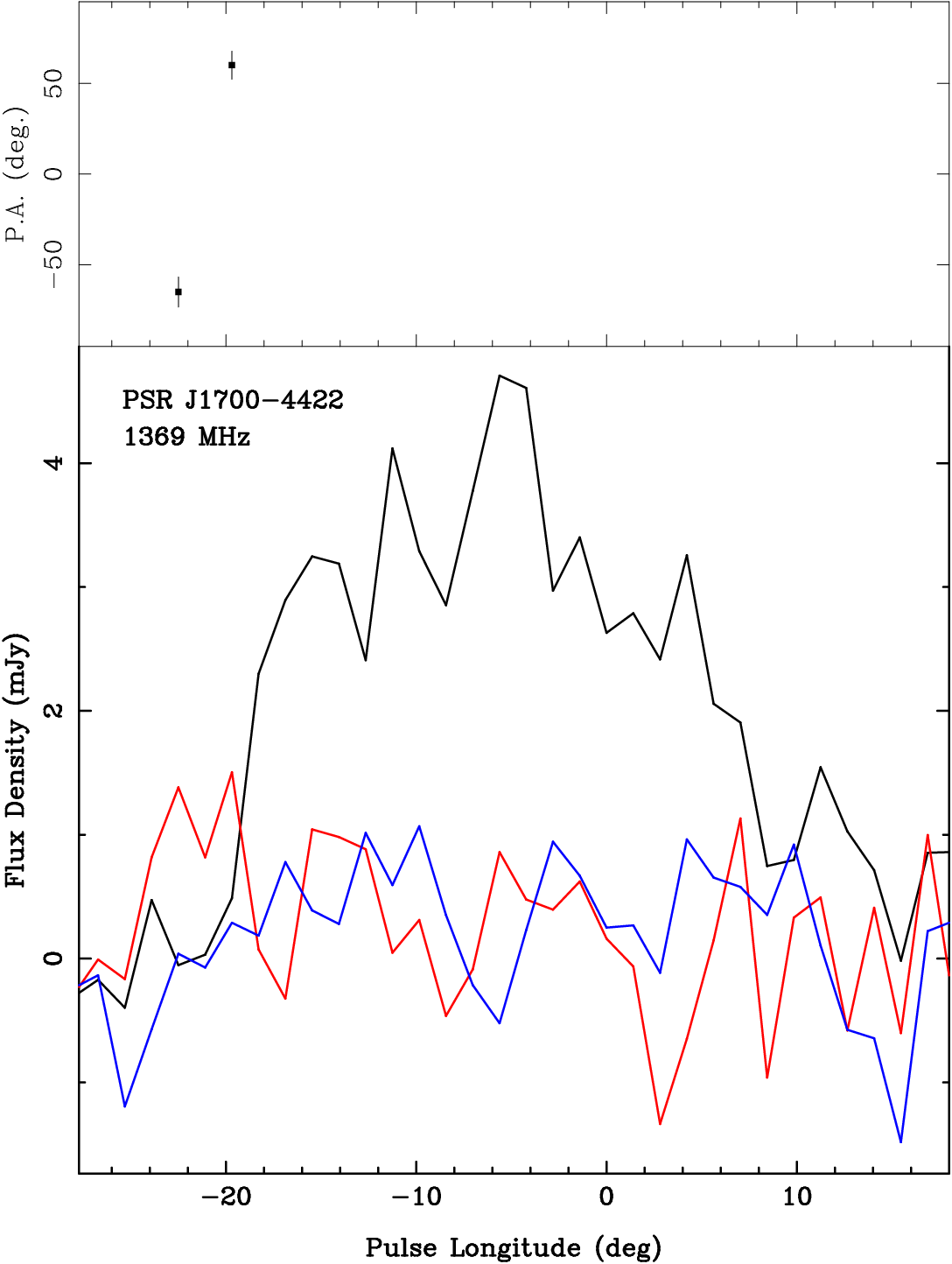}}&
\resizebox{0.3\hsize}{!}{\includegraphics{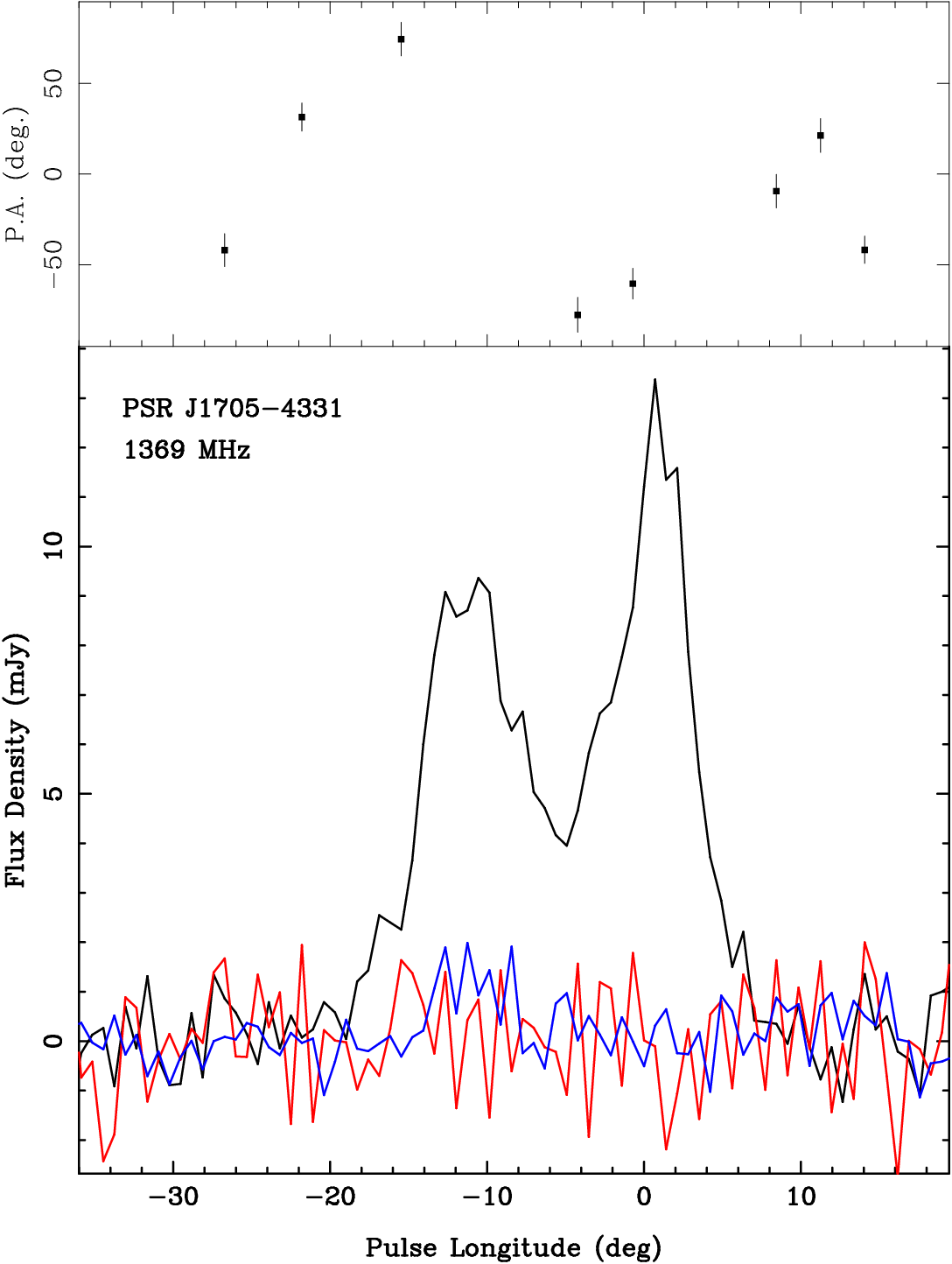}}\\
\end{tabular}
\end{center}
\caption[]{(continued)}
\end{figure*}

\addtocounter{figure}{-1}
\begin{figure*}
\begin{center}
\begin{tabular}{ccc}
\resizebox{0.3\hsize}{!}{\includegraphics{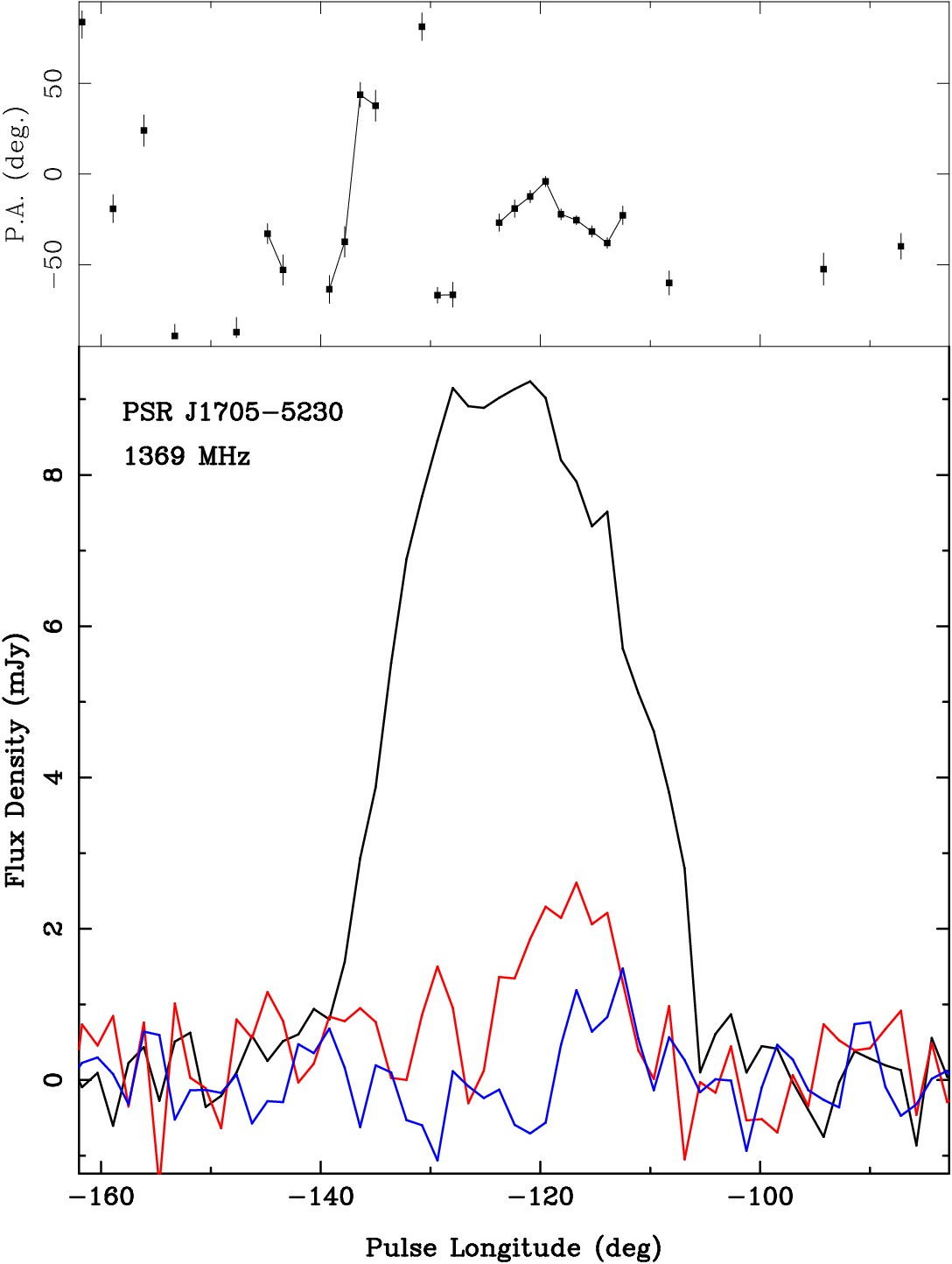}}&
\resizebox{0.3\hsize}{!}{\includegraphics{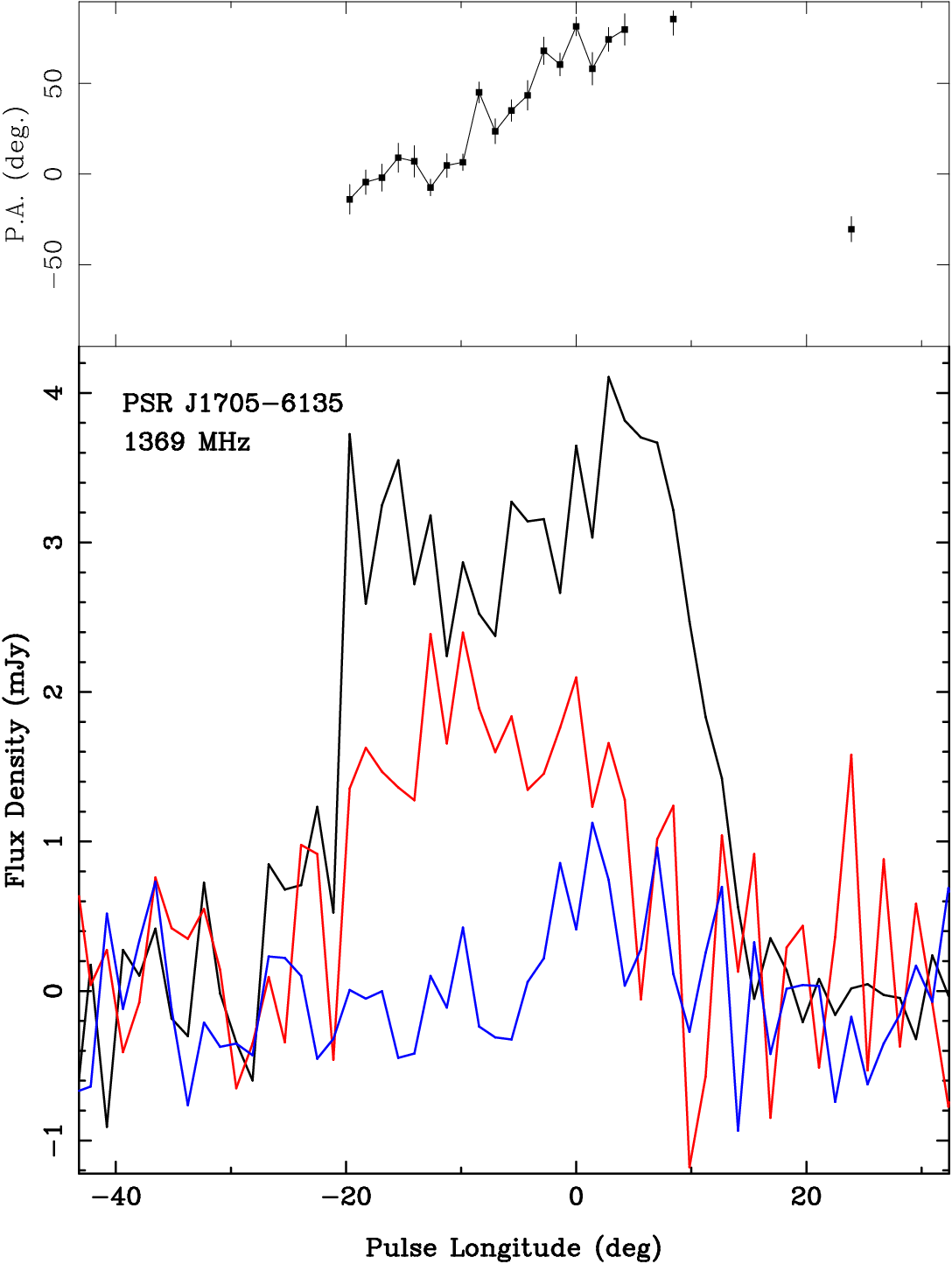}}&
\resizebox{0.3\hsize}{!}{\includegraphics{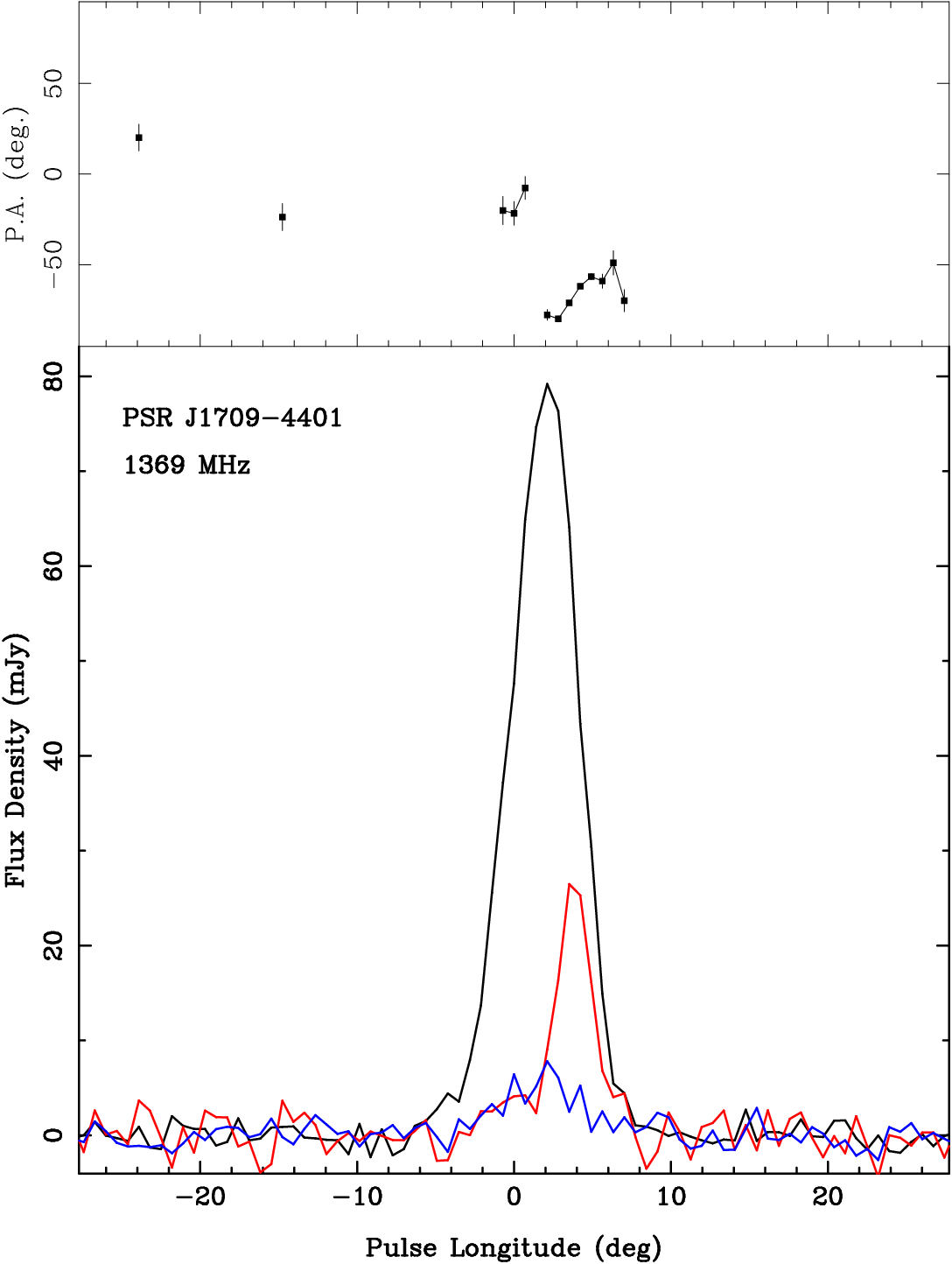}}\\
\resizebox{0.3\hsize}{!}{\includegraphics{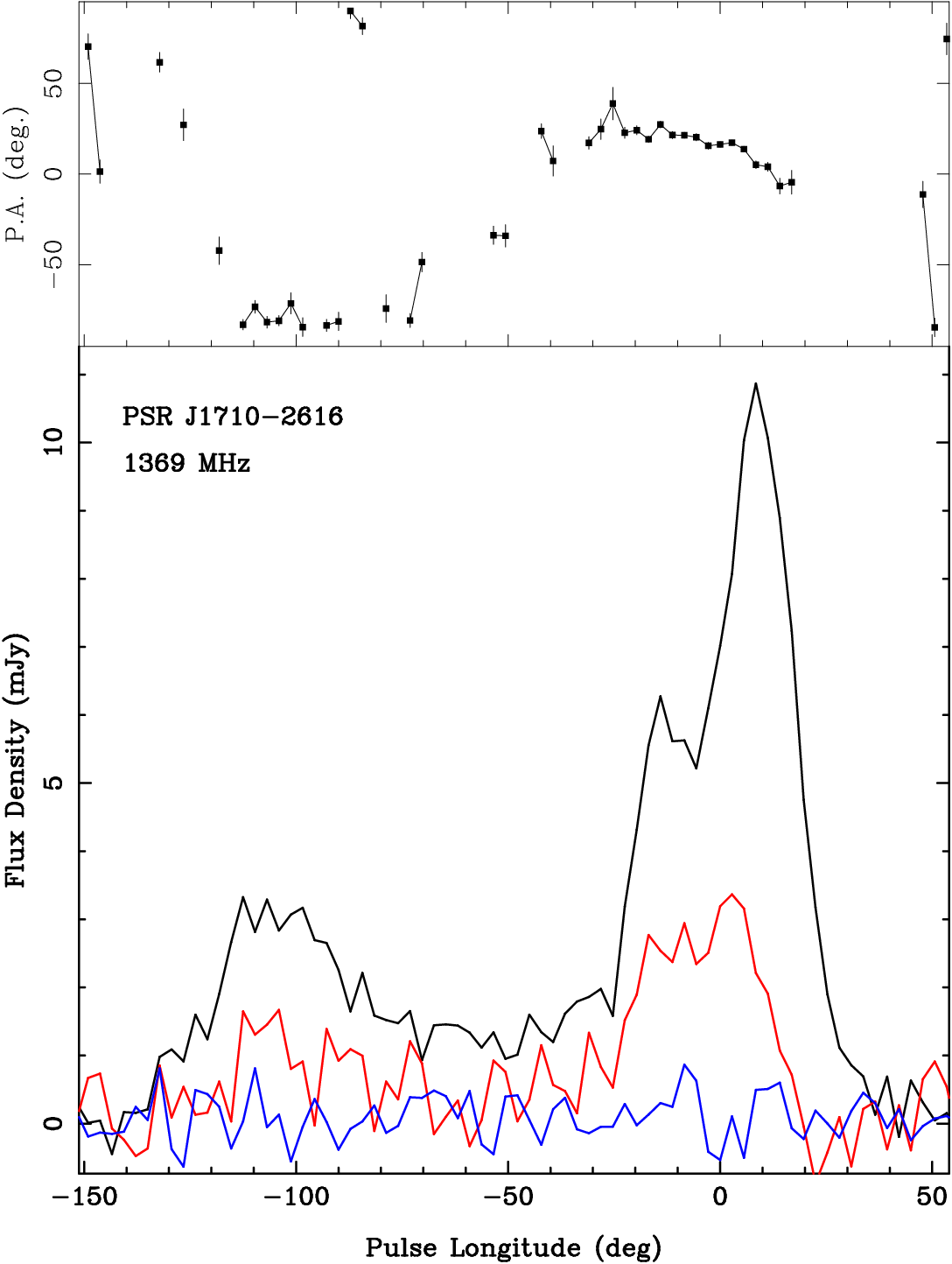}}&
\resizebox{0.3\hsize}{!}{\includegraphics{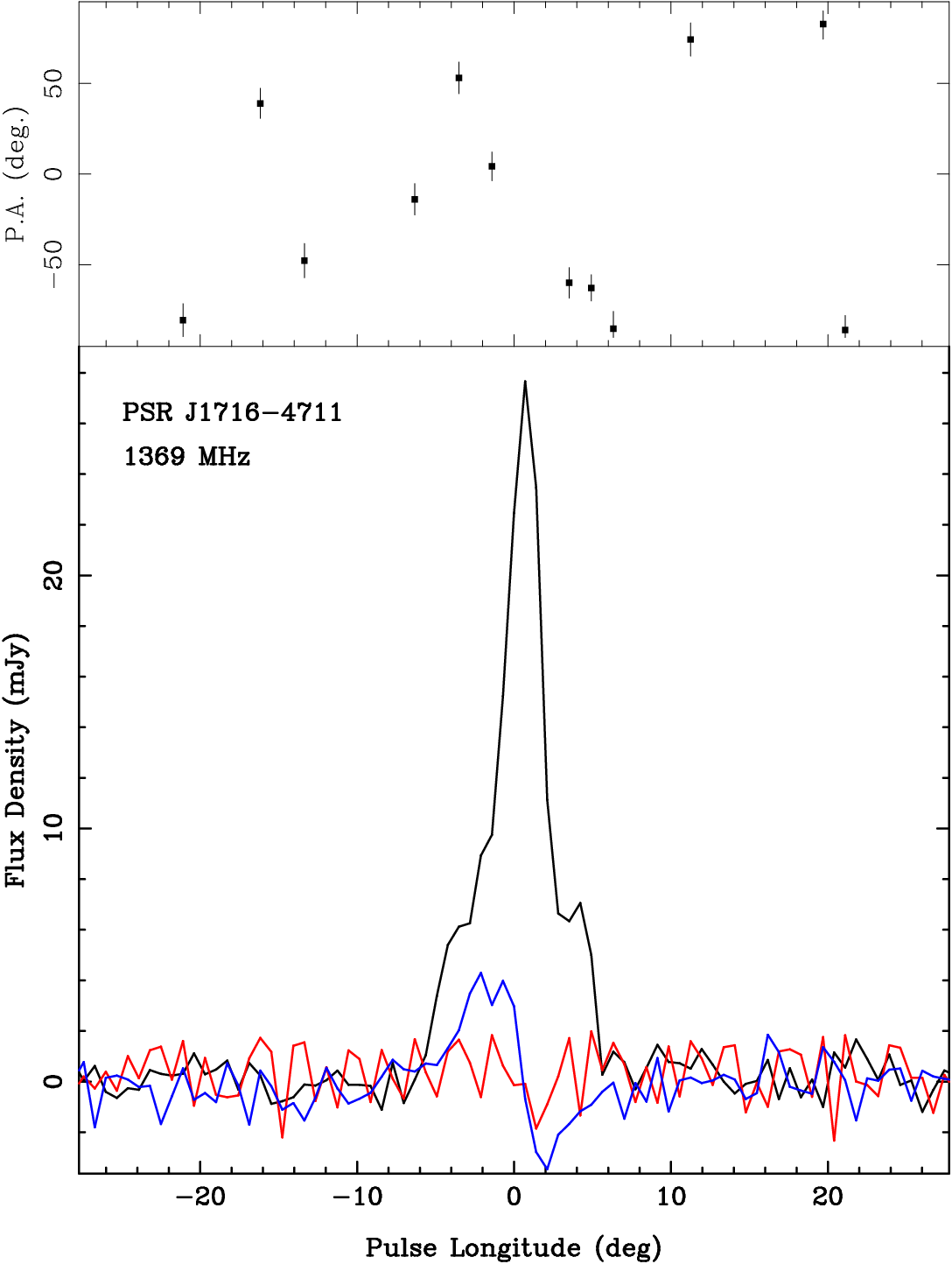}}&
\resizebox{0.3\hsize}{!}{\includegraphics{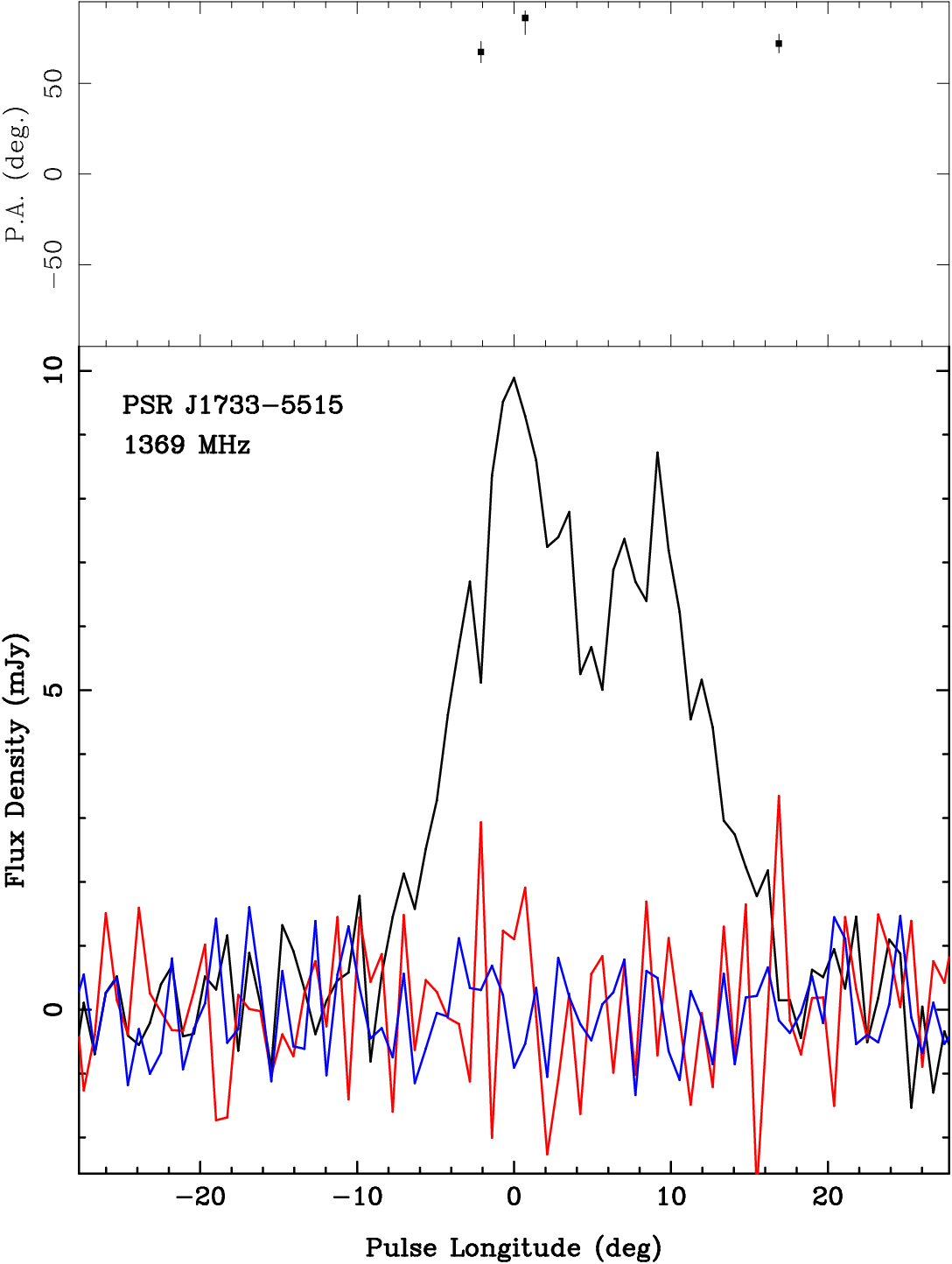}}\\
\resizebox{0.3\hsize}{!}{\includegraphics{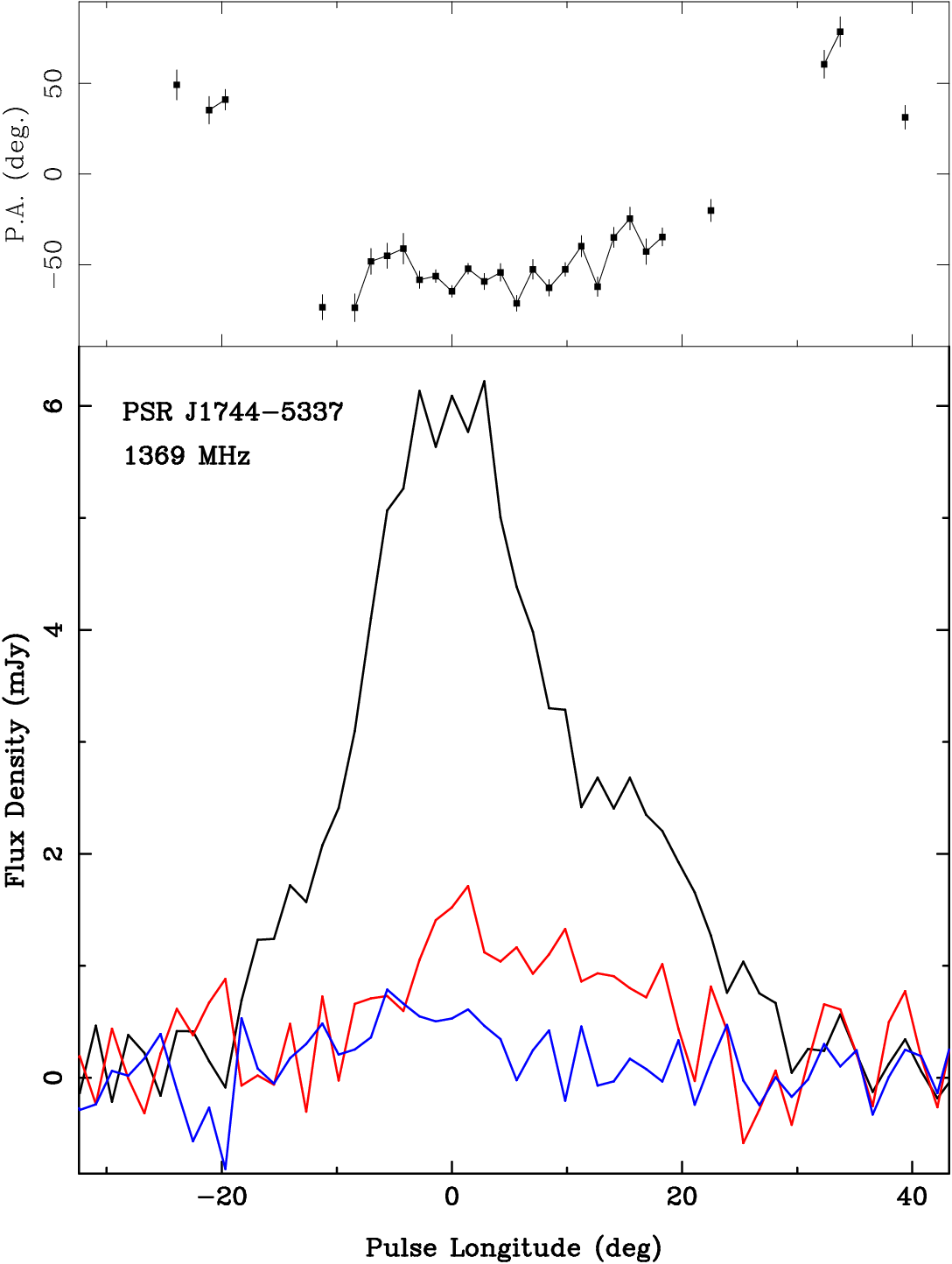}}&
\resizebox{0.3\hsize}{!}{\includegraphics{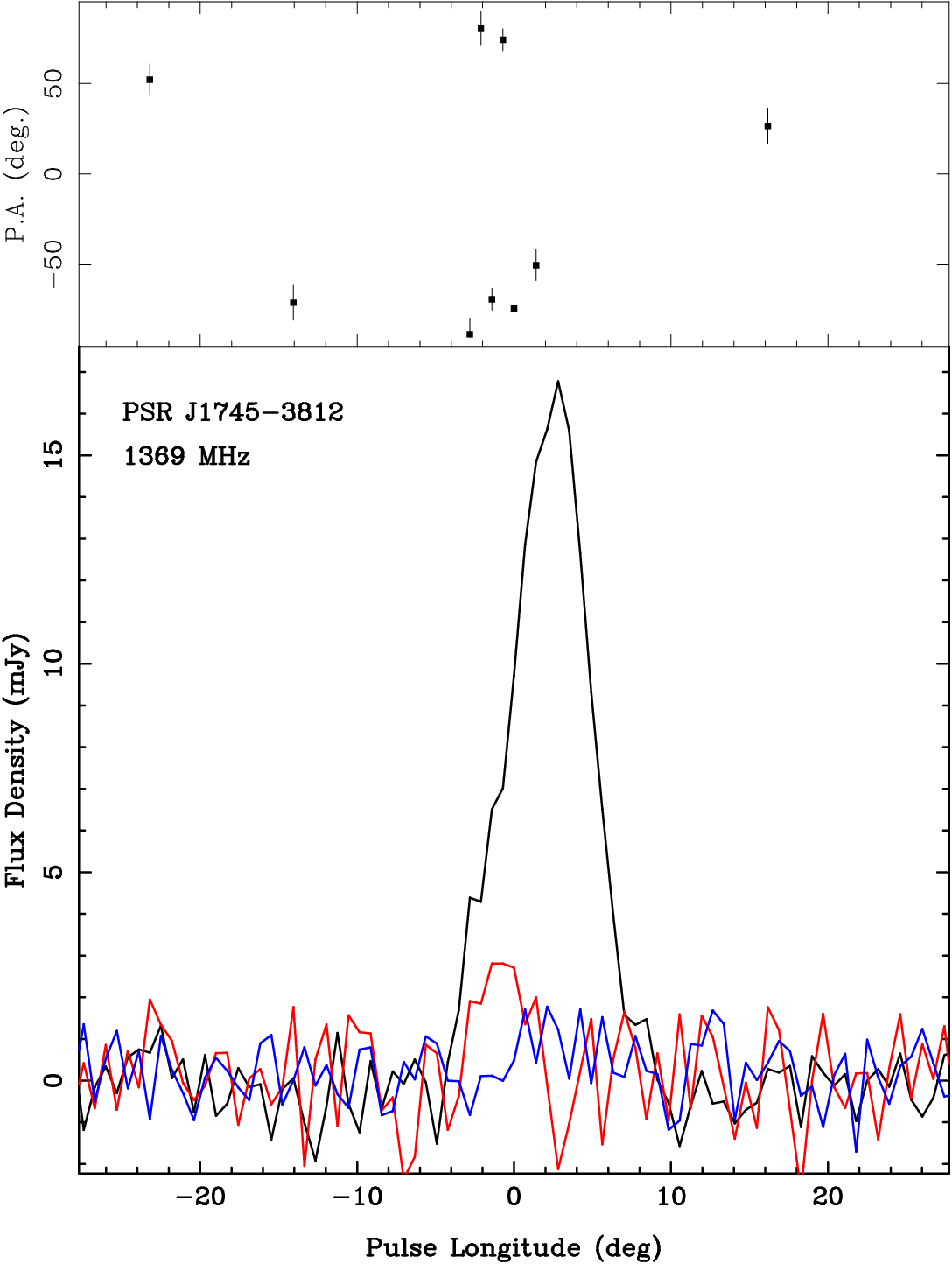}}&
\resizebox{0.3\hsize}{!}{\includegraphics{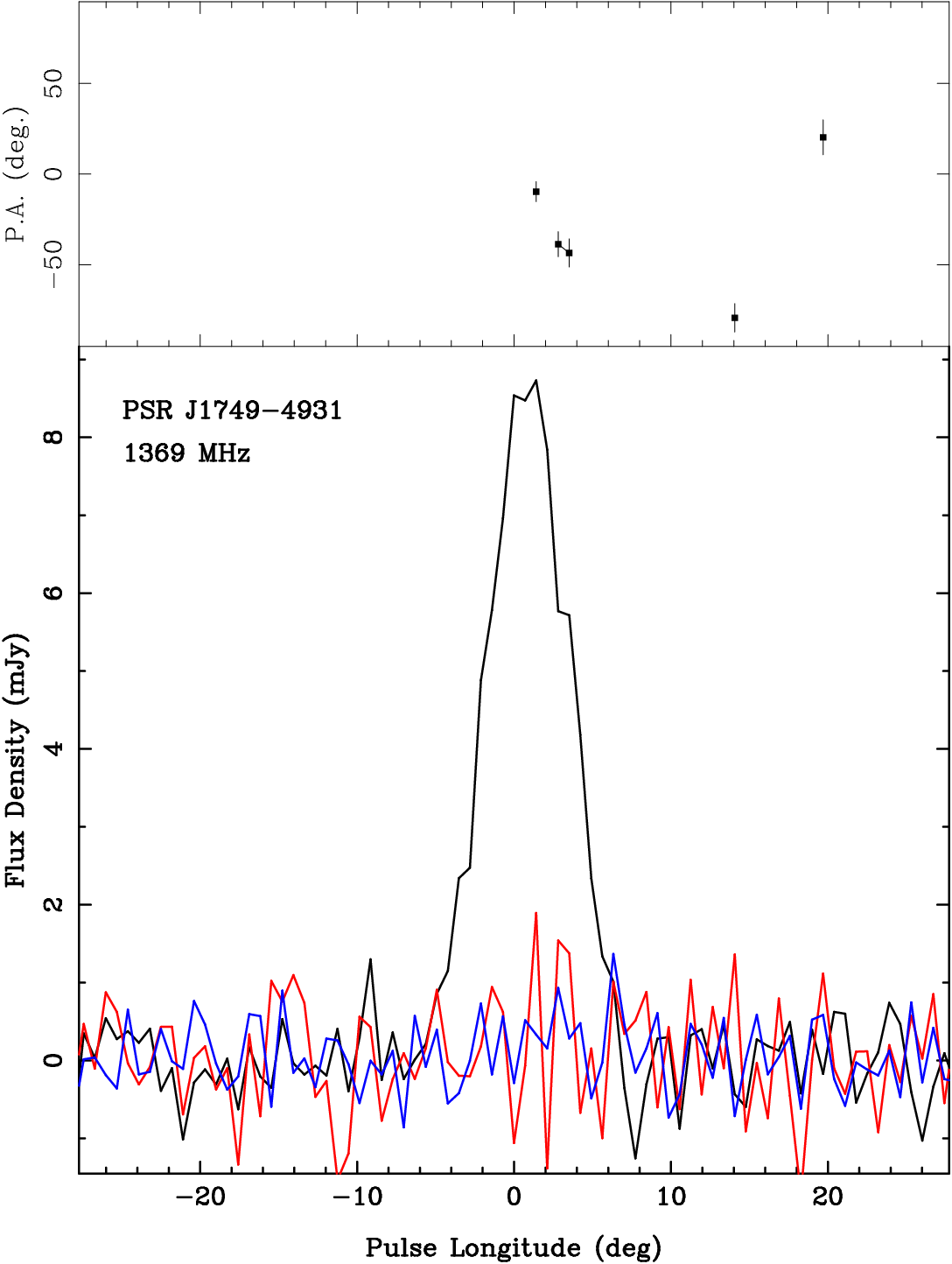}}\\
\end{tabular}
\end{center}
\caption[]{(continued)}
\end{figure*}

\addtocounter{figure}{-1}
\begin{figure*}
\begin{tabular}{cc}
\resizebox{0.3\hsize}{!}{\includegraphics{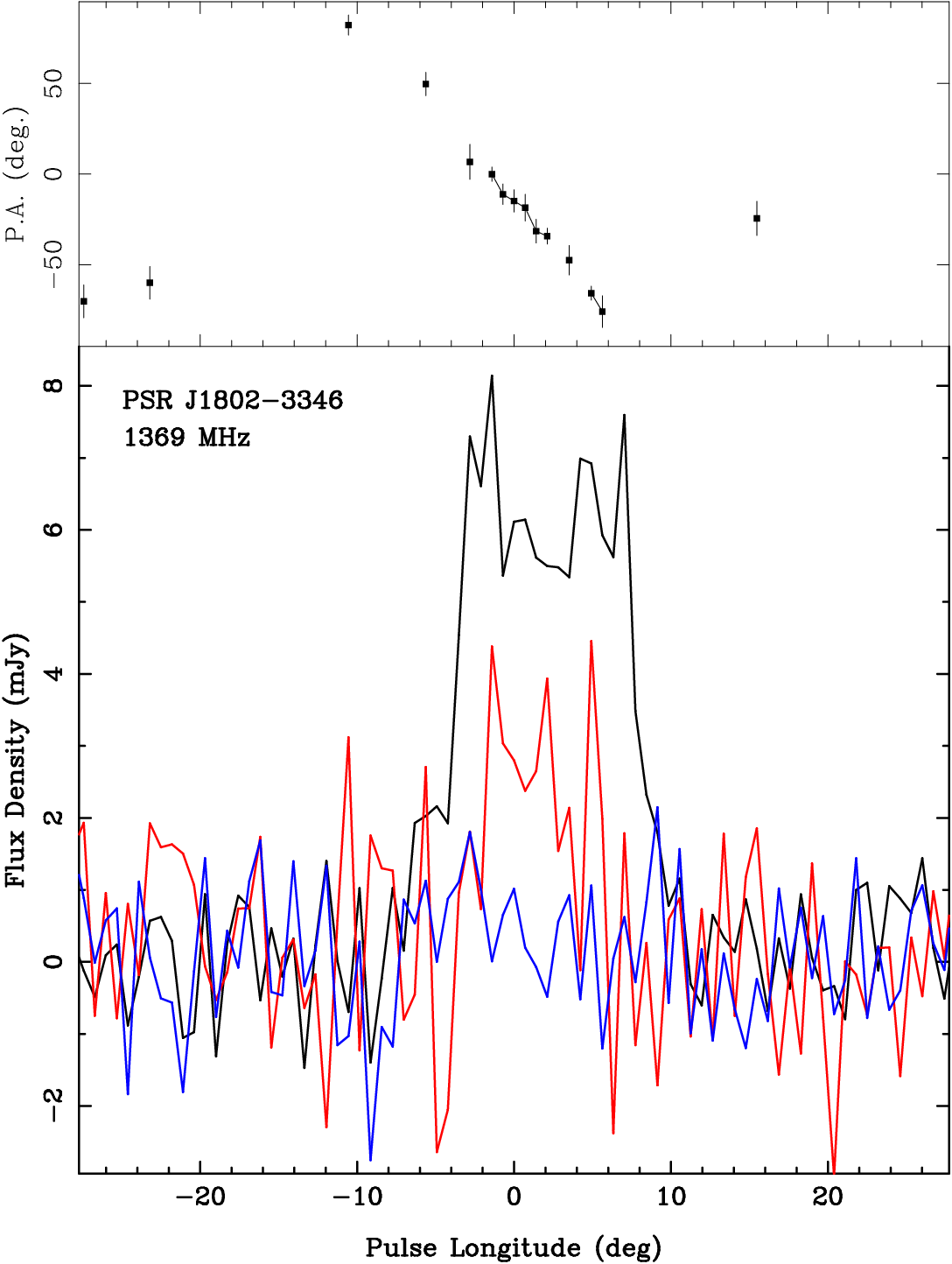}}&
\resizebox{0.3\hsize}{!}{\includegraphics{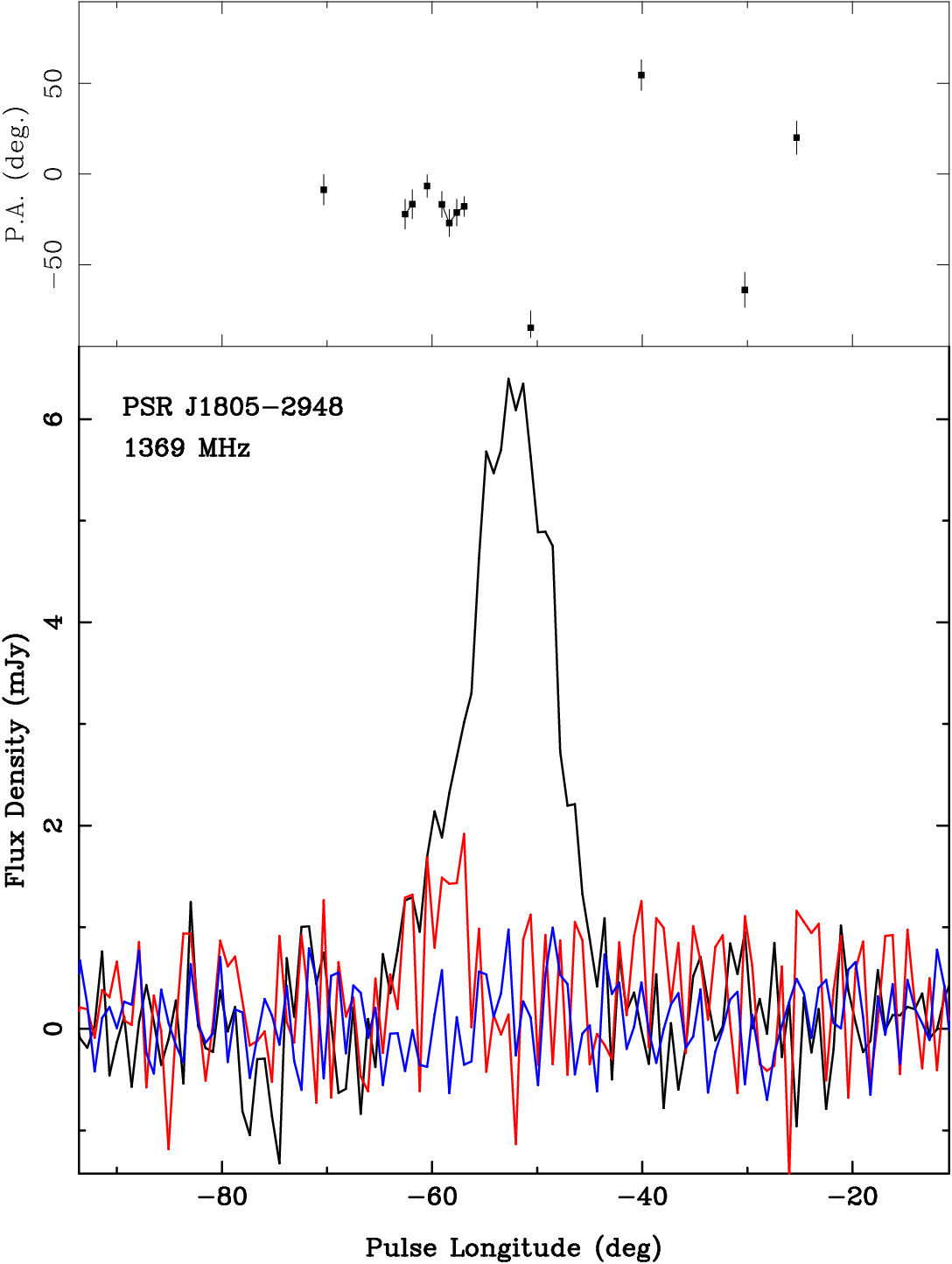}}\\
\resizebox{0.3\hsize}{!}{\includegraphics{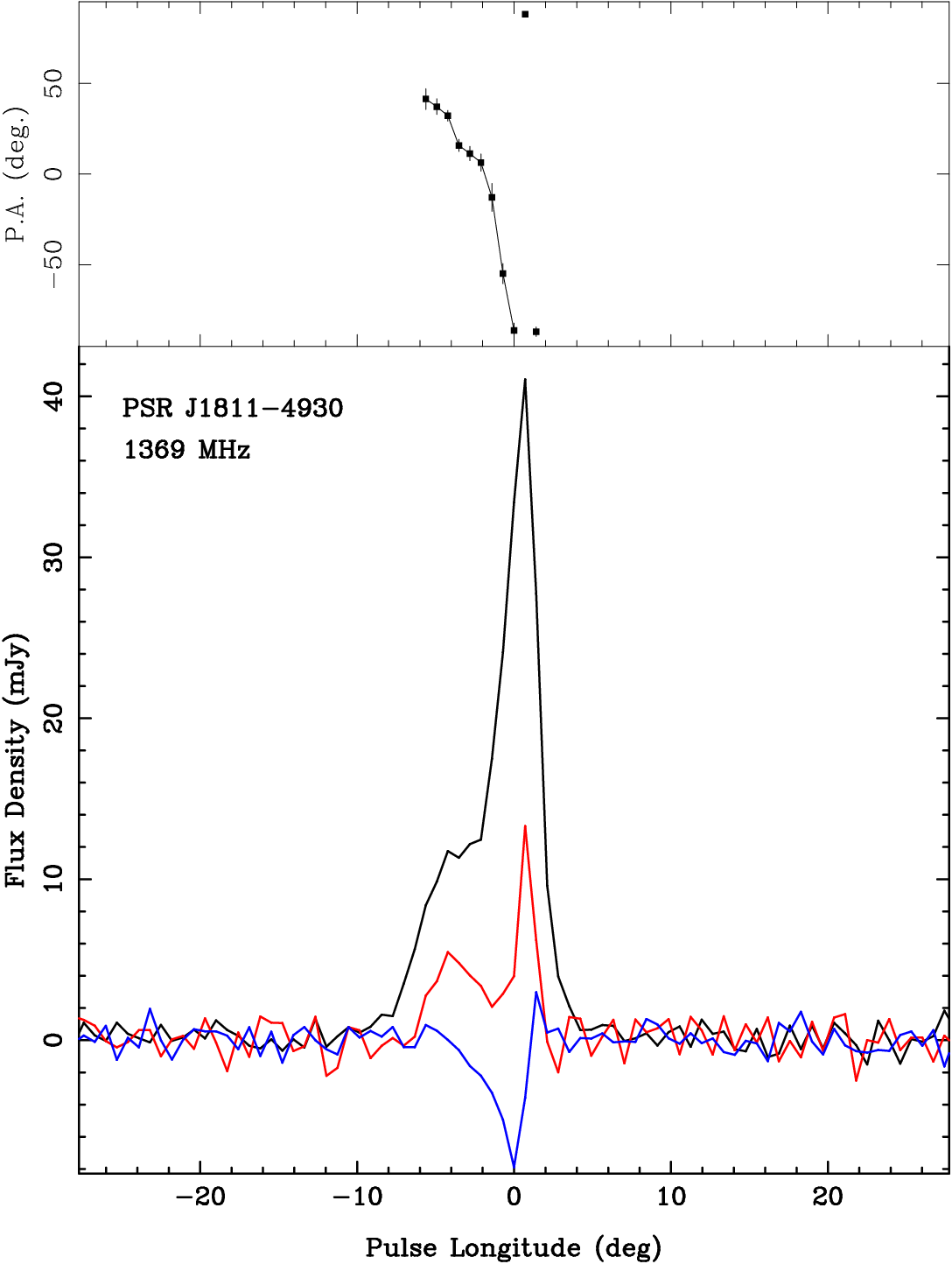}}&
\resizebox{0.3\hsize}{!}{\includegraphics{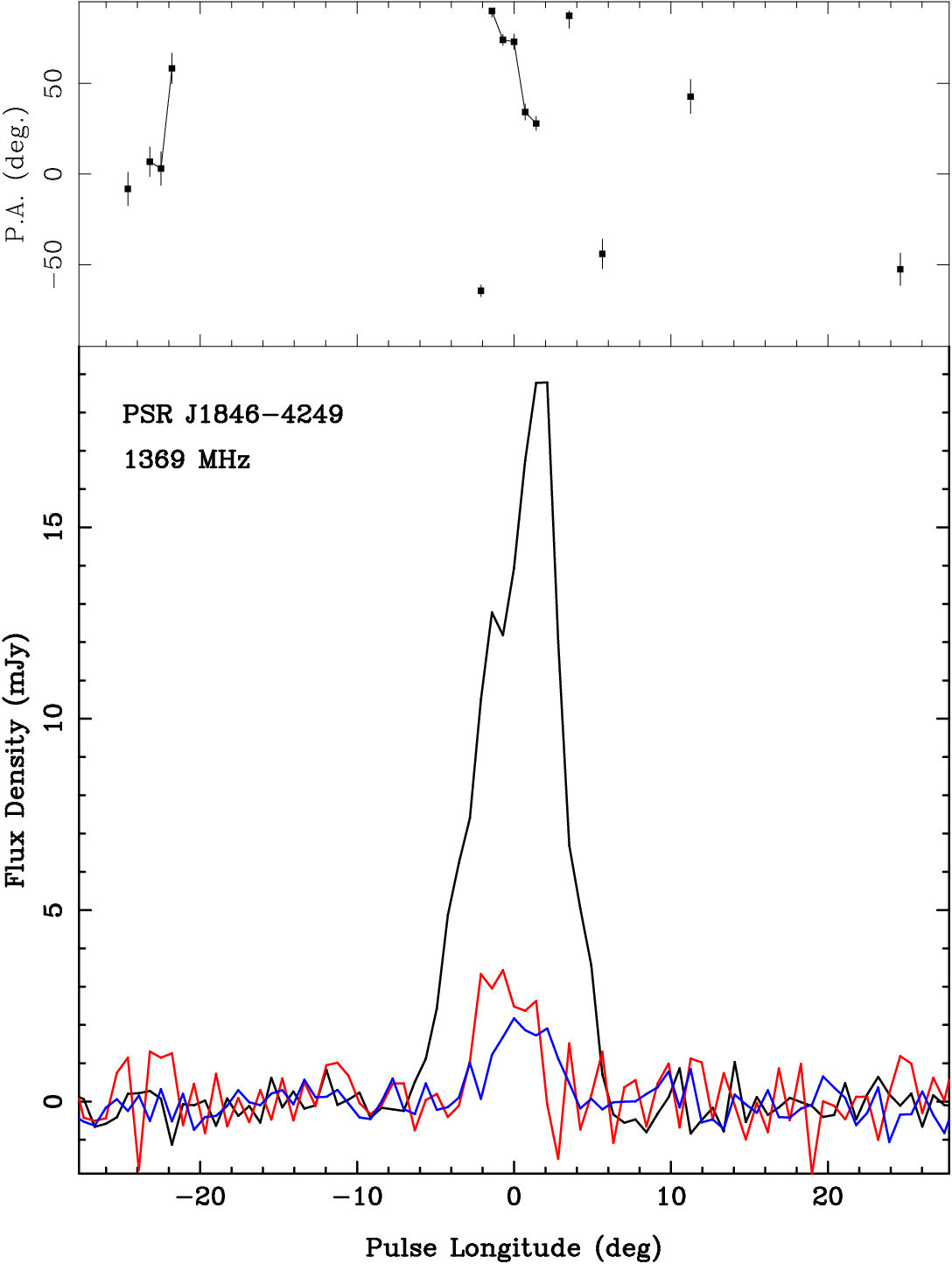}}\\
\end{tabular}
\caption[]{(continued)}
\end{figure*}

\noindent\textbf{J0807--5421}\\
The profile is relatively narrow, but the total intensity shows two clear, though blended components, with the trailing being the brightest. In contrast, the linear polarisation peaks in the centre of the profile and is significantly narrower than the total intensity. The circular polarisation displays a sign change towards the trailing component. The ${\rm PA}$ curve does not exhibit the swing expected from the RVM, but rather a sort of an arch.\\
\\
\\
\textbf{J0905--6019}\\
The profile is relatively narrow and it shows an asymmetric single peak. Although the linear polarisation is low, we were able to derive a ${\rm RM}$. The circular polarisation is faint and left-handed.\\
\\
\\
\textbf{J0912--3851}\\
The profile shows two distinct, narrow components, with the leading component being brighter than the trailing one. The linear polarisation also shows two peaks, narrower than in total intensity. The circularly polarised signal displays a sign change in the centre of the profile. We computed a ${\rm RM}$ value for each of the linear polarisation peaks and we found them to be compatible with only overlapping the extremes of the respective 1 $\sigma$ error bar.\\
\\
\\
\textbf{J0949--6902}\\
This bright integrated profile shows two almost completely blended, relatively narrow components. The linear polarisation is faint, and the circular polarisation exhibits a change of sign in the profile centre.\\
\\
\\
\textbf{J1036--6559}\\
The total intensity, the linear and the circular profiles all show a single component. The ${\rm PA}$ curve appears to increase with the phase longitude, and to decrease at its very end.\\
\\
\\
\textbf{J1105--4353}\\
The total intensity is noisy and single-peaked. The linear polarisation is noisy as well, and the circular polarisation is basically absent. The ${\rm PA}$ curve has no real pattern.\\
\\
\\
\textbf{J1237--6725}\\
The profile shows two blended components, with the leading component being the brightest. There is a faint signature of the presence of linear polarisation. Note that the observations were folded with a period that is half the real one, which was discovered at a later time. This can have affected the quality of the observations.\\
\\
\\
\textbf{J1251--7407}\\
The total intensity profile is narrow and asymmetric, made of at least three blended components.  Linear and circular polarisation appear to be significant under the trailing component. The ${\rm PA}$ profile presents three changes of slope in the first half of the pulse profile. After that, it follows a smooth swing with a positive slope that covers about $50\degree$ before changing the sense of the slope at its very end.\\
\\
\\
\textbf{J1331--5245}\\
This noisy profile shows at least two blended components, where the leading is the brightest. The linear and circular polarisation profiles follow the total intensity to a large extent. However, the linear polarisation is larger under the leading component whereas the circular polarisation is more significant under the trailing components. The ${\rm PA}$ profile is flat along the leading component, and it shows a steep swing with a negative slope across the trailing one covering about $100\degree$. An almost orthogonal jump occurs between these two parts of the ${\rm PA}$ profile.\\
\\
\\
\textbf{J1346--4918}\\
The profile shows a single, asymmetric component. The circular polarisation exhibits a sign change against the maximum of the total intensity profile. The ${\rm PA}$ profile presents a very smooth decrease in the first half of the pulse profile.\\
\\
\\
\textbf{J1409--6953}\\
This noisy total intensity profile is box-shaped. It is perhaps a blended double, although this could be an effect of the occurrence of more than two components. The linear polarisation is as well noisy and the ${\rm PA}$ values have no real pattern. The circularly polarised profile exhibits a right-handed maximum against the profile trailing component.\\
\\
\\
\textbf{J1416--5033}\\
This noisy profile shows at least two components, whereof the leading one is the brightest. The linear polarisation profile is noisy, and there is no hint of circularly polarised signal.\\
\\
\\
\textbf{J1432--5032}\\
The total intensity profile is box-shaped. The linear polarisation is noisy but significant, and the left-handed circularly polarised profile is mainly present close to the leading edge of the total intensity curve. The ${\rm PAs}$ exhibit a smooth swing across the profile covering about $70\degree$. Note that the observations were folded with a period that is half of the real one, discovered at a later time. This can have affected the quality of the observations.\\
\\
\\
\textbf{J1443--5122}\\
This noisy and relatively broad profile is asymmetric and shows a single component. The linearly polarised profile is significant when close to the leading edge, and another peak occurs at the centre of the profile. There is almost no circular polarisation. The ${\rm PAs}$ exhibit a smooth swing with a positive slope, covering about $120\degree$.\\
\\
\\
\textbf{J1517--4636}\\
The profile displays a narrow, single component. The linear polarisation largely follows the total intensity but it is narrower. The ${\rm PA}$ curve exhibits a change of slope close to the leading edge of the pulse profile, followed by a steep swing with a positive slope, that extends over $\sim50\degree$.\\
\\
\\
\textbf{J1530--6336}\\
The total power profile shows two principal components, with the leading being the brightest. The circular polarisation follows the total power, but it is narrower. On the contrary, the linear polarisation is characterised by at least three components. The first two of them are almost blended and occur before the trailing peak of the total intensity profile. The ${\rm PAs}$ show two swings with similar slopes under the two leading components of the linear polarisation. They are separated by an OPM jump. The third part of the ${\rm PA}$ profile is a swing with a flatter slope.\\
\\
\\
\textbf{J1534--4428}\\
The total intensity profile extends over more than $40\degree$ and consists of a bright leading component followed by a flat structure. There is a significant degree of linear polarisation which largely tracks the total intensity profile. If we interpret this structure as a zone of partially overlapping components, the depolarisation can be explained thanks to the fact that the linearly polarised profile is narrower than the total power one. The ${\rm PA}$ curve is largely flat but rises steeply in the middle of the profile before flattening off again.\\
\\
\\
\textbf{J1551--4424}\\
This profile is affected by interstellar scattering, and it shows a typical steep rising edge to a peak followed by a more gradual decay. The small linear polarisation fraction is concentrated towards the leading edge of the profile. The ${\rm PA}$ swing is remarkably flat, an effect that is induced by the scattering (\citet{lh03}).\\
\\
\\
\textbf{J1552--6213}\\
The total intensity profile is single-peaked and slightly asymmetric, with the trailing edge being steeper than the leading. The circular polarisation is barely visible and slightly right-handed in the second part of the profile. The linear polarisation shows two components, with the brightest roughly corresponding to the maximum of the total power. The first part of the ${\rm PA}$ profile is followed by a non-orthogonal jump. The second part shows a generally rising trend.\\ 
\\
\\
\textbf{J1607--6449}\\
The profile is made of at least two almost completely blended components. The linear polarisation is noisy. There is a significant occurrence of the right-handed, circularly polarised signal, that is mostly present in the first half of the pulse profile.\\
\\
\\
\textbf{J1612--5805}\\
The total intensity profile shows three features: a narrow, slightly asymmetric leading component and blended, fainter central and trailing components. The linearly polarised profile is mainly present beneath the leading component, and its peak almost coincides with the maximum of the total power. The circular polarisation shows a change of sign between the leading and the central component. The ${\rm PA}$ curve starts flat and exhibits a very steep swing with negative slope across almost the entire leading peak, followed by a slightly increasing curve toward the centre of the pulse profile. The two parts of the ${\rm PA}$ profile are separated by a jump of $\sim 110\degree$.\\
\\
\\
\textbf{J1614--3846}\\
The total intensity profile is noisy, box-shaped and symmetrical, and these attributes are largely mirrored by the linear polarisation. The circular polarisation is almost absent. The ${\rm PA}$ curve exhibits a smooth swing, with a positive slope that covers about $50\degree$.\\
\\
\\
\textbf{J1622--3751}\\
The profile is a blended double with the trailing component being the brightest. The linear polarisation, in contrast, is more significant in the leading component. The circular polarisation changes sign in the centre of the profile. Unusually amongst this sample, the ${\rm PA}$ profile shows the classic RVM signature: a flat beginning followed by a wide swing beneath the linearly polarised leading and trailing peaks, and flatter again at the end of the profile. The swing centre coincides with the minimum in the total intensity. It covers about $120\degree$.\\
Despite of the fact that the pulse profile is narrow, the steep swing of ${\rm PA}$ lends itself to the RVM fitting. We find that, although the angle between the spin and magnetic axis is unconstrained, the impact angle must be less then 4$\degree$. Interestingly also, the inflexion point of the RVM (the magnetic pole) aligns with the midpoint of the profile to within 0.5$\degree$. The lack of significant offset implies a low emission height of less than 100~km. Such a low emission height favours a non-orthogonal rotator with preferred values of $\alpha \lesssim 40\degree$.\\
\\
\\
\textbf{J1626--6621}\\
The profile shows two distinct, relatively narrow components, with the leading component being significantly brighter. The linear and the circular polarisations occur in correspondence of the profile's leading component. The ${\rm PA}$ profile exhibits a steep swing across this peak, which covers about $50\degree$.\\
\\
\\
\textbf{J1627--5936}\\
This broad profile extends over more than 100\degree of longitude. It shows an asymmetric, relatively narrow leading component followed by a central structure and a broader and fainter trailing component that is probably a blended double. The linear and circular polarisation profiles roughly follow the total power. However, while all three approximately peak at the same longitude in the leading component, in the trailing one the maxima of linear and circular polarisations are shifted. In the central structure, both circular and linear depolarisations occur. A change of handedness is displayed by the circular polarisation between the two main components. The ${\rm PAs}$ are mainly flat beneath the leading component, and show a swinging behaviour compatible with the RVM predictions in correspondence of the trailing component.\\
\\
\\
\textbf{J1629--3636}\\
The total power shows two peaks, with the asymmetric trailing being the brightest and narrowest. The linearly polarised profile mirrors the total intensity but it is narrower, while the circular polarisation is visible just in correspondence of the trailing component and it is left-handed. The ${\rm PA}$ profile is flat for both of the linearly polarised components.\\
\\
\\
\textbf{J1648--6044}\\
The profile has a single, asymmetric peak. However, the linear polarisation displays two clear components. The ${\rm PA}$ curve starts with a smooth swing that covers about $50\degree$, and continues with an almost orthogonal jump between the two components of the linearly polarised profile. The last part of the ${\rm PAs}$ value is practically flat.\\ 
\\
\\
\textbf{J1705--4331}\\
The profile shows a classic double structure with the trailing component slightly brighter than the leading component. There is some circular polarisation in the leading component.\\
\\
\\
\textbf{J1705--5230}\\
The profile is relatively broad and box-shaped, possibly a blend of several components. The linear polarisation shows a first, weak peak followed by a brighter one close to the total power trailing edge. In correspondence of the main, linearly polarised component, the ${\rm PAs}$ exhibit a practically flat trend.\\
\\
\\
\textbf{J1705--6135}\\
The profile is noisy, broad and box-shaped, and it is possibly a blended double. The fraction of linear polarisation is relatively high, particularly against the leading part of the profile. The ${\rm PA}$ curve exhibits a smooth swing across the profile, with a positive slope that covers $\sim 130\degree$.\\
\\
\\
\textbf{J1709--4401}\\
The profile of this intermittent pulsar shows a single, relatively narrow and pretty symmetrical peak. The linear polarisation has a main component close to the total intensity trailing edge, and it shows a hint of a minor peak on the leading side. The flux density is about one third of the total power one, and their maxima are misaligned. The circular polarisation is scarce and noisy. 
Beneath the weak leading component in the linearly polarised profile, the ${\rm PA}$ curve starts flat and follows a steep trend with a positive slope. The ${\rm PA}$ profile under the main linear polarisation peak is separated from the leading one by an almost orthogonal jump. It has a flat start too, followed by two swings with positive and negative slopes, respectively. \\
\\ 
\\
\textbf{J1710--2616}\\
This broad profile shows emission over nearly $180\degree$ of longitude. A broad leading component is followed by a bridge of emission linking it to a blended double. The linear polarisation mostly follows the total intensity but the circular polarisation remains low throughout. Although the low linear polarisation in the profile centre, the characteristic S-shape from the RVM is recognisable. In fact, the large longitude coverage of the pulse profile and the smooth ${\rm PA}$ swing lends itself well to RVM fitting. Results show that $\alpha$ must be less than 30$\degree$, with an impact parameter of $\sim$20$\degree$ or less. The location of the inflexion point of the RVM is coincident with the profile centre. The pulsar therefore appears to be an almost aligned rotator.\\
\\
\\
\textbf{J1716--4711}\\
The profile shows a single, relatively narrow component possibly flanked by two outriders. The circular polarisation displays a clear change of sign in correspondence of the profile centre.\\
\\
\\
\textbf{J1733--5515}\\
The profile shows two blended components of almost equal amplitude. Very small linear or circular polarisation can be discerned.\\
\\
\\
\textbf{J1744--5337}\\
The profile is affected by the interstellar scattering. It shows a broad and asymmetric leading component blended with a second one. The linear polarisation profile is significant especially in the second half of the pulse profile. The ${\rm PA}$ curve is flat.\\
\\
\\
\textbf{J1745--3812}\\
The profile shows a single and slightly asymmetric component with low circular polarisation. In spite of a moderate degree of linear polarisation, we were not able to obtain a ${\rm RM}$ value for this pulsar.\\
\\
\\
\textbf{J1749--4931}\\
This single-peaked profile shows no clear signs of circular polarisation, while the linear polarisation is present but weak.\\
\\
\\
\textbf{J1802--3346}\\
This noisy and box-shaped profile shows at least two blended components. The linear polarisation profile follows the total intensity but it is narrower, and the ${\rm PA}$ curve displays a swing with negative slope that covers $\sim 90\degree$.\\
\\
\\
\textbf{J1805--2948}\\
This noisy profile shows a single, relatively broad and asymmetric component. There appears to be a linearly polarised component on the leading edge of the profile with a flat ${\rm PA}$ swing.\\
\\
\\
\textbf{J1811--4930}\\
The profile of this intermittent pulsar is a blended double with the trailing component brighter. The linear polarisation follows the total power, though it is narrower. On the other hand, the circular polarisation peaks where $L$ is fainter. The ${\rm PA}$ profile shows a steep swing with negative slope in correspondence of the leading component that covers about $130\degree$, while it is flat beneath the trailing.\\
\\
\\
\textbf{J1846--4249}\\
This profile shows two blended peaks. The linear and circular polarisations, however, exhibit a single, box-shaped component at the centre of the pulse profile. The ${\rm PA}$ curve shows a steep swing spanning $80\degree$.
\begin{table*}
\caption{Pulsars for which ${\rm RM}$ can be determined. We show the spin period ($P$), the profile widths at 10\% (W$_{10}$) and 50\% (W$_{50}$) of the total intensity peak, the logarithm of the spin down luminosity (Log$\dot{E}$), the total intensity flux ($S0$), the percentages of the linear, the circular and the absolute value of the circular polarisations ($L\%$, $V\%$, $|V|\%$), the rotation and the dispersion measures (${\rm RM}$ and ${\rm DM}$), the ${\rm DM}$ derived distance from the Sun (via the NE2001 electron density model from \citealp{cl02}, that gives uncertanties up to about $30\%$), the average value of the magnetic field along the line of sight ($<B_{||}>$) and the logarithm of the characteristic age (Log~$\tau_{C}$). 1 $\sigma$ errors on the last digit(s) are reported in parentheses. 3 $\sigma$ errors are reported for $S0$.}\label{Tab:RMpar}
\begin{center}
\footnotesize
\begin{tabular}{| c  c  c  c  c  c  c  c  c  c  c  c  c  c|}
\hline
Name & P   &   W$_{10}$  &  W$_{50}$ &  Log~$\dot{E}$ &  S0 &  L\% &  V\% &  $|$V$|$\% &  ${\rm RM}$ & ${\rm DM}$ & Distance & $<B_{||}>$ & Log~$\tau_{C}$ \\
     & [s] &   [ms]     &   [ms]   &                & [mJy] &    &    &   & [rad m$^{-2}$] & [pc cm$^{-3}$] & [kpc] & [$\mu$G] & \\  
\hline
J0807$-$5421 &  0.527 & 17 & 11 & 32.0 & 0.35(1) & 14.5(7) & 3(1) & 5.8(6) & $-$65(3) & 165 & 0.26 & $-$0.48 & 7.3\\     
J0905$-$6019 &  0.341 & 14 & 6 & 32.7 & 0.36(1) & 5.8(7) & 1(1) & 1.3(7) & $-$63(23) & 91 & 2.9 & $-$0.85 & 7.0\\        
J0912$-$3851 &  1.526 & 48 & 38 & 31.6 & 0.14(1) & 22(1) & 0(1) & 10(1) & 94(15) & 63 & 0.5 & 1.85 & 6.8\\           
J0949$-$6902 &  0.64 & 10 & 4 & 32.0 & 0.31(1) & 6.5(8) & 2(1) & 3.6(7) & $-$58(14) & 93 & 2.9 & $-$0.77 & 7.2\\         
J1036$-$6559 &  0.534 & 16 & 9 & 32.5 & 0.27(1) & 12(1) & 3(1) & 5.2(9) & $-$88(20) & 158 & 4.0 & $-$0.69 & 6.8\\        
J1105$-$4353 & 0.351 & 22 & 12 & 33.4 & 0.17(2) & 23(2) & 2(3) & 4(2) & 17(1) & 38 & 1.4 & 0.56 & 6.3\\ 
J1237$-$6725 &  2.111 & 40 & 30 & 31.0 & 0.48(2) & 4.8(9) & 0(1) & 1.6(8) & 24(14) & 176 & 3.9 & 0.17 & 7.2\\        
J1251$-$7407 &  0.327 & 14 & 3 & 32.6 & 0.24(1) & 23(1) & 6(2) & 6(1) & $-$121(9) & 89 & 2.4 & $-$1.66 & 7.2\\           
J1331$-$5245 &  0.648 & 43 & 27 & 31.9 & 0.32(2) & 30(1) & 16(1) & 17(1) & 83(5) & 148 & 4.2 & 0.69 & 7.3\\          
J1409$-$6953 &  0.529 & 31 & 24 & 32.4 & 0.26(2) & 16(1) & $-$4(2) & 8(1) & $-$30(10) & 171 & 4.6 & $-$0.22 & 7.0\\        
J1432$-$5032 &  2.035 & 52 & 33 & 31.4 & 0.29(2) & 18(1) & 4(1) & 4(1) & 11(3) & 113 & 2.8 & 0.13 & 6.7\\            
J1443$-$5122 &  0.732 & 117 & 47 & 31.5 & 0.68(3) & 22(1) & 1(1) & 2.6(9) & 43(6) & 87 & 1.9 & 0.61 & 7.5\\          
J1517$-$4636 &  0.887 & 27 & 16 & 32.1 & 0.37(1) & 19.6(9) & 5(1) & 5.0(8) & $-$68(7) & 126 & 3.1 & $-$0.66 & 6.8\\      
J1530$-$6336 & 0.91 & 11 & 32 & 31.6 & 0.43(1) & 22.3(8) & 16(1) & 16.7(7) & 202(11) & 201 & 4.9 & 1.24 & 7.2\\ 
J1534$-$4428 &  1.221 & 178 & 14 & 30.6 & 0.55(3) & 28(1) & $-$2(1) & 3(1) & 24(6) & 137 & 3.9 & 0.22 & 8.0\\          
J1551$-$4424 &  0.674 & 129 & 27 & 31.4 & 1.14(3) & 17.1(6) & 2(1) & 3.2(6) & $-$32(5) & 66 & 2.4 & $-$0.6 & 7.8\\       
J1552$-$6213 & 0.199 & 7 & 3 & 32.1 & 0.34(2) & 24(1) & 0(1) & 1(1) & 42(14) & 122 & 2.66 & 0.43 & 8.1\\ 
J1612$-$5805 &  0.616 & 22 & 4 & 32.2 & 0.31(2) & 16(1) & 3(1) & 9(1) & $-$21(12) & 172 & 3.6 & $-$0.15 & 7.0\\          
J1614$-$3846 &  0.464 & 45 & 17 & 32.6 & 0.18(2) & 31(2) & 3(3) & 3(2) & 45(9) & 111 & 2.7 & 0.51 & 6.9\\            
J1622$-$3751 &  0.731 & 48 & 24 & 32.4 & 0.20(1) & 30(1) & 7(2) & 9(1) & 85(7) & 154 & 3.9 & 0.69 & 6.7\\            
J1626$-$6621 &  0.451 & 39 & 3 & 32.5 & 0.19(2) & 20(1) & 12(3) & 14(1) & 39(12) & 84 & 2.2 & 0.58 & 7.0\\           
J1627$-$5936 &  0.354 & 159 & 85 & 30.8 & 1.62(4) & 22.2(6) & 1.3(9) & 7.0(6) & 89(5) & 99 & 2.2 & 1.11 & 8.9\\      
J1629$-$3636 & 2.988 & 41 & 12 & 31.0 & 0.20(2) & 29(1) & 2(2) & 2(1) & 0(4) & 101 & 2.4 & $-$0.001 & 6.8\\ 
J1648$-$6044 &  0.584 & 31 & 12 & 31.9 & 0.66(2) & 19.7(7) & 0(1) & 0.3(6) & 59(3) & 106 & 2.6 & 0.69 & 7.3\\        
J1705$-$5230 & 0.231 & 14 & 21 & 32.2 & 0.60(2) & 15.3(9) & 1(1) & 4.8(8) & $-$38(9) & 170 & 3.95 & $-$0.28 & 7.9\\ 
J1705$-$6135 &  0.809 & 85 & 43 & 30.6 & 0.30(3) & 36(2) & 2(3) & 4(1) & 86(7) & 106 & 2.9 & 1.0 & 8.4\\             
J1709$-$4401 &  0.865 & 12 & 24 & 32.7 & 1.15(3) & 21.5(5) & 7.9(8) & 7.9(5) & $-$122(2) & 225 & 4.4 & $-$0.67 & 6.3\\   
J1710$-$2616 &  0.954 & 393 & 99 & 30.0 & 1.40(5) & 32.5(8) & 2(1) & 4.4(7) & $-$9(3) & 111 & 2.6 & $-$0.1 & 8.9\\       
J1744$-$5337 &  0.356 & 49 & 18 & 32.2 & 0.39(2) & 20(1) & 3(1) & 4.1(8) & 52(6) & 124 & 3.6 & 0.52 & 7.5\\          
J1749$-$4931 & 0.446 & 8 & 13 & 32.4 & 0.15(1) & 9(1) & 3(2) & 3(1) & 42(21) & 52 & 1.4 & 1.00 & 7.1\\ 
J1802$-$3346 &  2.461 & 77 & 120 & 30.5 & 0.20(2) & 28(2) & 1(3) & 1(2) & 236(17) & 217 & 5.4 & 1.35 & 7.5\\    
J1805$-$2948 &  0.428 & 22 & 10 & 32.4 & 0.18(1) & 16(1) & 1(2) & 1(1) & 23(21) & 167 & 3.77 & 0.17 & 7.2\\
J1811$-$4930 &  1.433 & 39 & 11 & 31.5 & 0.46(2) & 23.2(8) & $-$8(1) & 11.0(8) & 42(6) & 44 & 1.2 & 1.19 & 7.0\\      
J1846$-$4249 &  2.273 & 67 & 60 & 30.6 & 0.29(1) & 13.6(9) & 8(1) & 8.0(8) & 82(10) & 62 & 1.8 & 1.63 & 7.5\\        
\hline
\end{tabular}
\end{center}
\end{table*}

\begin{table*}
\caption{Pulsars for which no ${\rm RM}$ can be determined. Parameters and errors like in Table \ref{Tab:RMpar}.}\label{Tab:noRMpar}
\begin{center}
\footnotesize
\begin{tabular}{| c  c  c  c  c  c  c  c  c  c  c|}
\hline
Name & P   &  W$_{10}$  &  W$_{50}$ &  Log~$\dot{E}$ &  S0 &  L\% &  V\% &  $|$V$|$\% & ${\rm DM}$ & Log~$\tau_C$\\
    & [s]  &   [ms]   & [ms] &               & [mJy] &      &    &  &  [pc cm$^{-3}$]  & \\  
\hline
J0919$-$6040 &  1.217 & 52 & 26 & 29.3 & 0.23(1) & 2(1) & 7(1) & 7(1) & 82 & 9.3  \\        
J1054$-$5946 &  0.228 & 29 & 7 & 32.8 & 0.23(2) & 5(2) & 3(2) & 5(1) & 253 & 7.2  \\        
J1143$-$5536 &  0.685 & 24 & 12 & 31.8 & 0.25(1) & 0(1) & 2(1) & 2(1) & 185 & 7.3  \\       
J1346$-$4918 &  0.3 & 18 & 10 & 31.7 & 0.70(2) & 6.9(6) & 3.9(9) & 5.6(5) & 74 & 8.1\\    
J1416$-$5033 &  0.795 & 25 & 12 & 31.0 & 0.13(1) & 14(2) & 0(3) & 3(2) & 58 & 8.0 \\        
J1539$-$4835 &  1.273 & 91 & 18 & 31.4 & 0.21(2) & 0(1) & 2(3) & 8(1) & 118 & 7.2 \\        
J1607$-$6449 &  0.298 & 19 & 3 & 31.6 & 0.22(2) & 10(1) & $-$6(2) & 12(1) & 89 & 8.3 \\       
J1625$-$4913 &  0.356 & 23 & 9 & 33.8 & 0.22(2) & 1(2) & 2(3) & 2(2) & 720 & 5.9 \\         
J1634$-$5640 &  0.224 & 15 & 8 & 32.2 & 0.24(2) & 2(1) & 1(2) & 3(1) & 149 & 7.9\\          
J1647$-$3607 &  0.212 & 15 & 5 & 32.7 & 0.17(2) & 11(2) & $-$2(4) & 2(2) & 222 & 7.4 \\       
J1700$-$4422 &  0.756 & 72 & 45 & 30.6 & 0.24(3) & 6(2) & 9(4) & 11(2) & 413 & 8.5 \\       
J1705$-$4331 &  0.223 & 23 & 6 & 32.4 & 0.43(2) & 3(1) & 3(1) & 3(1) & 185 & 7.7\\          
J1716$-$4711 &  0.556 & 15 & 4 & 32.3 & 0.31(2) & 5(1) & 6(1) & 19.3(9) & 287 & 7.0\\      
J1733$-$5515 &  1.011 & 69 & 45 & 31.3 & 0.38(3) & 6(1) & 0(2) & 1(1) & 84 & 7.5 \\       
J1745$-$3812 &  0.698 & 24 & 11 & 32.4 & 0.28(2) & 11(1) & 5(2) & 5(1) & 160 & 6.7\\ 
\hline
\end{tabular}
\end{center}
\end{table*}

\section{Discussion}

\begin{figure}
\begin{center}
\includegraphics[scale=0.49]{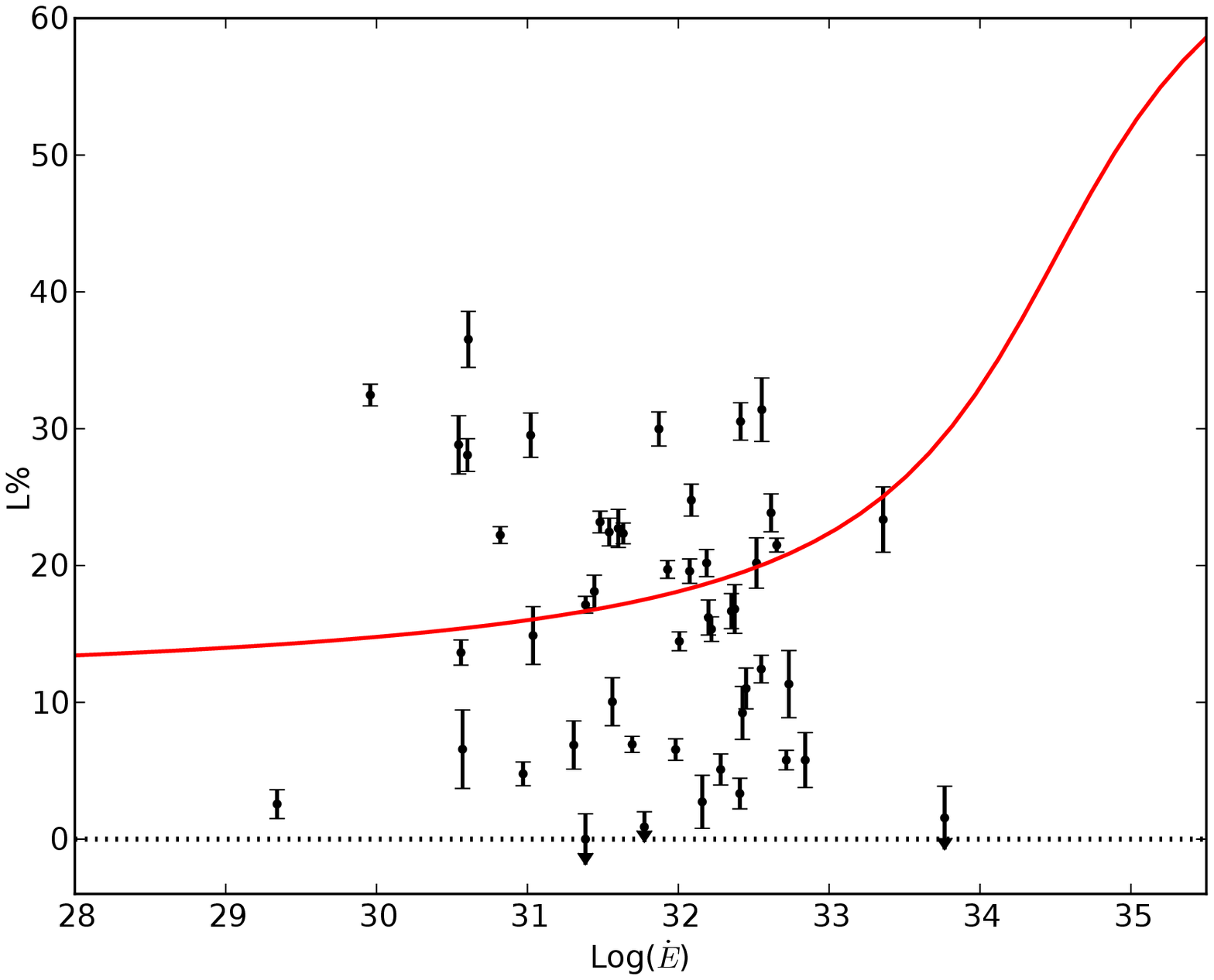}
\end{center}

\caption{Percentage of linear polarisation against the spin-down luminosity $\dot{E}$. The black points represent the individual pulsars of our sample with 1 $\sigma$ error bars (the arrows imply an upper limit), while the \textbf{red} line is the fit reported in \citet{wj08}.}
\label{edot}
\end{figure}

\begin{figure}
\begin{center}
\includegraphics[scale=0.49]{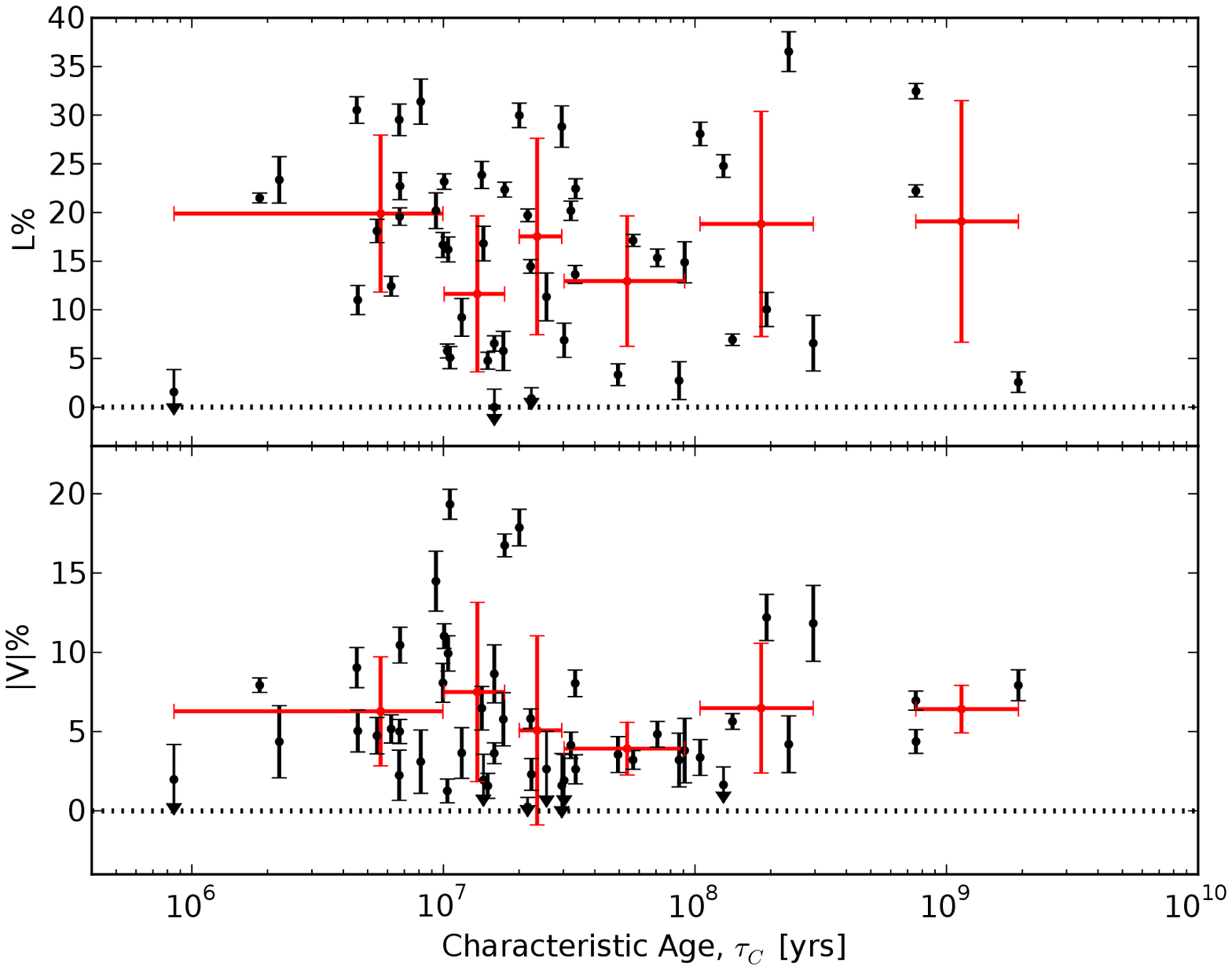}
\end{center}
\caption{Percentage of linear polarisation against the characteristic age in the top panel, and percentage of absolute circular polarisation against the characteristic age in the bottom panel. The black points represent the individual pulsar of our sample with 1 $\sigma$ error bars (the arrows imply an upper limit), while the \textbf{red} points and the vertical and horizontal bars represent the average over suitable groups of pulsars, the scatter and the range of age involved in the computation of the mean, respectively.}
\label{tau}
\end{figure}

In our sample, the percentage of the linear polarisation, $L\%$, ranges from a few  percent to almost 40\%. However, only two of the sources (PSR~J1614--3846 and PSR~J1705--6135) approach the aforementioned upper limit: the mean of $L\%$ is $\sim$16. \\
The dependence of $L\%$ on the pulsar spin-down luminosity is:
\begin{equation}
\rm{\dot{E}} \simeq 3.95 \times 10^{31} erg~s^{-1} \left(\frac{\dot{P}}{10^{-15}}\right) \left(\frac{P}{s}\right)^{-1}
\end{equation}
where $\dot{P}$ is the spin period derivative \citep{lk05}, is reported by \citet{hx97a}, \citet{cmk01} and \citet{jw06}. These authors noticed that higher values of $\dot{E}$ gave higher values of $L\%$. This trend was better modelled by \citet{wj08}, who found that the correlation between the two quantities is not linear. 
They identified two main regions, a low $\dot{E}$ (less than $5\times 10^{33}$~erg~s$^{-1}$), low $L\%$ (less than $50\%$) area and a high $\dot{E}$ (more than $2\times10^{35}$~erg~s$^{-1}$), high $L\%$ (exceeding $50\%$) one, divided by a narrow transition zone. As can be seen from Figure~\ref{edot}, the results derived from our low $\dot{E}$ sample do not conflict with \citet{wj08}: all of the pulsars (except one) belong to the low $\dot{E}$ interval and show $L\%$ smaller than $40\%$. Moreover, no clear correlation of $L\%$ vs. $\dot{E}$ is present over the low $\dot{E}$ sample.
\\[5pt]
\indent In Figure \ref{tau} we show the characteristic age, $\tau_C$:
\begin{equation}
\tau_C = \frac{P}{2 \dot{P}} s
\end{equation}
plotted versus $L\%$ and $|V|\%$ (that is the percentage of the absolute circular polarisation) and compare our results with \citet{gl98}.
Since $\tau_C$ of the pulsars in our sample (except PSR~J1625--4913) exceeds 1~Myr,
according to the results of \citet{gl98}, values of $L\%$ around $20\%$ are expected.
Although a large degree of scatter is present in the sample, the average values of $L\%$ are in fact between $10\%$ and $20\%$.
Regarding the percentage of $|V|$, we find a less pronounced degree of scatter in the data, and generally lower values of the average $|V|\%$ with respect to the results of \citet{gl98}.
We expected values around $8\%$ for $10^6~\rm Myr\leq\tau_C\leq10^7~\rm Myr$, and slightly higher results for older ages. We instead find a generally flat trend when the values of $|V|\%$ are averaged over the six bins in $\tau_C$, into which our sample has been split. In particular, the average $|V|\%$ is $\sim6\pm3$ for pulsars with $\tau_C\geq10^7~\rm Myr$ yrs, fully compatible with the value of $\sim6\pm5$ for pulsars with $\tau_C\leq10^7~\rm Myr$ yrs.
\\[5pt]

\indent For the majority of the pulsars in our sample, we can recognise the presence of more than one component in the profiles. This is not unexpected since pulsars of an advanced age typically have more complicated profiles than younger objects \citep{ran83,lm88,ran93,jw06,kj07}.
\citet{kj07} attribute this evidence to the location of the emitting regions crossed by the line of sight, assuming that each of them corresponds to one component in the profile. In particular, at a fixed observing frequency, the radio emission in young pulsars should be produced from a limited range of altitudes above the neutron star surface. This range widens and descends to lower heights in the magnetosphere with increasing age of the pulsar. According to the model presented by \citet{kj07}, this naturally increases the number of emitting regions crossed by the observer line of sight, and hence the number of components in the profile. A large fraction of the profiles in our sample show a blended double, i.e. the superposition of two main components that ranges from barely distinguishable (as in PSR~J0807--5421) to well (as in PSR~J0912--3851) visible. There is also a tendency for the trailing component to be brighter than the leading one. According to literature (i.e. \citealt{ran83}), a double component profile should indicate a mainly conal emission. Emission structures that are not well-defined are also observed, the clearest example of which is for PSR~J1534--4428. Emission bridges are also exhibited among otherwise separated components, as in PSRs~J1627--5936 and J1710--2616. Given the relatively small ${\rm SNR}$ of the majority of the pulsars, it is not easy to distinguish the occurrence of multiple components from the case of pure double profiles. Nevertheless, some objects certainly show at least three components, e.g. PSRs~J1251--7407, J1607--6449 and J1802--3346.
\\[5pt]
\indent The linear polarisation, when present, follows the total intensity in the majority of the cases, although it often shows a general edge depolarisation that causes a narrowing in the polarisation profile, as illustrated in PSRs~J1517--4636 and J1811--4930. The phenomenon of the linear depolarisation is usually explained via the superposition of two emission modes that are in competition in pulsars \citep{scr84}.
\\[5pt]
\indent The circular polarisation profiles are often barely visible, but show some cases of change in handedness between the components (as in PSRs~J0807--5421, J1612--5805, J1627--5936) or across the profile (as in PSRs~J1346--4918 and J1716--4711).
\\[5pt]
\indent As mentioned in \S2, in those pulsars (9 over the total sample of 34 objects for which ${\rm RM}$ has been determined) that show more than one peak in the linear polarisation profile, we separately fit for the ${\rm RM}$ component by component. In the majority of the cases, we obtained fully compatible (at 1 $\sigma$) ${\rm RM}$ values. In PSRs~J0912--3851 and J1629--3636 the agreement is marginally accomplished only by the overlap of the extremes of the respective 1 $\sigma$ uncertainties intervals. However, this is expected on a statistical bases given the available sample of 9 sources.

\subsection{Pulsars ${\rm RM}$ and Galactic Magnetic Field}

\begin{figure}
\begin{center}
\includegraphics[angle=0,scale=0.31]{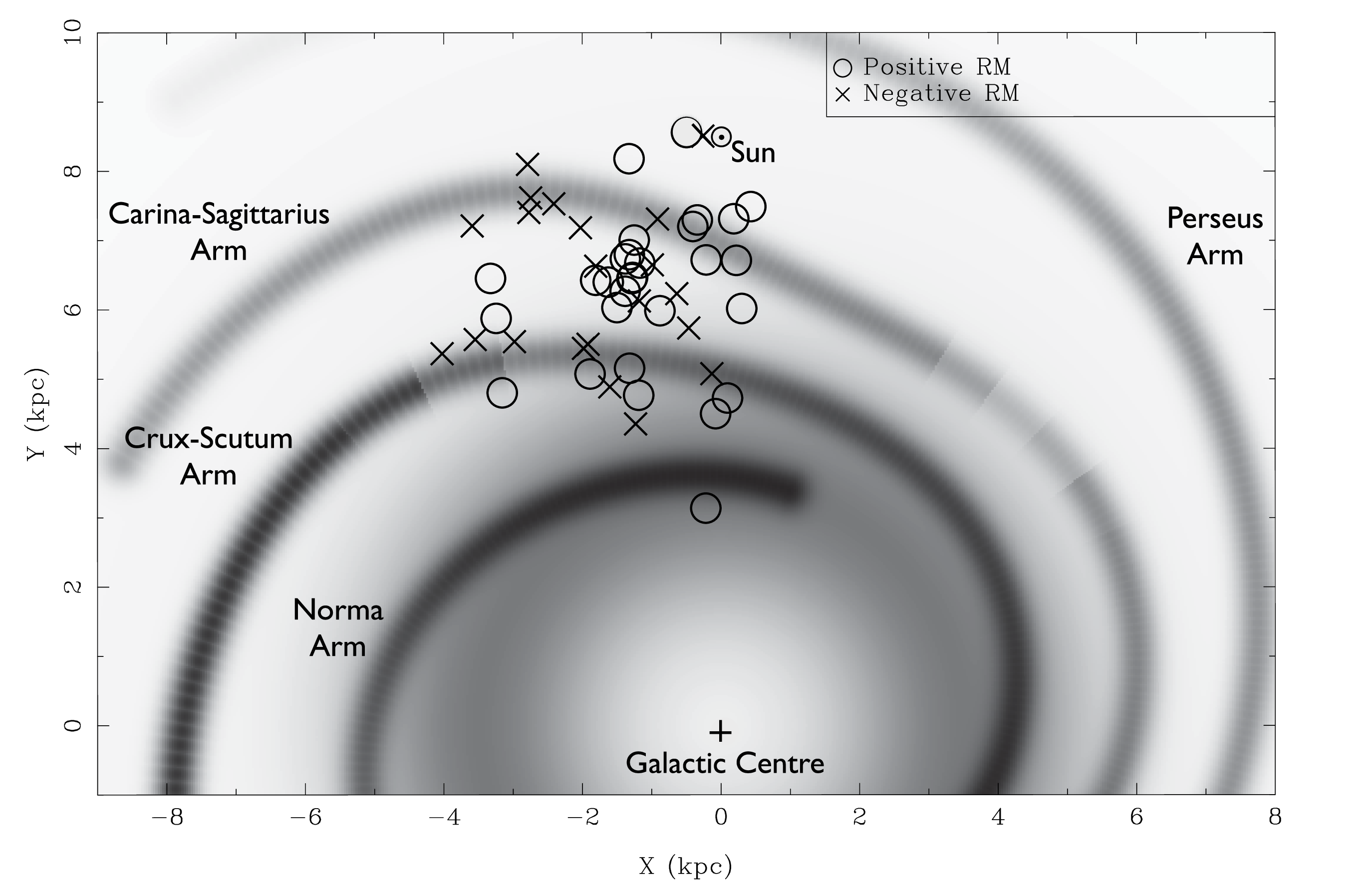}
\end{center}
\caption{A scheme of the Milky Way seen from the North Galactic Pole. In dark grey are shown the galaxy arms as from \citet{tc93}. The symbol $\odot$ indicates the Sun, the circles and the crosses indicate the pulsars of our sample with positive and negative values of ${\rm RM}$, respectively. The distances of the pulsars have been computed using the NE2001 electron model \citep{cl02}, prone to errors on the calculated distance up to the $30\%$}
\label{Fig:rm}
\end{figure}

We have collected all the sources discovered so far by the HTRU southern survey and having a measured value of ${\rm RM}$. The list totals 51 pulsars, resulting from the present work, as well as from \citet{htruplan,htruIV} and \citet{htruVII}. These values can be used to obtain an estimate of the average intensity and sign of the projection of the Galactic magnetic field vector ($<\rm{B}_{||}>$) along the 51 lines of sight to the pulsars. In fact, the ${\rm RM}$ is defined as:
\begin{equation}
\rm{RM}=\frac{\it{e}^3}{2\pi \it{m_e}^2 \it{c}^4} \int_{0}^{d}\it{n_e}(\it{l})\it{B}_{||}(\it{l})d\it{l}
\end{equation}
where $e$ is the electron charge, $m_e$ is the electron mass, $d$ is the distance between the emitting object and the observer, $n_e$ is the electron column density and B$_{||}$ is the projection of the magnetic field vector along the line of sight. Since the ${\rm DM}$ is defined as:
\begin{equation}
\rm{DM}=\int_{0}^{d} \it{n_e}(\it{l})d\it{l}
\end{equation}
is it possible to obtain $<\rm{B}_{||}>$ as:
\begin{eqnarray}
<\rm{B}_{||}>&& = 1.232 \frac{\int_{0}^{d}n_e(l)\rm{B_{||}}(l)dl}{\int_{0}^{d}n_e(l)dl}\nonumber \\
&&=1.232\left(\frac{\rm{RM}}{\mathrm m^{-2} \mathrm {rad}}\right)\left(\frac{\rm{DM}}{\mathrm {cm}^{-3} \mathrm {pc}}\right)^{-1}\mu \mathrm G.
\end{eqnarray}\label{eq:avB}
\noindent
The resulting values of $<B_{||}>$ are reported in the second last column of Table \ref{Tab:RMpar}. For each of the considered objects, we also derived a measurement of the distance (see Table \ref{Tab:RMpar}) using the ${\rm DM}$ value of each object and the NE2001 electron density model \citep{cl02}. Assuming these distances, all the selected pulsars are located within $2$ kpc in height from the Galactic plane and thus the lines of sight to all of them are expected to be useful to investigate the behaviour of the Galactic magnetic field in the proximity of the Galactic disk \citep{nj08}. In Figure \ref{Fig:rm} we have reported the positions - projected onto the Galactic plane - of the objects of our sample.
\\[5pt]
\indent Our sample does not support the hypothesis suggested by \citet{val05} of a prevailing counter-clockwise direction of the Galactic magnetic field in an annulus included between $4$ and $6$ kpc from the Galactic centre and a prevailing clockwise direction outside the annulus. First, looking at Figure  \ref{Fig:rm} it is evident the occurrence of opposite signs for the values of ${\rm RMs}$ for many pairs of pulsars which are very close to each other. As already pointed out by other authors (e.g. \citealt{nj08}), this is an indication for variations of intensity and direction of the Galactic magnetic field also over small scales. To be more quantitative, we also computed (as first suggested by \citealt{ls89}), the average intensity of the magnetic field in the intermediate region between pairs of pulsars:
\begin{equation}
<\rm{B_{||}>_{d_1 - d_2}} = 1.232 \frac{\rm{\Delta RM}}{\rm{\Delta DM}} \mu \mathrm G
\end{equation}
where d$_1$ and d$_2$ are the distances of the two sources from the Sun and $\Delta{\rm RM}$ and $\Delta{\rm DM}$ are the differences between the ${\rm RM}$ and the ${\rm DM}$ values of the two considered pulsars, respectively. In doing that, we followed the prescription of  \citet{nj08}, i.e. investigating pairs of pulsars the projected positions of which are closer than $5$\degree in Galactic longitude. For the limited range in distances and Galactic longitudes of our sample, a counter-clockwise direction for the Galactic magnetic field would correspond to a prevalence of positive values of $\mathrm{<B_{||}>_{d_1 - d_2}}$ for pairs located in the first Galactic quadrant (Galactic longitudes between $0$\degree and $90$\degree) and a prevalence of negative values of $\mathrm{<B_{||}>_{d_1 - d_2}}$ for pairs in the fourth Galactic quadrant. At variance with the expectations of the model of \citet{val05}, no trend is recognisable in our sample. In particular, within the annulus mentioned above, the values of $\mathrm{<B_{||}>_{d_1 - d_2}}$ for 6 pairs of pulsars are compatible with a counter-clockwise direction of the Galactic magnetic field, whereas a clockwise direction is preferred on the basis of 6 other pairs. Similarly, the results for 24 pairs of pulsars would favour a clockwise direction for the region outside the annulus, whilst the consideration of 26 other pairs would suggest the opposite direction\footnote{The total number of pairs is larger than the number of pulsars of our sample since few pulsars of the sample enter more than one pair.}.
\\[5pt]
\begin{figure}
\begin{center}
\includegraphics[angle=0,scale=0.42]{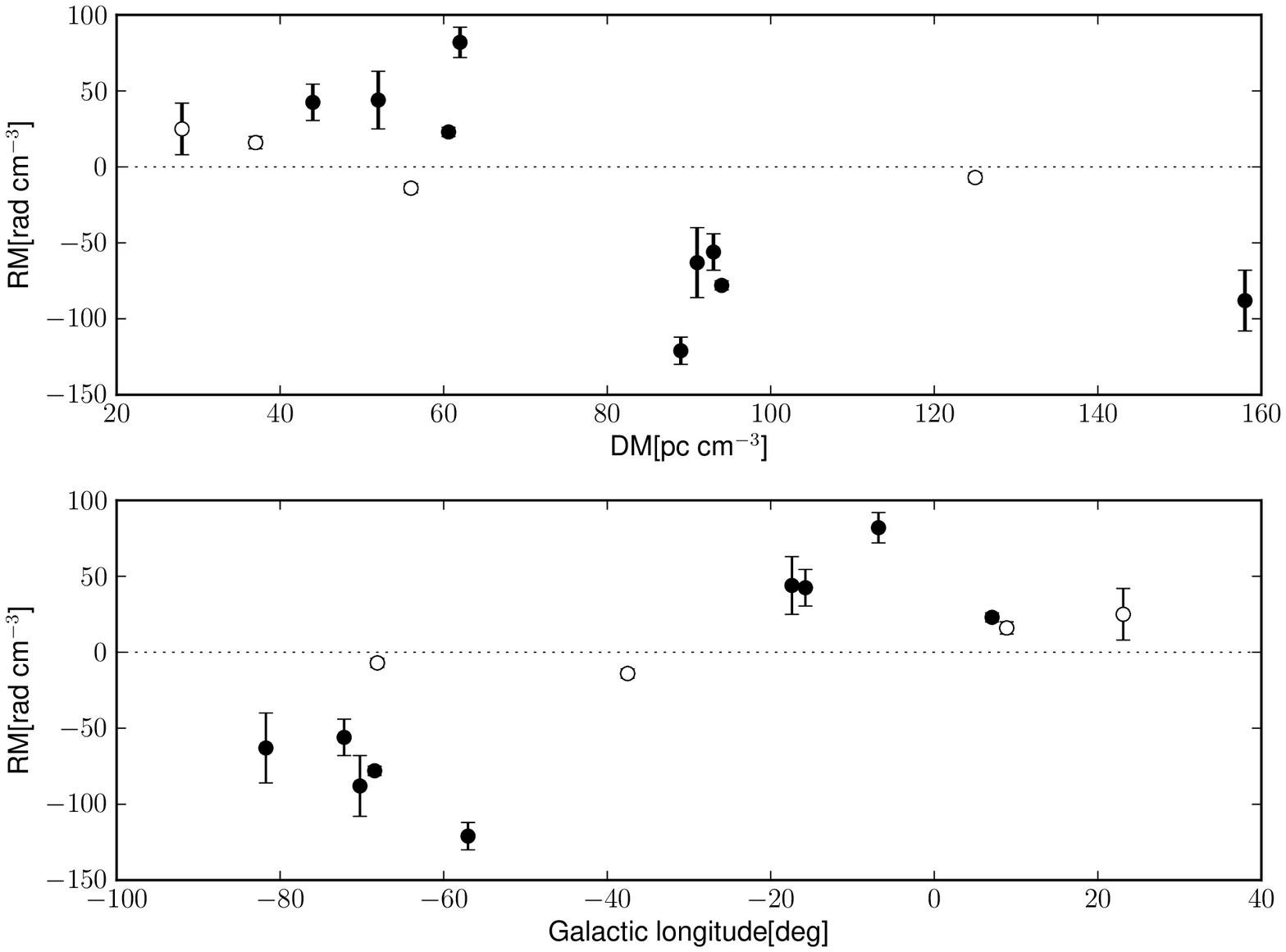}
\end{center}
\caption{In the top panel are shown the ${\rm RM}$ values for the pulsars the projected positions of which are in agreement with their belonging to the Carina-Sagittarius arm, plotted in function of their ${\rm DM}$. In the bottom panel are shown the ${\rm RM}$ values for the same pulsars plotted in function of their Galactic longitude. Empty circles indicate pulsars with positive Galactic latitude, whereas filled circles are associated with pulsar located at negative Galactic latitude. 1 $\sigma$ error bars are overlapped to the data points. Note that the error bar associated with some of the data points are too small to be visible.}
\label{Fig:rmdm}
\end{figure}

\indent We also performed a preliminary investigation of the compatibility of our sample with the model of \citet{hml+06},  which states the occurrence of a  counter-clockwise direction for the Galactic magnetic field along the arms and an opposite direction of that in the inter-arm regions. Given the distances of the pulsars in our sample (typically spanning the range $1-4$ kpc) and the relatively small number of available objects, our investigation focused on the case of the closest arm, i.e. the Carina-Sagittarius arm (see Figure \ref{Fig:rm}).  We then selected those pulsars whose projected position is compatible with them belonging to the area of the Carina-Sagittarius arm or close (within 0.5 kpc) to that. That choice left us with 13 objects, whose Galactic longitudes span the range between $-82$\degree and $23$\degree.  Adopting the same criteria as mentioned above, 8 pairs of pulsars can be selected in this region and the related values of $\mathrm{<B_{||}>_{d_1 - d_2}}$ measured. It results a prevalence of pairs (6 vs 2) indicating a counter-clockwise direction of the Galactic magnetic field along the Carina-Sagittarius arm, nominally in agreement with the model of \citet{hml+06}. Figure 5 indicates that some large scale ordered component of the Galactic magnetic field can indeed be present in the Carina-Sagittarius arm, being reflected in the overall trend for the ${\rm RM}$ values, which change from positive to negative values with increasing values of ${\rm DM}$. Unfortunately our sample is not suitable to test the detailed dependence of ${\rm RM}$ vs ${\rm DM}$ inferred by \citet{hml+06} for the objects belonging to the Carina arm and having $\rm{DM}<200$ pc cm$^{-3},$ i.e. $\rm{RM}\propto-0.6~\rm{DM}$, for the pulsars with Galactic longitudes between $-76$\degree and $-68$\degree. In fact Han et al. used only pulsars at Galactic latitude less than $|8|$\degree, which are too rare in our sample (resulting from a survey at intermediate and high Galactic latitudes) for a meaningful comparison. However, Figure 5 also shows that the status of the magnetic field in the Carina-Sagittarius arm is more complex than described by the relatively simple model of Han et al., with a large scatter of values of ${\rm RM}$ for similar values of ${\rm DM}$ and the trend in Figure 5 which is much more evident for the pulsars below the Galactic plane than for the ones at positive Galactic latitudes. As a consequence, additional components in the Galactic magnetic field are likely needed, like those investigated by \citet{nj08}. A significant improvement in the modelling is expected when the sample presented here will be complemented by the discoveries resulting from the low-latitude part of the HTRU survey (Ng et al., in preparation).\\

\section{Summary}

We have presented a polarimetric analysis of 49 long-period pulsars discovered as part of the HTRU southern survey. We were able to compute the ${\rm RM}$ for 34 of them, while 9 objects show almost no polarised signal.\\[5pt]
\indent We found that the percentage of $L$ among the pulsars in the sample is mainly around $15-20\%$, in agreement with previous studies \citep{gl98,wj08} for sources with $\dot{E}$ lower than $5\times10^{33}$erg~s$^{-1}$ and a characteristic age larger than 1 Myr. In addition, the mean degree of $|V|$ is roughly compatible with expectations, although it does not show any sign of a minimum in the range of ages between $10^6$ and $10^7$ years as in \citet{gl98}. This can be due to the smaller number of pulsars in our sample. However, we believe that these differences are not significant.\\[5pt]
\indent For the majority of the total power profiles, we recognised the presence of more than one component, as expected for a sample of ``old'' pulsars. In particular, we note the frequent occurrence of blended-double shaped profiles. According to the literature (i.e., \citealt{ran83}), this is an indication of a conal emission. The linear polarisation profiles often mirror the total intensity shape, although the former are almost always narrower than the latter, as already noticed in \citet{ran83}  while the fainter circular polarisation profiles show a handedness reversal in a few cases. The ${\rm PA}$ swings vary from flat behaviours to mode jumps and some occurrences of RVM-like swings. For two of the analysed pulsars, the fit for the swing yields some geometrical constraints on the radio-beam. Both appear to be almost aligned ($\alpha<<45\degree$) rotators.\\[5pt]
\indent We have also carried out a preliminary analysis of the Galactic magnetic field resulting from the available sample of pulsars discovered so far in the HTRU southern survey that have a computable ${\rm RM}$ value, and we studied the implications of the results we obtained. The data do not support the model presented by \citet{val05}, whereas there is some agreement with the one proposed by \textbf{\citet{hml+06}} and \citet{nj08}. In contrast with \citet{val05}, \citet{hml+06} and \citet{nj08} claim that the Galactic magnetic field has a counter-clockwise direction in the arms and a clockwise direction in between. However, given the limited number of pulsars in our sample and their proximity to the Sun, it is difficult to put significant constraints on more complicated large scale models for the Galactic magnetic field for the time being. 

\section*{Acknowledgements}
The Parkes radio telescope is part of the Australia Telescope which is funded by the Commonwealth of Australia for operation as a National Facility managed by CSIRO. Part of this research was carried out at the Jet Propulsion Laboratory, California Institute of Technology, under a contract with the National Aeronautics and Space Administration. CT also thanks Delphine Perrodin for her help.

\bibliographystyle{mn2e}
\bibliography{journals,modrefs,psrrefs,crossrefs}

\bsp
\label{lastpage}
\end{document}